\documentclass[aps,amsmath,twocolumn,amssymb,floatfixng,showpacs,superscriptaddress,footinbib,longbibliography]{revtex4-1}
\pdfoutput=1
\usepackage[dvips]{graphics}
\usepackage{bm}
\usepackage{float}
\usepackage{epsfig}
\usepackage{enumerate}
\usepackage{subfigure}
\usepackage{amsmath}
\usepackage{color}
\usepackage{braket}
\usepackage{graphicx}
\usepackage{float}
\usepackage[colorlinks=true,linktoc=page,linkcolor=red,citecolor=blue,urlcolor=magenta]{hyperref}

\newcommand\bea{\begin{eqnarray}}
\newcommand\eea{\end{eqnarray}}
\newcommand\beq{\begin{equation}}  
\newcommand\eeq{\end{equation}}

\begin{document} 

\title{Adiabatic charge transport in extended SSH models} 
 \author{Dharana Joshi}
%\email{xxx@hyderabad.bits-pilani.ac.in}
\affiliation{Department of Physics, BITS Pilani-Hyderabad Campus, Telangana 500078, India}
\author{Tanay Nag}
\email{tanay.nag@hyderabad.bits-pilani.ac.in}
\affiliation{Department of Physics, BITS Pilani-Hyderabad Campus, Telangana 500078, India}

%-------------------------------------------------------------------------------------------------------------------------------------------------
\begin{abstract}
We explore the topological properties of extended SSH  models, considering four sub-lattices in a unit cell and  second-nearest-neighbor intercell hopping for SSH4 and SSH long-range (SSHLR) models, respectively. The additional tuning parameters cause the
SSH4 (SSHLR) model to host chiral symmetry protected two (two and four) zero-energy modes producing a richer phase diagram that we characterize by momentum space, periodic-bulk and open-bulk real space winding numbers. We introduce time to study charge transport in the  periodically driven SSH4 and SSHLR models under the adiabatic limit. We remarkably find that the whole parameter space turned topological for a certain choice of the remaining parameters
leading to always finite quantized value of pumped charge at the end of a complete cycle. Considering time as another variable,  we characterize these new phases of the driven models by  momentum space Chern number, periodic-bulk and open-bulk real space Bott index. We also investigate the time evolution of pumped charge for these models and connect it with the intriguing windings of the mid-gap energy levels with time. Interestingly, the maximum value of Chern number or Bott index for the driven models is more than that of the winding number associated with the static model indicating the fact that there exist more zero-energy modes during the full course of a driving cycle compared to the underlying static models. We further extend our study to the quantum metric where the fluctuations in the above quantity can identify the presence of a topological phase boundary. 
\end{abstract}
%---------------------------------------------------------------------------------------------------------------------------------------------------

\maketitle

%-------------------------------------------------------------------------------------------------------------------------------------------------
\section{INTRODUCTION}
\label{s1:intro}
%-------------------------------------------------------------------------------------------------------------------------------------------------
%%%% symmetry preserved topological phase %%%%
The mathematical idea of topology where a given topological class of an object does not change under 
continuous deformations without puncturing the surface,  is recently promoted to the field of condensed matter. Two phases can be topologically distinct (identical) if the bulk gap closes (remains finite) under continuous variation of the band parameter \cite{nakahara2018geometry,mukhi1990introduction,Shou-Cheng11,bernevig2013topological}. In this way,  
the paradigm of symmetry-broken phases, following the Landau theory of phase transition for local order parameters, is revolutionized by the  symmetry-protected topological phases  hosting boundary modes within the bulk gap. Interestingly, there exist topological phases in one-dimensional (1D) model while the long-range order is not supported following the  Mermin-Wigner theorem \cite{Halperin2019,Guo2016}. This leads to the fact that symmetry is a key resource to protect the topological order which is a non-local order.  
Going deep into the symmetry aspect, the  topological matters are classified in ten-fold ways depending on the time-reversal, particle-hole and chiral symmetries while the different types of topological invariants are employed to characterize them in different dimensions \cite{Altland97,Chiu16, Ryu_2010}.

%%%%%%%%%%  historical background and brief summary of various topological phases %%%%%%
The experimental discovery of integer quantum Hall effect provides a foundation for topological   phases of matter where two-dimensional (2D) electron gas under a perpendicular magnetic field is shown to exhibit insulating bulk but conducting edges
with quantized transverse Hall conductivity \cite{Klitzing86,Thouless82}. In order to explain the integer quantum Hall effect, 
time reversal symmetry broken  Chern insulator models \cite{jotzu2014chern} are introduced theoretically among which Haldane model stands as the most earliest model \cite{Haldane88}.  This leads  to furhter interesting phenomena such as quantum spin Hall effect theoretically \cite{Kane05} and experimentally \cite{brune2012spin} where the time reversal symmetry protected spin-polarized edge modes are observed in the presence of intrinsic spin-orbit coupling. Going beyond the above mentioned first-order topological phases, the crystalline symmetries play major role in exploring the  higher-order topological phases \cite{Benalcazar17,schindler2018higher,ghosh2023generation, Ghosh21,Arouca24}.    The experimental demonstration of the topological phase of matter is not only restricted to the solid state system \cite{Stuhl15,Meier2016}, rather it is vastly explored in the meta-material platforms \cite{Rechtsman2013,Peano15,Lu2016,PhotonicChen,XueAcousticKagome,Xiao2015}.

%%%%%%%%% SSH model and winding number %%%%%%%%%%%%%
To this end, we discuss the wide variety of topological phases in odd spatial dimension where zero-energy end localized modes are protected by chiral symmetry of the model \cite{Liu2023,Matveeva2023}.  The Su-Schrieffer-Heeger (SSH) model \cite{SSH79,Meier2016}, describing electrons hopping on a one-dimensional lattice with staggered hopping amplitudes, is a foundational model for understanding 1D topological insulators \cite{hasan2010colloquium,qi2011topological,Asb_th_2016,jin2024topological,Nava23,Cinnirella24}. In this direction,  $p$-wave Kitaev chain is another tight-binding model yielding topological superconductivity \cite{Kitaev_2001,DeGottardi13,DeGottardi13b,Rajak14}. The bulk boundary correspondence 
serves as a tool to predict the number of zero-energy edge states based on the bulk invariant which is the winding number for the above models \cite{jackiw1976,Rhim18,atala2013direct}. 
The topological characterization is based on the relative winding between the two components of a $2$-band Hamiltonian in a parametric space. For a multiband model having more than $2$-bands, one has to project the Hamiltonian in the chiral symmetry basis first and then compute the relative winding of the anti-diagonal block of the projected Hamiltonian \cite{Lin2021}. Apart from the momentum space analysis, the real space winding number using the position operator method has been computed in the presence of disorder \cite{Lin2021,Ghosh2024,Zhang_2022,Tang2020,Song2019,He2020,Kivelson82}. It is important to note that the winding number is inherently related to the bulk polarization, obtained from  Wilson loop using  the  Bloch wave functions   \cite{Mondragon-Shem14,Chiu16,Marzari12}. 
The experimental realization of winding numbers includes the interferometric architectures \cite{Flurin17},
scattering measurements \cite{Barkhofen17, Eric18}.

%%%%%%%%%% Charge transport and Bott index and quantum metric

The 1D SSH model is found to be very useful to understand the charge transport when it is adiabatically evolved with time \cite{NIU91,Citro_2023,Kumar2021}. The quantized charge transport is a fascinating feature of the driven SSH model where the on-site potential and staggered hopping terms are varied; this is referred to Rice-Mele charge pump \cite{shun2018topological}. The introduction of time converts the 1D circular parameter space into a 2D torus enabling us to characterize the topology of the quantized charge pump in terms of Chern number \cite{Fukui2005,Agrawal2022,Saha21}. It is an interesting phenomena that conductance becomes quantized in the absence of an external magnetic field
in the Rice-Mele charge pump while the 2D Chern insulator phases require an external magnetic field such as Haldane model.  Importantly, the time reversal symmetry is broken for both the above cases validating the Chern number to be able to characterize the phase. Going beyond the momentum space version of Chern number,  the  disordered 2D Chern insulators are characterized  in terms of Bott index where the Berry curvature is represented in the real space  \cite{bellissard1995noncommutative,Prodan_2010,Prodan_2011,Bianco11,Loring_2010,hastings2010almost}. It is important to note that the Berry curvature is anti-symmetric part of quantum geometric tensor while the symmetric part known as,  quantum metric is a useful tool to measure the quantum distance in the parametric space \cite{Tan2019,Panahiyan2020,Ma_2013,Ma2014,Ozawa2021,Cheng_2024,Zeng_2024,Zhang_mar}.

%%%%%%%%%%%%%% motivations, questions and our findings%%%%%%%%%%%%%%
Given the fact that static SSH model and adiabatically driven SSH model are extensively studied, we here focus on the extended version of the model with additional tuning parameters \cite{Maffei_2018,Han_2021}. For the SSH and extended SSH models, much has been studied about the interacting part \cite{Schobert21,Padhan24,Feng2022,Jin23}, but little has been studied about the charge transport in the non-interacting part. This yet simple instant conceives various  interesting aspect and we seek the answers to the following questions.  What are the roles of additional parameters in charge transport in the presence of adiabatic drive? How can one understand the underlying phases in terms of the Bott index? Does quantum metric help in identifying the phase boundaries?  Our work sheds lights on the new topological phases  mediated by the  additional parameters  as compared to the regular SSH model. We characterize these phases using momentum space as well as real space winding numbers. The adiabatic drive introduces quantized charge transport after a complete cycle for a certain choice of parameters resulting in only topological phases for the remaining parameter space. We further validate this finding by computing the momentum space Chern number and real space Bott index where time plays the role of another cyclic parameter. We further explore the behavior of quantum metric in different phases while its fluctuations identify  the phase boundaries. 

%%%%%%%%%%%%%% organization %%%%%%%%%%%%%%%%%%
The rest of the manuscript is organized in the following manner. We discuss the static version of the extended SSH models in Sec. \ref{s2:model}. We explore their topological phases in terms of the winding numbers in Sec. \ref{s3}. The time-dependent extended SSH models are demonstrated in Sec. \ref{s4}. We explore the charge transport and topological phases using Chern number and Bott index in Sec. \ref{s5}. In Sec. \ref{s6} we present concluding remarks. 

%%%%%%%%%%%%%%%%%%%%%%%%%%%%

%-------------------------------------------------------------------------------------------------------------------------------------------------
\section{Model Hamiltonian}
\label{s2:model}
%-------------------------------------------------------------------------------------------------------------------------------------------------
\begin{figure}
    \centering
    \includegraphics[width=0.5\textwidth]{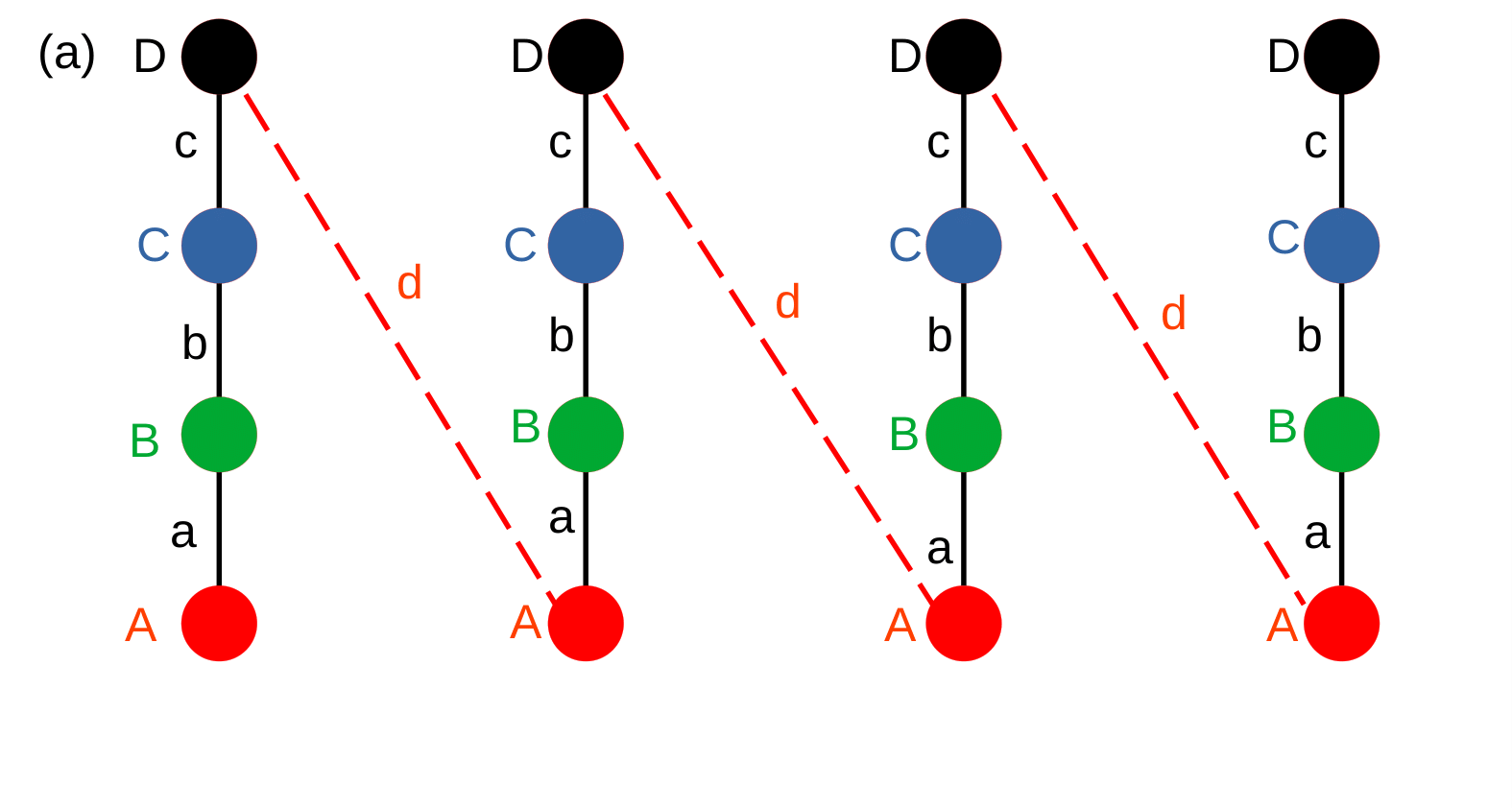}
    \includegraphics[width=0.5\textwidth]{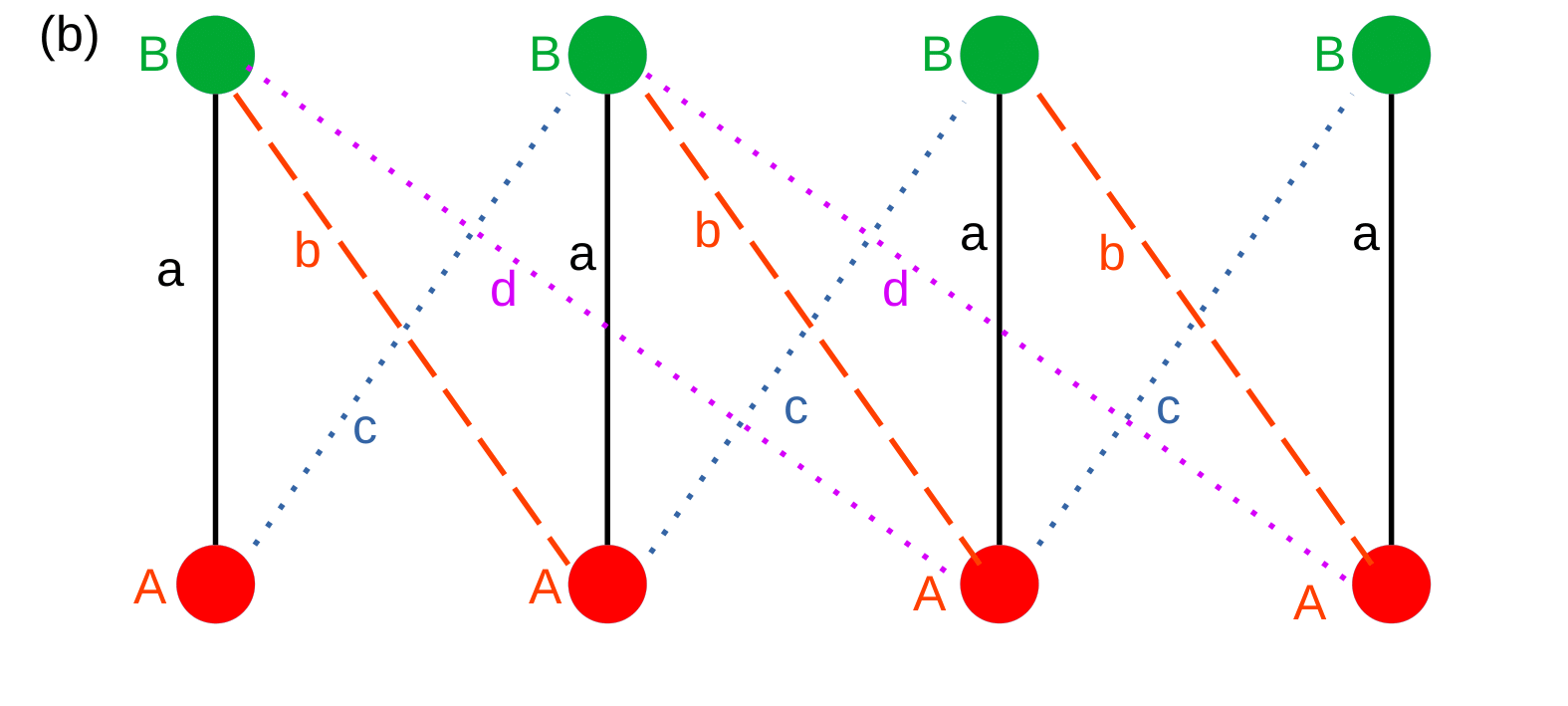}
    \caption{Schematic of the SSH4 (a) and the SSHLR (b) models with  four and two sub-lattices, respectively. The intra (inter)-cell hoppings are designated by solid (dashed and dotted) lines.}
    \label{fig:ssh}
\end{figure}

\subsection{SSH4 Model}
\label{ssec:ssh4}
Going beyond the realm of $2$-band SSH model, we consider $4$-band SSH model, referred to as SSH4 model in the literature, with unequal hopping amplitudes among four different sub-lattices \cite{Obana19,LEE202296,Maffei_2018}. One unit cell comprises of $A$, $B$, $C$, and $D$ sub-lattices where
$a$, $b$, and $c$, ($d$) represent 
intra (inter)-cell nearest neighbour hoppings taking place between $A_n \leftrightarrow B_n$, $B_n \leftrightarrow C_n $, and $ C_n\leftrightarrow D_n$ ($A_{n+1} \leftrightarrow D_n$), here $n$ represents the unit cell index, see Fig. \ref{fig:ssh} (a). The momentum space Hamiltonian, describing the SSH4 model in the basis $(C^{\dagger}_{A,k},C^{\dagger}_{B,k},C^{\dagger}_{C,k},C^{\dagger}_{D,k})$, is given by 
\begin{equation}
H(k)=
\begin{bmatrix}
 0 & a & 0 & de^{-ik} \\
 a & 0 & b & 0 \\
 0 & b & 0 & c \\
 de^{ik} &0 & c & 0  
  \end{bmatrix}
\label{eq:ssh4}  
\end{equation}
The energy of the bands are given by $E^{1,2,3,4}_{k}=\pm (a^2+b^2+c^2 +d^2 e^{ik} \pm (-4 a^2c^2 -4b^2d^2 e^{ik}+ (a^2+b^2+c^2+d^2 e^{ik})^2+ 4 abcd (e^{-ik}+e^{2ik}))^{1/2})^{1/2}$. For the ease of analysis, we consider $a=b=1$. We find that the gap closes at $k=0$ and $\pi$ leading to the phase boundaries $c=d$ and $c=-d$, respectively. In general,  this model Eq. (\ref{eq:ssh4}) supports a topological (trivial) phase when $ac < bd$ ($ac > bd$).  Note that for $a=c$ and $d=b$ the SSH4 reduces to the usual SSH model.

Similar to the SSH model, SSH4 model respects the 
chiral symmetry $\mathrm{ \Gamma} = \sigma_0 \sigma_z$ such that
$\mathrm{\Gamma} H(k) \mathrm{\Gamma} = -H(k)$
and falls into BDI class.  The Hamiltonian acquires block anti-diagonalized  form once it is written in the eigen-basis of 
$\Gamma$ as follows:  \begin{equation}
\tilde{H}(k)=
\begin{bmatrix}
  0 & h^\dagger  \\
  h & 0 
  \end{bmatrix} =
\begin{bmatrix}
 0 & 0 & a & de^{-ik} \\
 0 & 0 & b & c \\
 a & b & 0 & 0 \\
 de^{ik} &c & 0 & 0  
  \end{bmatrix}
\label{eq:ssh4_block}  
\end{equation}
The chiral symmetry allows us to characterize the topological phase in terms of the winding number which we will examine the following sections.

%-------------------------------------------------------------------------------------------------------------------------------------------------
\subsection{Long Range SSH Model}
\label{s2:ss2}
%-------------------------------------------------------------------------------------------------------------------------------------------------

Having demonstrated the SSH4 model, now we discuss another 
variant of $2$-band SSH model in the presence of hopping beyond the first nearest neighbour\cite{Maffei_2018,Beatriz2018}. Allowing the hopping between two different sub-lattices, this SSH long-range (SSHLR) model hosts hoppings between $B_n \leftrightarrow A_{n+1}$,  $A_n \leftrightarrow B_{n+1}$ and $B_n \leftrightarrow A_{n+2}$ as represented by hopping amplitudes $b$, $c$ and $d$, respectively, see Fig. \ref{fig:ssh} (b). The momentum space Hamiltonian, describing the SSHLR model in the basis $(C^{\dagger}_{A,k},C^{\dagger}_{B,k})$, is given by  
\begin{equation}
H(k)=
\begin{bmatrix}
 0 & a+ be^{-ik}+c e^{ik}+d e^{-2ik} \\
 a+be^{ik}+ce^{-ik}+de^{2ik} & 0    
  \end{bmatrix}
\label{eq:sshlr}  
\end{equation}
The energy of the bands are given by $E^{\pm}_{k}=\pm (a^2+b^2+c^2+d^2+2(ab+ac+bd)\cos k+ 2(bc+ad)\cos 2k + 2cd \cos 3k)^{1/2}$. For the ease of gap analysis, we consider $a=b=1$. The gap closes for momentum modes $k=0$, $2\pi/3$ and $\pi$  leading to critical lines $c=-2-d$, $c=1-d$ and $c=d$, respectively. 
This model Eq. (\ref{eq:sshlr}) has chiral symmetry $\Gamma H(k) \Gamma= -H(k)$ with $\Gamma=\sigma_z$. The   Hamiltonian $\tilde{H}(k)$ in the chiral basis acquires the same form as $H(k)$ which is already block anti-diagonalized. 
This model reduces to the original  SSH model once the long-range hopping terms $c$ and $d$ are set to zero. 
This model exhibits richer topological phases than the previous one due to its long-range nature which we will explore below. 

%-------------------------------------------------------------------------------------------------------------------------------------------------
\section{topological phases of time-independent models}
\label{s3}
%-------------------------------------------------------------------------------------------------------------------------------------------------
\subsection{Momentum space winding number}
\label{s3ss1}

In this section, we examine the phase diagrams, associated with SSH4 and SSHLR models in Eqs. (\ref{eq:ssh4}) and (\ref{eq:sshlr}), respectively. The chiral symmetry of the problem enables us to obtain the anti-diagonal blocks $h$ in $\tilde{H}(k)$ staring from a multi-band Hamiltonian. The winding number\cite{Maffei_2018,He2020} is defined in terms of the winding of the anti-diagonal block $h$ as follows
\begin{equation}
W =\frac{dk}{2\pi i} \int_0^{2\pi} dk \, {\rm Tr}\big[ h^{-1} \partial_k h\big]= 
\frac{dk}{2\pi i} \int_0^{2\pi} dk \, \partial_k \log\big[\det(h)\big]
\label{eq:winding1}
\end{equation}
One can also use $h^\dagger$ while computing the winding number that results in the change of sign only.  Instead of using the chiral basis, one can also alternatively compute the winding of $q$, obtained from the flattened Hamiltonian $Q=\sum_{E_n(k)>0}\ket{u_n(k)}\bra{u_n(k)}-\sum_{E_n(k)<0}\ket{u_n(k)}\bra{u_n(k)}$. Note that  $\tilde{H}(k)\ket{u_n(k)}=E_n(k)\ket{u_n(k)}$.  Winding number is given by 
\begin{eqnarray}
W &=& \frac{dk}{4\pi i} \int_0^{2\pi} dk \, {\rm Tr}\big[ \Gamma Q \partial_k Q\big] \nonumber \\
&=& \frac{dk}{2\pi i} \int_0^{2\pi} dk \, \big[ q^{-1} \partial_k q\big] \nonumber \\
&=&
\frac{dk}{2\pi i} \int_0^{2\pi} dk \, \partial_k \log\big[\det(q)\big] 
\label{eq:winding2}
\end{eqnarray}
with 
\begin{equation}
Q = \begin{pmatrix} 
0 & q^\dagger \\ 
q & 0 
\end{pmatrix}. \ 
\nonumber \\
\end{equation}
Interestingly, $Q$ is unitary as well as Hermitian matrix leading to the fact that $q^\dagger=q^{-1}$.  
There exist a connection between $q$ and $h$ which is the following: $q= U_A U^{\dagger}_B$ and $U_{A,B}$ are obtained from singular value decomposition of $h = U_A \Sigma U_B^\dagger$.

\begin{figure}
\includegraphics[width=0.48\textwidth]{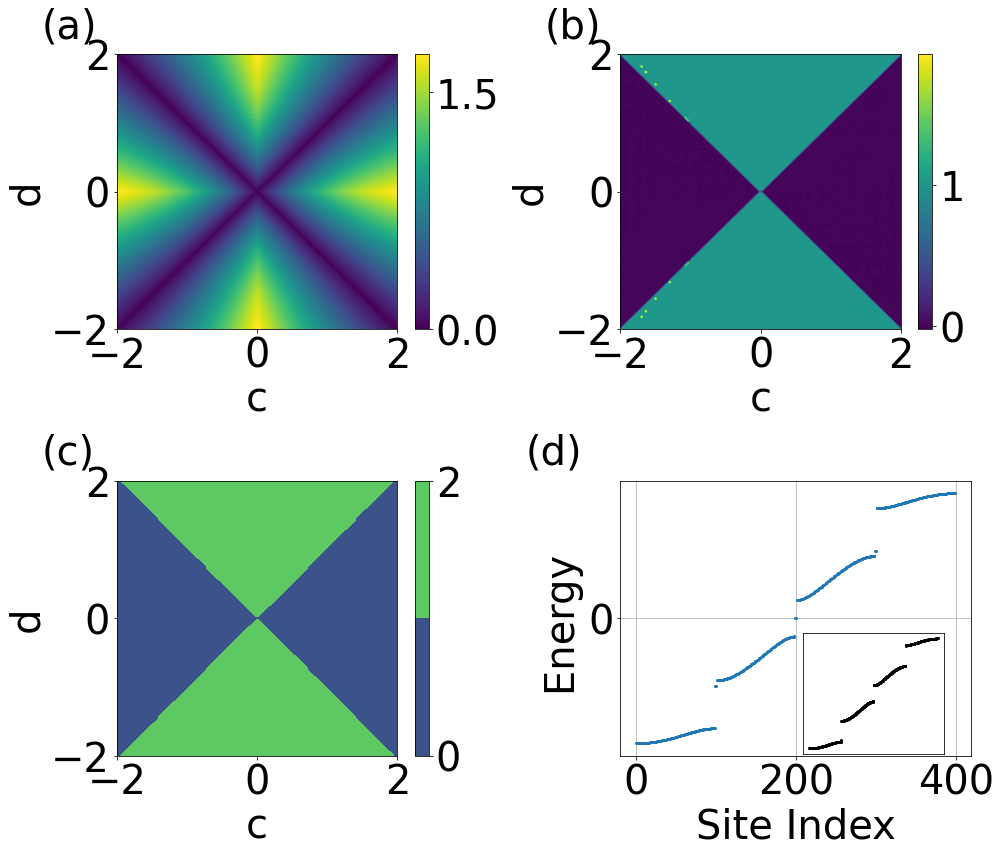}
\caption{We show (a) gap closing in momentum space, (b) winding number phase diagram,
(c) the number of zero energy modes,  and (d) the spectrum associated with single particle energy level for the SSH4 model Eq. (\ref{eq:ssh4}). We consider momentum space Hamiltonian for (a,b) and  OBC in real space for (c,d). We use $a=b=1$ for (a,b,c). We consider $(a,b,c,d)=(1,1,1,2)$ and $(1,1,2,1)$  for (d) and its inset, respectively.
}
\label{fig:ssh4_gap}
\end{figure}

\begin{figure}    
\includegraphics[width=0.48\textwidth]{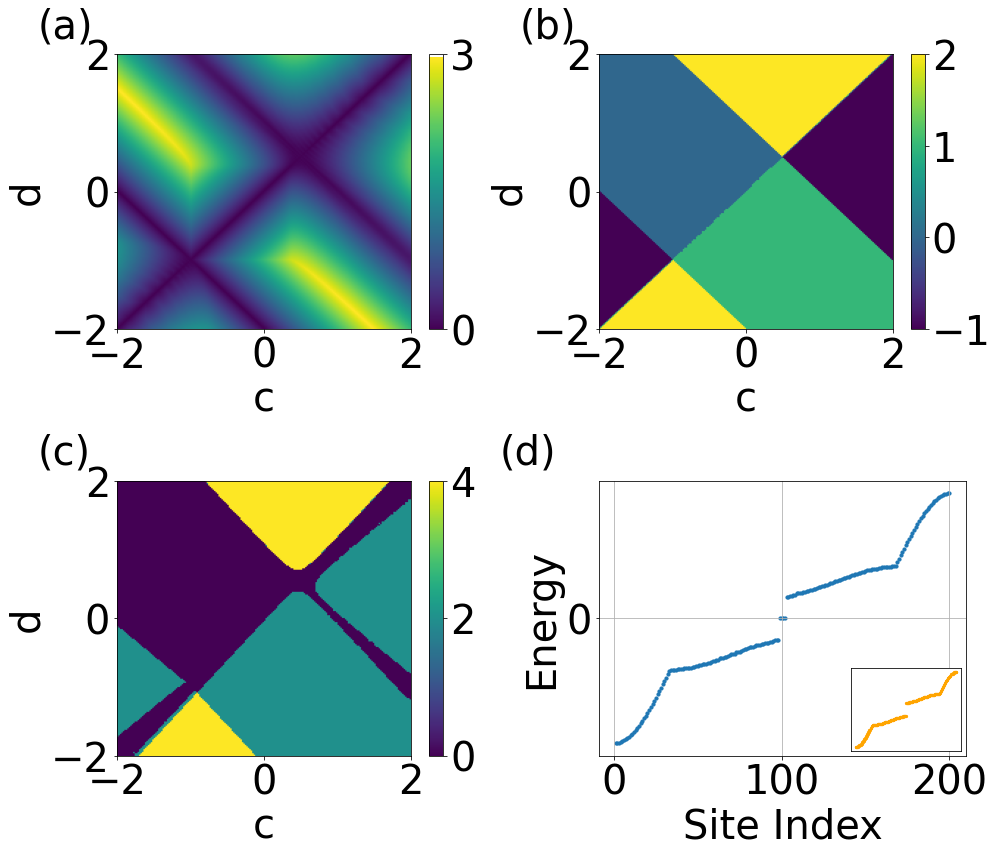}
\caption{We show (a) gap closing in momentum space, (b) winding number phase diagram,
(c) the number of zero energy modes,  and (d) dispersion associated with single particle energy level for the SSHLR model Eq. (\ref{eq:sshlr}). We consider momentum space Hamiltonian for (a,b) and  OBC in real space for (c,d). We use $a=b=1$ for (a,b,c). We consider $(a,b,c,d)=(1,1,1,2)$ and $(1,1,-1,0)$  for (d) and its inset, respectively.
}
\label{fig:sshlr_gap}    
\end{figure}

We demonstrate the profile of the bulk gap, associated with $H(k)$, over the $d$-$c$ plane in Fig. \ref{fig:ssh4_gap} (a) for SSH4 model where gapless lines indicate the separation between two gapped phases. In order to identify the topological nature of the gap, we compute the winding number, following Eq. (\ref{eq:winding2}),  from which the region $|d|> |c|$ (keeping $a=b=1$) is found to exhibit $W=1$, see Fig. \ref{fig:ssh4_gap} (b). We verify these findings from the real space Hamiltonian by plotting the number of zero energy modes under OBC as shown in Fig. \ref{fig:ssh4_gap} (c). Across the phase boundaries, the number of zero energy modes changes abruptly from $2$ to $0$ while $W$ shows a jump from $1$ to $0$. We depict one representative  energy dispersion of the topological [trivial] phase where the mid-gap [no mid-gap] zero energy states appear, see Fig. \ref{fig:ssh4_gap} (d) [(d) inset]. 
We repeat the same exercise for SSHLR model where we obtain a rich gap profile leading to 
a wide variety of topological phases with $W=-1,0,1$ and $2$ as shown in Figs. \ref{fig:sshlr_gap} (a,b). The phase boundaries are determined by the lines $d=-c+1$, $d=-c-2$ and $d=c$. 
The number of zero energy mode is found to be $4$ for $W=2$ in addition to $2$ and $0$ with $W=1$, and $0$, respectively, as observed in Fig. \ref{fig:sshlr_gap} (c). We demonstrate the mid-gap [no mid-gap] states  at zero energy for  the SSHLR model residing in a topological [trivial] phase, see Fig. \ref{fig:sshlr_gap} (d) [(d) inset].

% -------------------------------------------------------------------------------------------------------------------------------------------------

%-------------------------------------------------------------------------------------------------------------------------------------------------

%-------------------------------------------------------------------------------------------------------------------------------------------------

%-------------------------------------------------------------------------------------------------------------------------------------------------

\begin{figure}
    \centering
    \includegraphics[width=0.48\textwidth]{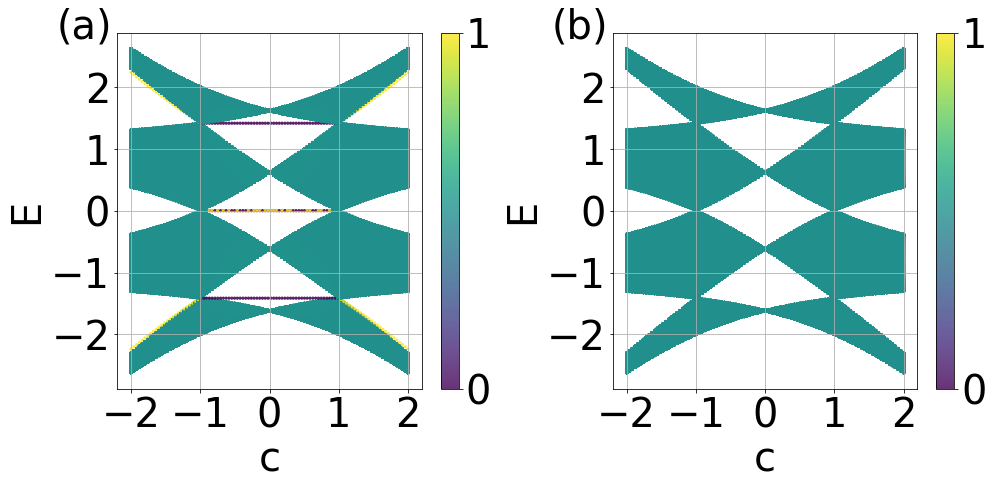}
    \includegraphics[width=0.48\textwidth]{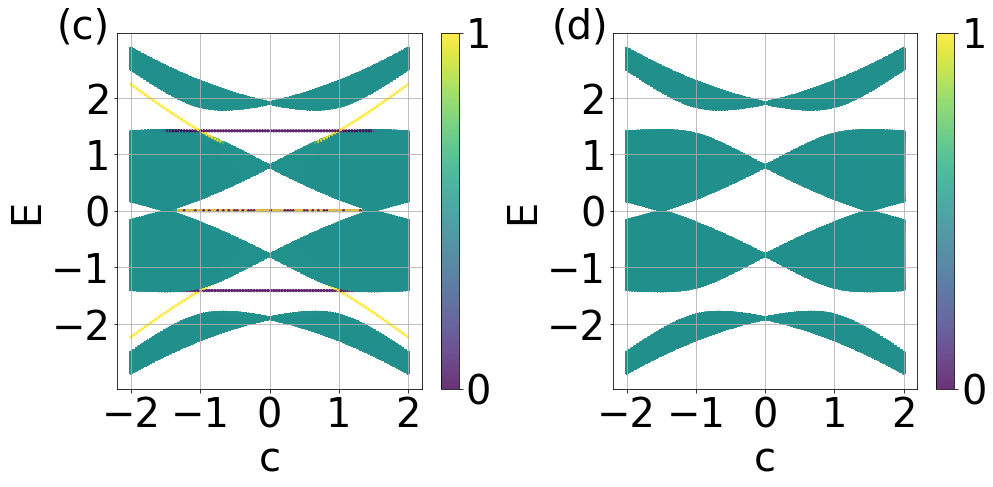}
    \caption{Evolution of energy levels of SSH4 model as a function  $c$ in (a) [(b)] and (c) [(d)] for $d=1$ and $1.5$ under OBC [PBC]. The values of $d$ are chosen from the dotted line in Fig. \ref{fig:ssh4_rwinding} (a).  The color bar represents the average position of the state: $1$  ($0$) corresponds to the left (right) edge.  At $E=0$, these edge modes are simultaneously present leading to an overlap in terms of their color code. }
\label{fig:ssh4_eng}    
\end{figure}

\begin{figure}
    \centering
    \includegraphics[width=0.48\textwidth]{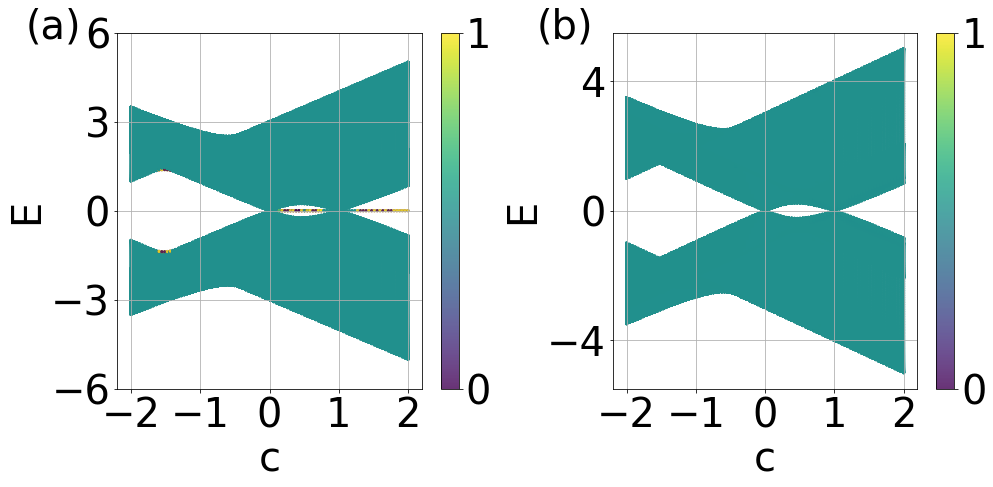}
    \includegraphics[width=0.48\textwidth]{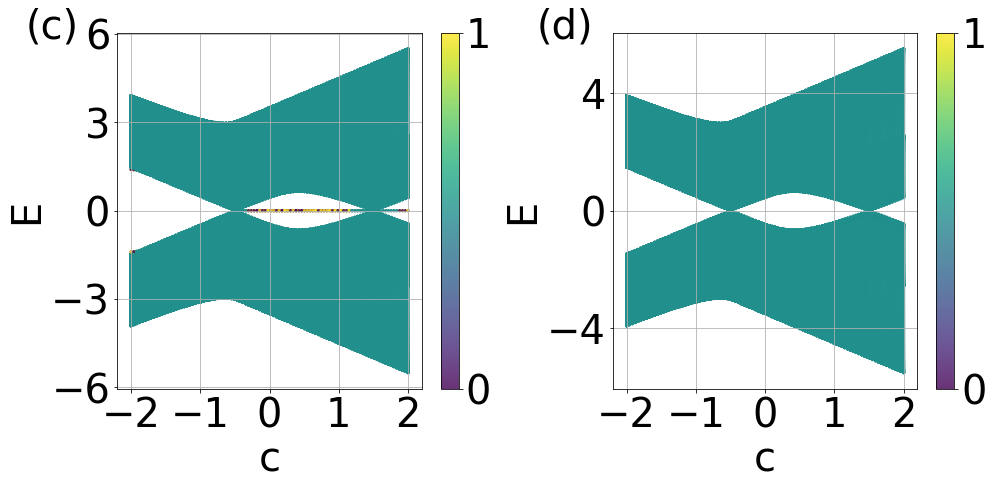}
    \caption{ Evolution of energy levels of SSHLR model as a function  $c$ in (a) [(b)] and (c) [(d)] for $d=1$ and $1.5$ under OBC [PBC]. The values of $d$ are chosen from the dotted line in Fig. \ref{fig:sshlr_rwinding} (a). }
\label{fig:sshlr_eng}    
\end{figure}

Interestingly, there exist finite energy mid-gap states between the sub-bands in SSH4 model and this leads to a  markedly different energy spectrum as compared to the SSHLR model. In order to investigate these modes under OBC, we show the evolution of energy levels by varying $c$ for SSH4 model in Fig. \ref{fig:ssh4_eng} (a,c) with $d=1$ and $1.5$, respectively. These parameters are chosen from the red and blue dotted lines in Fig. \ref{fig:ssh4_rwinding} (a).  Unlike the zero-energy modes, the finite energy modes do not vanish once the system enters into a trivial phase from a topological phase. However, the energies of these modes do not change within a given topological phase even when the parameters change.  The evolution of the energy levels under PBC conditions clearly show the gap-closing at zero energy is important to have the topological transitions, see  Fig. \ref{fig:ssh4_eng} (b,d) for $d=1$ and $1.5$, respectively. 
Therefore, the zero energy modes are found to be essential to determine the topology of the SSH4 model and which is why the winding number remains bounded between $W=\pm 1$. 
Now, moving to the SSHLR model, we show the emergence of $2$ and $4$ zero energy modes by varying $c$ in Fig. \ref{fig:sshlr_eng} (a,c) with $d=1$ and $1.5$, respectively. These parameters are chosen from the red and blue dotted lines in Fig. \ref{fig:sshlr_rwinding} (a).
 Unlike the SSH4 model, the long-range nature of the hopping yields $4$ zero energy modes leading to winding number $W=2$ in the SSHLR model.  
The topological phase transitions, associated with the gap closing at the zero energy, in the SSHLR model are nicely captured under PBC, see  Fig. \ref{fig:sshlr_eng} (b,d). This is in complete agreement with the red and blue dotted lines as depicted in  Fig. \ref{fig:sshlr_rwinding} (a). Importantly, the long-range hopping in SSHLR modifies the phase boundary $d=-c$, as noticed in the phase diagram for SSH4 model, leading to two new phase boundaries $d=-c+1$ and $d=-c-2$ as noticed in the phase diagram, see Fig. \ref{fig:sshlr_gap} (b). On the other hand, $d=c$ phase boundary remains unaltered for both of the models indicating the connection to the phase diagram of the pristine SSH model.

%-------------------------------------------------------------------------------------------------------------------------------------------------

%-------------------------------------------------------------------------------------------------------------------------------------------------

%-------------------------------------------------------------------------------------------------------------------------------------------------

%-------------------------------------------------------------------------------------------------------------------------------------------------

%-------------------------------------------------------------------------------------------------------------------------------------------------
\subsection{Real Space Winding Number}
\label{s3ss2}
%-------------------------------------------------------------------------------------------------------------------------------------------------

It would be an interesting exploration to validate the momentum space winding number through the real space methods where chiral symmetry again plays a pivotal role. In this case, we first discuss these methods using open boundary condition (OBC) and periodic boundary condition (PBC) such that phases of SSH4 and SSHLR can be revisited \cite{Lin2021,Ghosh2024,Song2019}. The Hamiltonian is expressed in real space on a 1D lattice having $L$ unit cells leading to $Ld \times Ld$ matrix with $d=4$ and $2$ for SSH4 and SSHLR models, respectively.

\begin{figure}[ht]
    \includegraphics[width=0.23\textwidth]{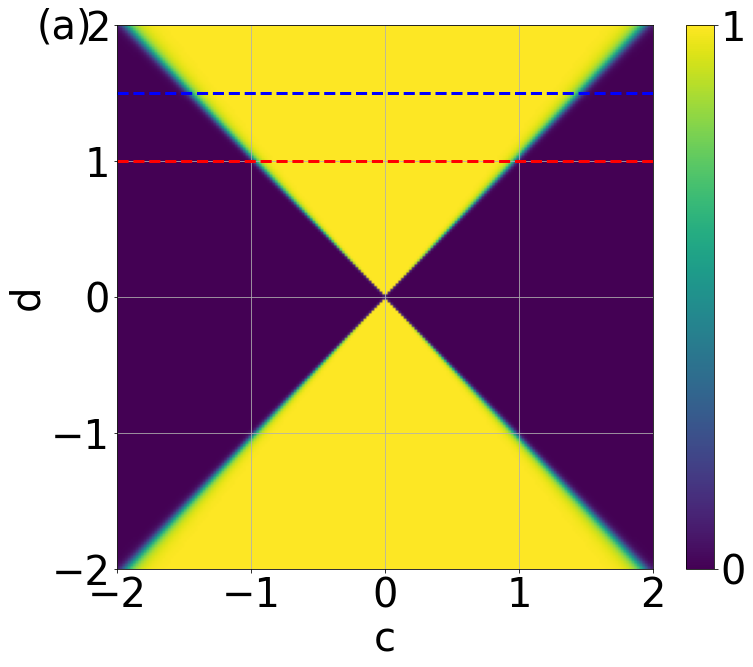}
    \includegraphics[width=0.23\textwidth]{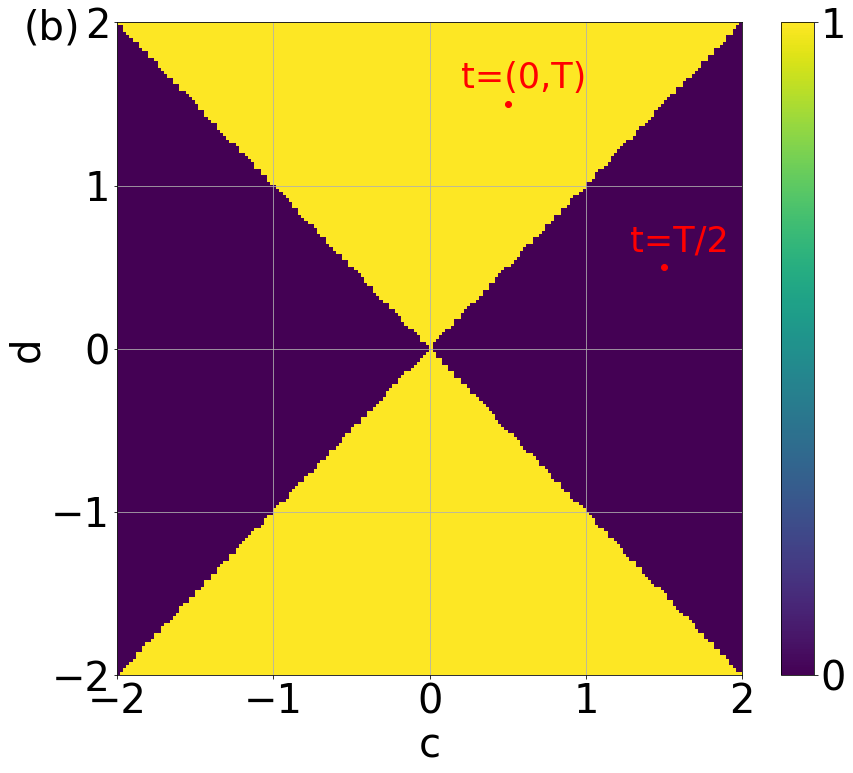}   
    \caption{We show the winding number phase diagram in the $c$-$d$ plane under OBC in (a) and PBC in (b) for SSH4 model ($a=b=1$), following Eqs. (\ref{eq:winding_robc}) and (\ref{eq:winding_rpbc}), respectively. We identify two-points by red dots in (b) which are relevant for the discussion on the time-dependent models in the next section. Here for (a) $N=100$, $L=70$ and $l=15$ for (b) $N=200$. }
\label{fig:ssh4_rwinding}    
\end{figure}

\begin{figure}
    \includegraphics[width=0.23\textwidth]{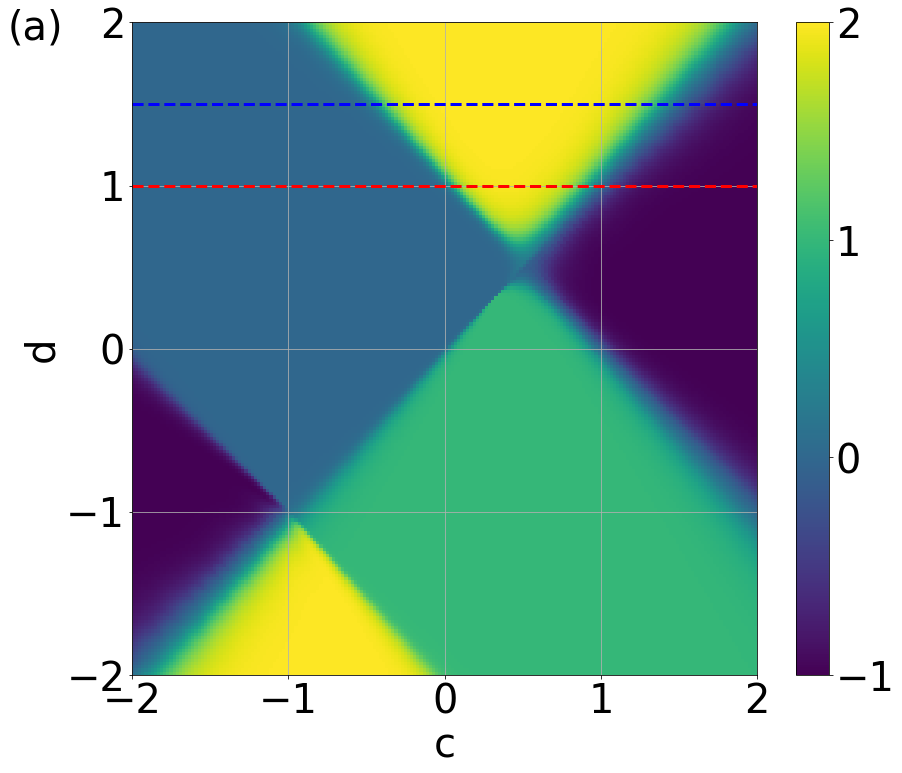}
    \includegraphics[width=0.23\textwidth]{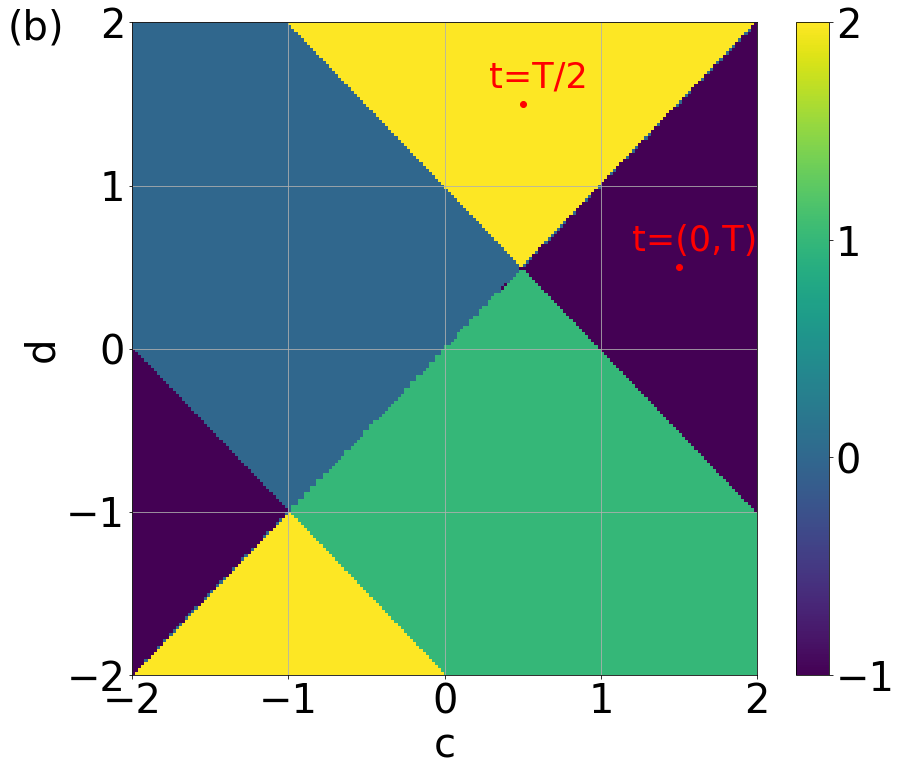}
    \caption{We show the winding number phase diagram in the $c$-$d$ plane under OBC in (a) and PBC in (b) for SSHLR model ($a=b=1$), following Eqs. (\ref{eq:winding_robc}) and (\ref{eq:winding_rpbc}), respectively. Here for (a) $N=100$, $L=70$ and $l=15$ for (b) $N=200$ .}
\label{fig:sshlr_rwinding}        
\end{figure}

\textit{Winding Number Using OBC or open-bulk winding number:} Going beyond the momentum space formalism, 
one can define the OBC winding number in real space \cite{Song2019,He2020} by using the lattice version of chiral symmetry operator $\Gamma$, position operator $X_L$, and eigenstates $\ket{u_n}$, corresponding to $n$-th energy level, of the Hamiltonian. This winding number is usually referred to as open-bulk winding number in the literature.  To be precise,  the real-space winding number with system having $L$ unit cells, is then expressed as 
\begin{equation}
W=\frac{1}{2L'}\mbox{Tr}'\Big(\tilde{\Gamma} Q[Q,X]\Big).
\label{eq:winding_robc}
\end{equation}
Here, $X$ is an extended position matrix of dimension $(Ld\times Ld)$. $X=X_L\otimes I_d$ and $(X_L)_{mn}=m\delta_{mn}$, and   $I_d$ denotes the identity matrix of size ($d\times d$), $d$ denotes the sub-lattice degrees of freedom in that unit cell, and ${\rm max}(m,n)=L$ denote the unit-cell index. $Q$ denotes a ($Ld \times Ld$) matrix which is unitary as well as Hermitian,  satisfying $Q^2=I$, $Q=\sum_{E_n>0}\ket{u_n}\bra{u_n}-\sum_{E_n<0}\ket{u_n}\bra{u_n}$ with  $H\ket{u_n}=E_n\ket{u_n}$.
$\tilde{\Gamma}=I_L\otimes\Gamma$ with $I_L$ being the $L\times L$ identity matrix and $\Gamma$ is the generator of the chiral symmetry, represented by a $d\times d$ unitary matrix. Note that Tr$'$ represents the partial trace, restricting the summation to the middle segment of the system $L'$ while excluding $2l$ number of boundary sites from both the ends such that $L' + 2 l =L$. This method of computing  the real space winding number under OBC yields quantized results when $l$ is sufficiently large to avoid boundary effects.

\begin{figure}[ht]
    \includegraphics[width=0.23\textwidth]{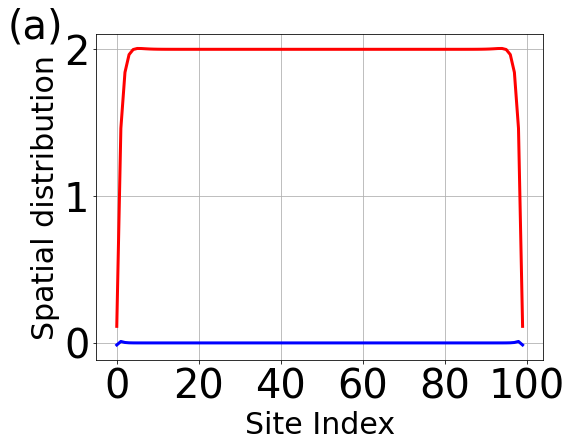}
    \includegraphics[width=0.23\textwidth]{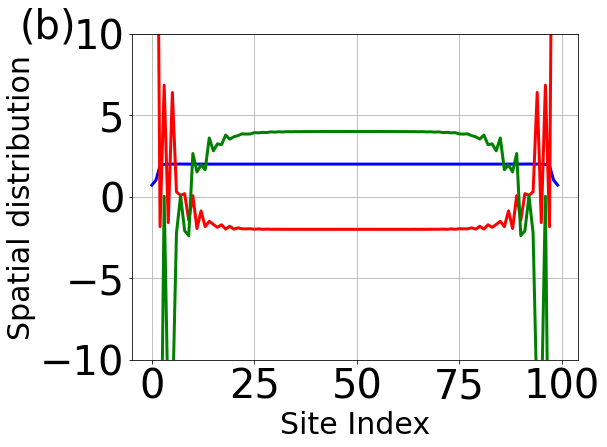}   
    \caption{Spatial distribution of winding number at each lattice site before executing the partial trace in Eq. (\ref{eq:winding_robc})  
     in (a, b) for SSH4 and SSHLR model, respectively. 
     In (a) blue and  and red lines correspond to $d=0.5$ and $1.5$ for $W=1$ and $0$, respectively.  In (b) blue, red, and green lines correspond to $d=-1.5$, $0.5$, and $1.5$ for $W=+1$, $-1$, and $+2$,  respectively. We consider $a=b=c=1$.}
\label{fig:orbital_sum}    
\end{figure}

Figure \ref{fig:orbital_sum} (a,b) represent the spatial distribution of winding numbers with lattice site index for SSH4 and SSHLR models, respectively. 
The site-resolved winding number is obtained from the diagonal elements of 
$\tilde{\Gamma} Q[Q,X]$ as mentioned in Eq. (\ref{eq:winding_robc}).
The boundary fluctuations are clearly seen from the spatial  distribution of the winding number
while it remains quantized inside the bulk. This site-resolved winding number helps to determine how many number of intermediate sites $L'$ to include and how many boundary sites $l$ to discard to compute the quantized winding number accurately.

The expression of OBC winding number Eq. (\ref{eq:winding_robc}) stems from the momentum space version of the winding number \cite{Song2019}. The derivative term $\partial_k Q$ in Eq. (\ref{eq:winding2}) can be converted to a commutator $[Q,X]$, owing to the fact that $\partial_t A= [A,H]$, where $k$ and $X$ are conjugate variables. Therefore, $ \Gamma Q \partial_k Q$ reduces to the form $ \Gamma Q [Q,X]$ as mentioned in Eq. (\ref{eq:winding_robc}). The important aspect in the present case is that $X$ is the position matrix  which can be obtained upon discretizing $\partial_k$ in the real space.   Now, executing the above method for SSH4 and SSHLR, we find phase diagrams that are in complete accordance with that of the momentum space phase diagrams, see Figs. fig \ref{fig:ssh4_rwinding} (a), \ref{fig:sshlr_rwinding} (a), \ref{fig:ssh4_gap} (b) and \ref{fig:sshlr_gap} (b). One can find that the phase boundaries, obtained from OBC winding number, are not as sharp as observed from the  PBC winding number study. This can be attributed to the finite size effects.

\textit{Winding number using PBC or periodic-bulk winding number:}
Having demonstrated the OBC real space winding number, we here describe the
PBC winding number using the projection onto chiral basis \cite{Lin2021}. This winding number is usually referred to as periodic-bulk winding number in the literature. The chiral symmetry generator $\Gamma$ follows \(U_\Gamma \Gamma U_\Gamma^\dagger = \pm 1\) which causes a division of the system into two sub-spaces namely, \(A\) and \(B\), corresponding to eigenvalues \(+\) and \(-\). To be precise, $U_\Gamma$ is  a unitary matrix obtained from the spectral decomposition of $\Gamma$ such that $U_\Gamma = U_\Gamma^A - U_\Gamma^B$ where \(U_\Gamma^A = \sum_{\alpha \in A} |\alpha\rangle \langle \alpha|\) and \(U_\Gamma^B = \sum_{\beta \in B} |\beta\rangle \langle \beta|\).
%Note that $U_\Gamma$  in the present case is  similar to the unitary matrix $Q$ as described before for the case of OBC winding number. 
Upon using the singular value decomposition of the anti-diagonal block $h$, one obtains two sets of unitary matrices: $h = U_A \Sigma U_B^\dagger $ where \(U_A\) and \(U_B\) are unitary matrices, and \(\Sigma\) is a diagonal matrix containing singular values. The anti-diagonal blocks of the flattened Hamiltonian \(Q\), having  eigenvalues \(\pm 1\), are unitary matrix $q$ and $q^{\dagger}$ where $q=U_A U^{\dagger}_B$. Under PBC, the position operator is defined as
\begin{equation}
\mathcal{X} = \exp\left(\frac{2i\pi X_{L}}{L}\right) \otimes I_d \ 
\nonumber\\
\end{equation}
where the quantity $\mathcal{X}$ is a matrix of size $(Ld \times Ld)$ and  $X_L$ and $I_d$ are  mentioned previously for the open-bulk winding number.

One has to first project the  position operator in the chiral basis in accordance with the subspaces $\sigma$ leading to $\tilde{\mathcal{X}}=U_\Gamma^\sigma \mathcal{X} U_\Gamma^\sigma$ with $\sigma=A,B$. Here  $\tilde{\mathcal{X}}$ is a  block-diagonal unitary matrix of size ($Ld \times Ld$) whose first [second] ($Ld/2 \times Ld/2$) diagonal block  corresponds to sub-space $A$ [$B$].    In order to compute the  sub-space  dipole operator \(\mathcal{X}_\sigma\), the projection onto the occupied bands is required and takes the form $\mathcal{X}_\sigma= U_\sigma^{\dagger} \tilde{\mathcal{X}} U_\sigma$ where $\mathcal{X}_A$ ($\mathcal{X}_B$) is obtained from  first (second) diagonal block of $\tilde{\mathcal{X}}$. Under PBC condition, the winding number is given by
\begin{equation}
\nu = \frac{1}{2\pi i} \text{Tr}\left[\log\left(\mathcal{X}_A \mathcal{X}_B^\dagger\right)\right]. \ 
\label{eq:winding_rpbc}
\end{equation} 
Now,  we execute the above method for SSH4 and SSHLR and the  phase diagrams are shown in Figs. \ref{fig:ssh4_rwinding} (b), \ref{fig:sshlr_rwinding} (b), respectively. These phase diagrams match very well with that of the momentum space phase diagrams, see Figs.  \ref{fig:ssh4_gap} (b) and \ref{fig:sshlr_gap} (b).  The important point here is that the finite size effects are almost absent unlike the previous open-bulk case leading to sharp phase boundaries for the periodic-bulk case.

It is noteworthy that position operator projected on the chiral sub-space of the occupied energy bands is related to the Wilson loop line-element matrices \cite{Lin2021}. The Tr operation mimics the Wilson loop itself, constructed out of the path ordering of the line elements. Therefore, Eq. (\ref{eq:winding_rpbc}) is intrinsically connected to the bulk polarization obtained from the ${\rm log [det]}$ of the Wilson loop.

%-------------------------------------------------------------------------------------------------------------------------------------------------
\section{Time-dependent extended SSH models}
\label{s4}
%-------------------------------------------------------------------------------------------------------------------------------------------------

Having investigated the static model $H(k)$, we now study the time-periodic Hamiltonian such that $H(k, t + T) = H(k, t)$ where  \(T\) represents the time period associated with the evolution. We consider adiabatic time evolution where the system follows instantaneous ground state and the bulk gap does not close during the evolution. This ensures that the bands of the system evolve through their instantaneous eigenstate discarding the non-equilibrium dynamics. This mechanism would lead to quantized transport of electron charge as demonstrated in Rice-Mele charge pump \cite{Asb_th_2016}. In what follows, we consider SSH4 and SSHLR model to investigate the adiabatic charge pumping.  Borrowing the concept from Rice-Mele charge pump, we will employ sine and cosine functions suitably in SSH4 and SHHLR models which we discuss below. 

%-------------------------------------------------------------------------------------------------------------------------------------------------

\subsection{Driven SSH4 Model}
\label{s4ss1}
%-------------------------------------------------------------------------------------------------------------------------------------------------

We introduce time-periodic function in the intra-sublattice as well as inter-sublattice terms for SSH4 Hamiltonian as given by  
\begin{equation}
H(k,t)=
\begin{bmatrix}
 -h_{\rm st} & a & 0 & (t_0+d_{\rm st}) e^{-ik} \\
 a & h_{\rm st} & b & 0 \\
 0 & b & -h_{\rm st} & t_0-d_{\rm st} \\
 (t_0+d_{\rm st})e^{ik} &0 & t_0-d_{\rm st} & h_{\rm st}
\end{bmatrix}
\label{eq:ssh4t}
\end{equation}
where intra-sublattice term $h_{\rm st}$ can be thought of a staggered on-site term, exhibiting alternative signs between adjacent sub-lattices. On the other hand, $d_{\rm st}$ denotes the hopping between sub-lattices $A \leftrightarrow D$ and $C \leftrightarrow D$; this hopping also has staggered nature. Comparing with the static model Eq. (\ref{eq:ssh4}), $d$ is modified with $t_0 + d_{\rm st}$, $c$ is replaced with  $t_0 - d_{\rm st}$, while the  diagonal elements no longer remain zero rather acquire finite value $h_{\rm st}$ for the on-site contributions. We consider the following time-periodic protocols
\begin{equation}
h_{\rm st} = h_0 \sin\left(\frac{2 \pi t}{T}\right),
\label{eq:hst}
\end{equation}
\begin{equation}
d_{\rm st} = d_0 \cos\left(\frac{2 \pi t}{T}\right)
\label{eq:dst}
\end{equation}
Comparing Eqs. (\ref{eq:ssh4t}) and (\ref{eq:ssh4}), one can find that the modulation of the $c$ and $d$-hopping takes place around the value $c=d=t_0$ while the on-site term modulates around $0$.  The $c$ and $d$ hopping  in this time-dependent model  have a relative phase difference $\pi$ owing to the $\pm d_{\rm st}$ factor.  
At specific instances of time  \( t = 0, T/2, T, \dots \), this time-dependent model simplifies and can be mapped onto a static SSH4 model yielding various topological and trivial phases, see red dots in Fig. \ref{fig:ssh4_rwinding}(b).

%-------------------------------------------------------------------------------------------------------------------------------------------------

\subsection{Driven SSHLR Model}
\label{s4ss2}
%-------------------------------------------------------------------------------------------------------------------------------------------------
We here demonstrate the time-periodic SSHLR model where in addition to the staggered on-site intra-sublattice terms, we consider time-periodicity in the long-range hopping inter-sublattice terms. This results in the following Hamiltonian, elevated from the underlying static model Eq. (\ref{eq:sshlr}), as given by   
\begin{widetext}
\begin{equation}
H=
\begin{bmatrix}
 -h_{\rm st} & a+ be^{-ik}+(t_0-d_{\rm st})e^{ik}+(t_0+d_{\rm st})e^{-2ik} \\
 a+be^{ik}+(t_0-d_{\rm st})e^{-ik}+(t_0+d_{\rm st})e^{2ik} & h_{\rm st}   
 \end{bmatrix}
 \label{eq:sshlrt}
\end{equation}
\end{widetext}
where $h_{\rm st}$ and $d_{\rm st}$ are taken from Eqs. (\ref{eq:hst}) and (\ref{eq:dst}), respectively as discussed previously for SSH4 model. 
Comparing Eqs. (\ref{eq:sshlrt}) and (\ref{eq:sshlr}), one can find similar modulation
profiles of $c$, $d$ and diagonal terms as compared to the driven SSH4 model. The phases obtained from the above model Eq. (\ref{eq:sshlrt}) at different time instances are shown by the red dots in Fig. \ref{fig:sshlr_rwinding}(b).  

%-------------------------------------------------------------------------------------------------------------------------------------------------
\section{topological phases of time-dependent models}
\label{s5}
%-------------------------------------------------------------------------------------------------------------------------------------------------

Having demonstrated the time-periodic models, we now focus on the topological aspects of these models. Thanks to the adiabatic evolution, time is treated as a parameter and there is no non-equilibrium dynamics taking place. 
We, therefore, have two periodic variable in these models namely, time and momentum. Therefore, one can use Chern number to characterize the emergent topological phase. Interestingly, we can study the evolution of quantized charge as a function of time which would enable us to better understand the Chern number. We will investigate the Bott index using the spatio-temporal Hamiltonian to verify the topological phases of these model using the real space formalism.

%During one complete cycle of this evolution, the charge pumped across the insulator is always quantized as an integer. This integer corresponds to a topological invariant that characterizes the system. This invariant reflects the topological properties of the underlying energy bands and remains robust against perturbations, as long as the system's energy gap does not close during the evolution.

%-------------------------------------------------------------------------------------------------------------------------------------------------
\subsection{Chern Number}
\label{s5ss1}
\begin{figure}
    \centering
    \includegraphics[width=0.23\textwidth]{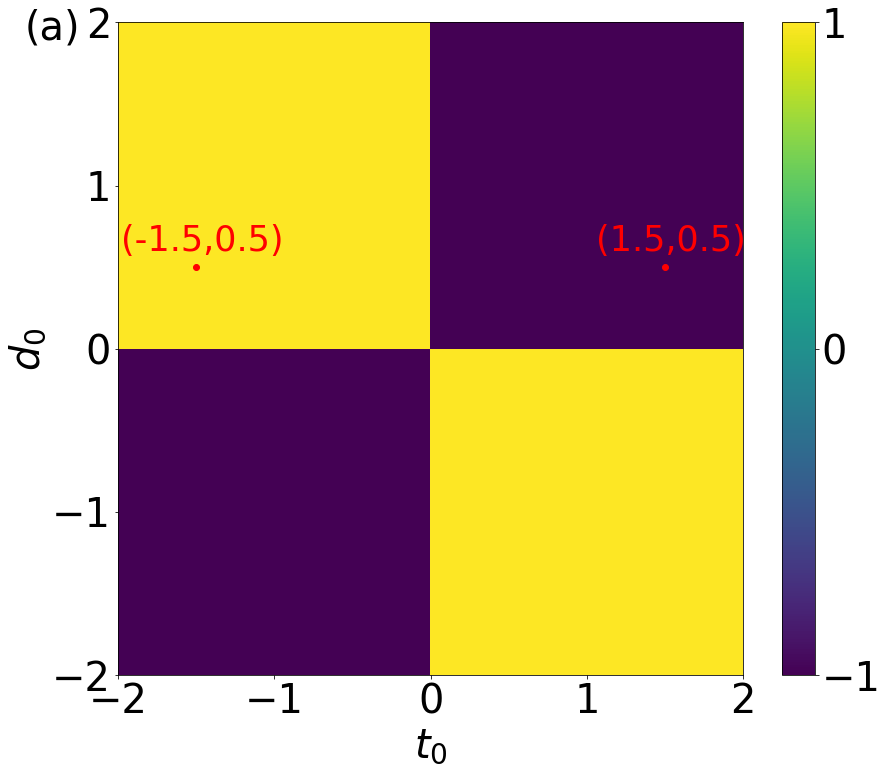}
    \includegraphics[width=0.23\textwidth]{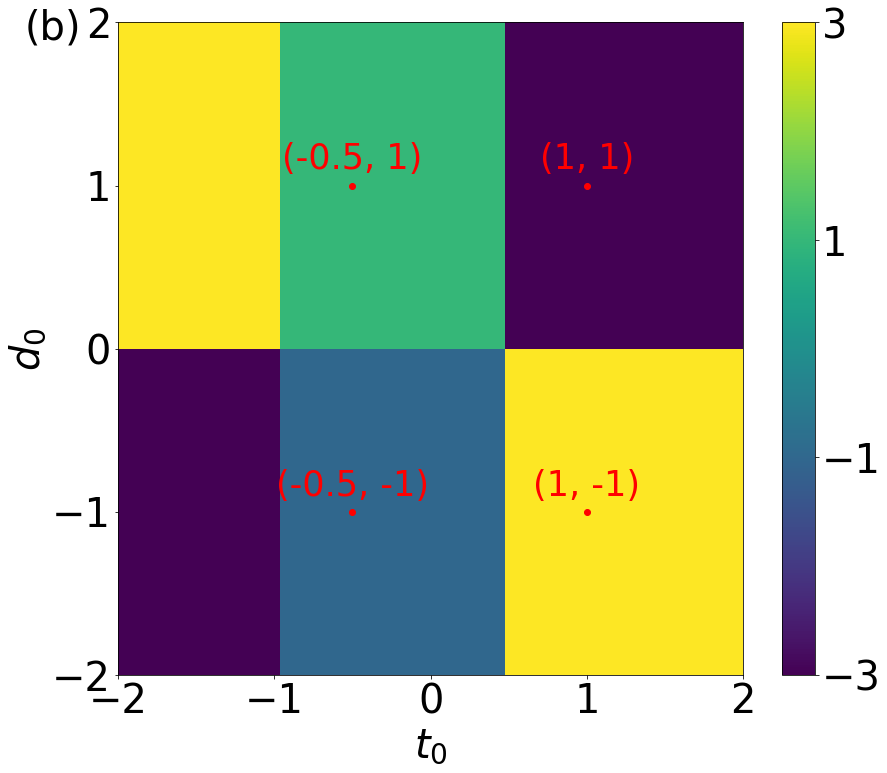}
    \caption{Topological phase diagram for driven SSH4 model and SSHLR model are depicted in (a) and (b), respectively. The color bar denotes the value of Chern number. We show two and four representative points in (a,b), respectively that we will be relevant for Figs. \ref{fig:ssh4_eng_t} and \ref{fig:sshlr_eng_t}.     Here $h_0=a=b=T=1$.}
\label{fig:chern_pd}
\end{figure}

%-------------------------------------------------------------------------------------------------------------------------------------------------
For the driven model, the time reversal symmetry is broken $H^*(-k,-t)\ne H(k,t)$ allowing us to characterize the topological phase in terms of the Chern number. The time-dependent bulk spectrum is always gapped that enables us to compute the Chern number of individual bands.  The Chern number for the \(n\)th band can be expressed as \cite{Fukui2005}
\begin{equation}
c_n = \frac{1}{2\pi i} \int_{0}^{2\pi} \int_{0}^{T} dk \, dt \, F_{12,n}(k,t),
\label{CheNumCon}
\end{equation}
where $F_{12,n}(k,t) = \partial_1 A_{2,n}(k,t) - \partial_2 A_{1,n}(k,t)$ ($1,2=k,t$) denotes the Berry curvature and $A_{\mu,n}(k,t) = \bra{n(k,t)} \partial_\mu \ket{n(k,t)}$ represents the Abelian Berry connection. The Chern number of the phase is found as the sum of the Chern numbers associated with filled bands below the fermi level is given by $C=\sum_n C_n$.  The phase diagram for SSH4 and SSHLR are demonstrated in  Figs. \ref{fig:chern_pd} (a,b), respectively, where $C$ acquires the values $\pm 1$ and $\pm 3$, $\pm 1$ with the parameters $a=b=1$.

In the driven case, we find a checkerboard-like phase diagram for both the models where all the phases  happen to be topological for the chosen parameter space $-2<t_0, d_0<2$. The phase boundaries for SSH4 model is given by $t_0=0$ and $d_0=0$ lines dividing phase diagram in four equal quadrants for $-2<t_0, d_0<2$. On the other hand, the phase boundaries are given by $t_0=-1$, $0.5$ and $d_0=0$ leading to six divisions  for $-2<t_0, d_0<2$ in the case of SSHLR model. The interesting point is that $d_0=0$ phase boundary is observed for both the driven models which is similar to the commonly observed $d=c$ phase boundary in their static analogs, see Figs. \ref{fig:ssh4_gap} (b) and \ref{fig:sshlr_gap} (b). This correlation can be understood from the adiabatic  connectivity between the driven and static models such as $d=c$ refers to a situation $d_{\rm st}=0$ i.e., $d_0=0$. Similarly, $t_0=0$ ($t_0=-1$, $0.5$) phase  boundary (boundaries) in the driven SSH4 (SSHLR) model is (are) originated from $d=-c$ boundary ($d=-c-2$ and $d=-c+1$ boundaries). While we find trivial regions in the static phase diagrams with $a=b=1$, we do not find any trivial phase in the driven model due to the fact that adiabatic drives turn the trivial gap into topological for the above set of parameters. The model transits through the topological as well trivial regions when time-periodic terms are introduced. To be precise, if $t=0$ and $T/2$ phases are topological/trivial then $t=T/4$ and $3T/4$ phases are most likely to be modified. For any value of $t$ other than $t=(0,T/2,T)$ the diagonal elements $h_{st}$ acquire finite value, and hence the model no longer remains analogous to the static SSH model.

This results in topological phases appearing in the phase diagram.  Interestingly,  Chern numbers take finite values  $C=\pm 3$ $(\pm 1)$  that are not expected from the static SSHLR and SSH4 model. The $|C|= 3$ ($1$) phase is caused by the fact that driven SSHLR (SSH4) model transits between 
$|C|=2$ $(0)$ at $t=T/2$ and $|C|=1$ $(1)$ at $t=0,T$  phases of static SSHLR (SSH4) model during the course of adiabatic dynamics, see red dots in Figs. \ref{fig:sshlr_rwinding} (b) and \ref{fig:ssh4_rwinding} (b).  Therefore, the adiabatic drive causes intriguing topological characters that are not usually observed in the underlying static model. The adiabatic drive leads to an admixture of static topological phases.  More interestingly, the phase diagram of the driven models always hosts topological phases when $a=b$. This is because the driving cycle embeds the parameter space partially within which the static model becomes topological. Once $a\ne b$, 
we can also get trivial phase ($C=0$) for the driven models. This is because of the fact that at $t=(0,T/2)$ system always resides in the same phase trivial/topological phase associated with the underlying static models.

\begin{figure}
    \centering
    \includegraphics[width=0.23\textwidth]{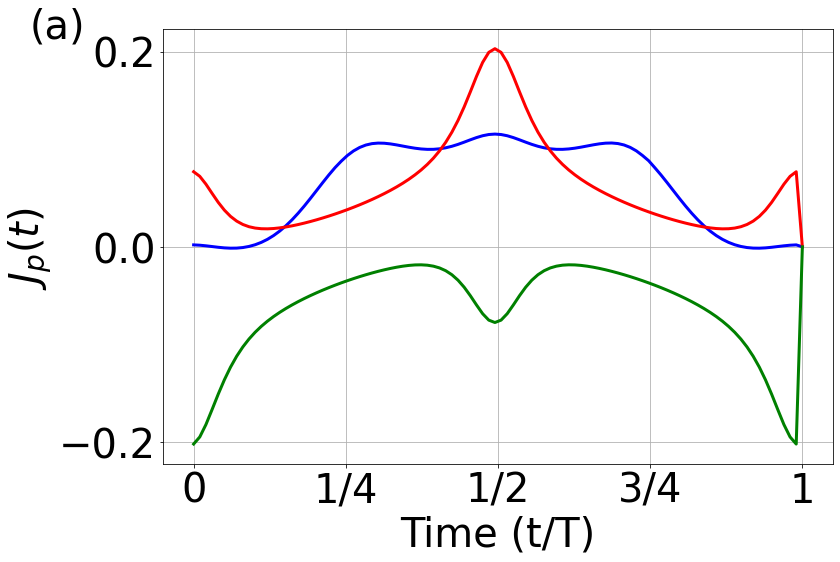}
    \includegraphics[width=0.23\textwidth]{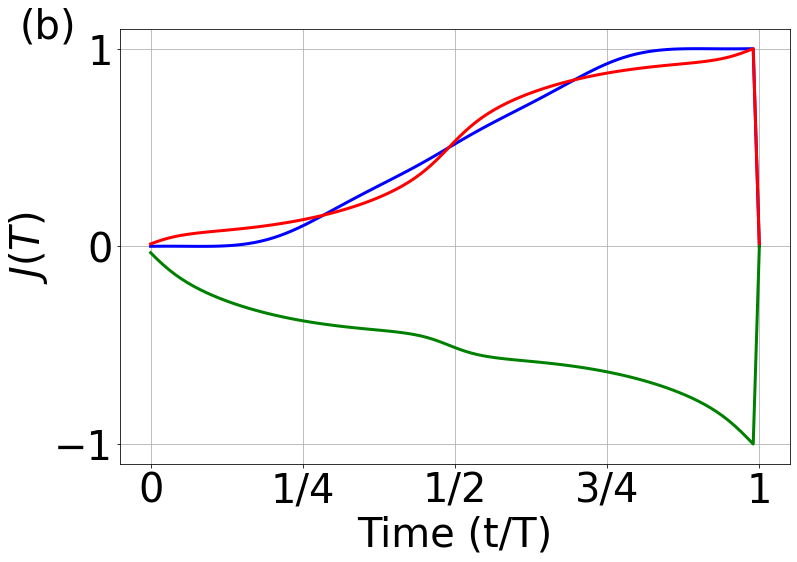}
    \includegraphics[width=0.23\textwidth]{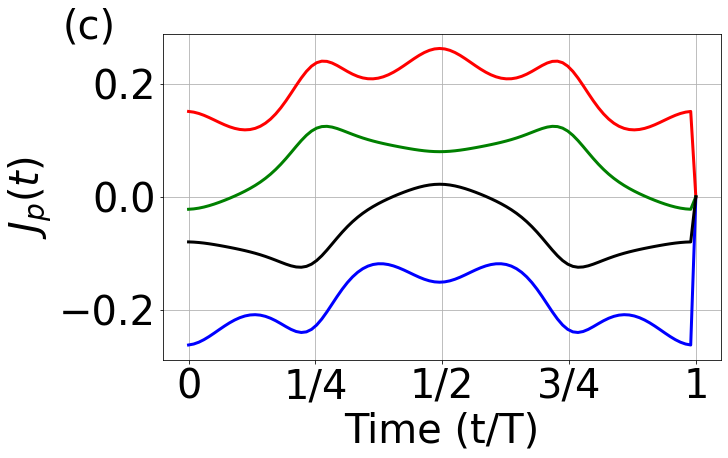}
    \includegraphics[width=0.23\textwidth]{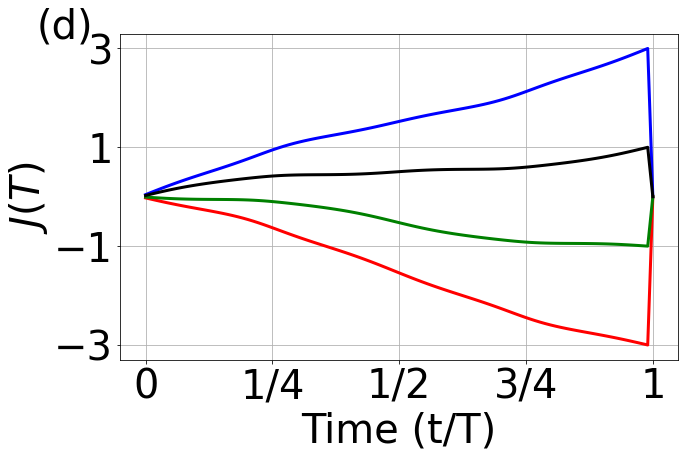}
    \caption{The time evolution of particle current and pumped charge in (a) [(b)] and (c) [(d)], respectively for driven SSH4 [SSHLR] model. For (a,b), we choose $(t_0, d_0)=(1.5, 0.5),(-1.5,0.5)$, and $(1,-1.2)$, represented by green, red, and blue,  respectively. For (c,d), we consider $(t_0, d_0 )=(1,1), (1,-1),(-0.5,1)$ and $(-0.5,-1)$ represented by blue, red, green, and black, respectively.}
\label{fig:current_t}    
\end{figure}

\begin{figure}
    \centering
    \includegraphics[width=0.23\textwidth]{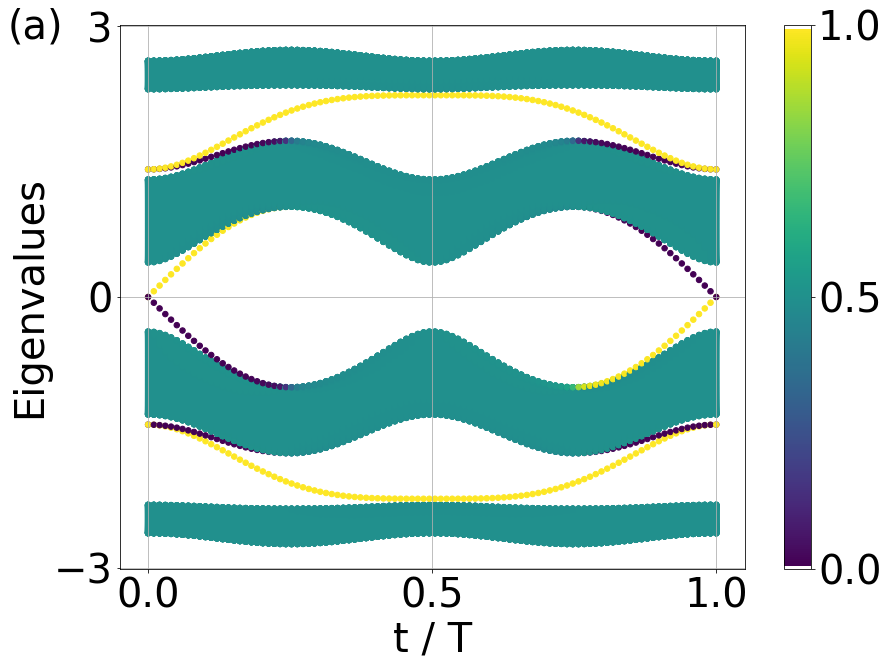}
    \includegraphics[width=0.23\textwidth]{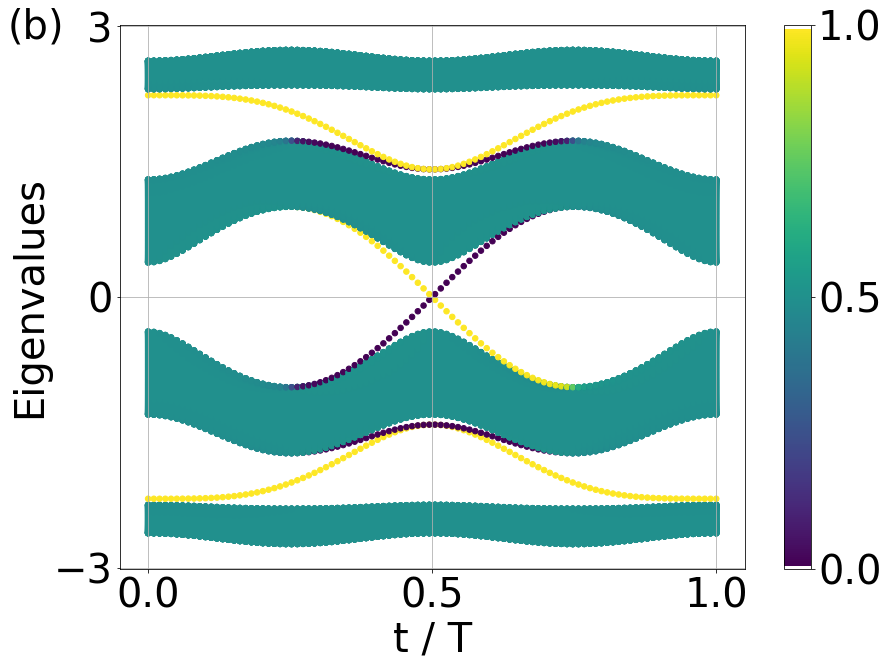}
    \caption{The variation of energy levels, computed from real space time-dependent SSH4 model Eq. (\ref{eq:ssh4t}), are shown as a function of time for $(t_0,d_0)=(1.5,0.5)$ and $(-1.5,0.5)$ in (a) and (b), respectively. These choices of parameter are depicted by red dots in Fig. \ref{fig:chern_pd} (a). The initial, final, and mid points of quench path of the adiabatic drive are shown in Fig. \ref{fig:ssh4_rwinding}.  The color bar represents the average position. }
\label{fig:ssh4_eng_t}
\end{figure}

\begin{figure}
    \centering
    \includegraphics[width=0.23\textwidth]{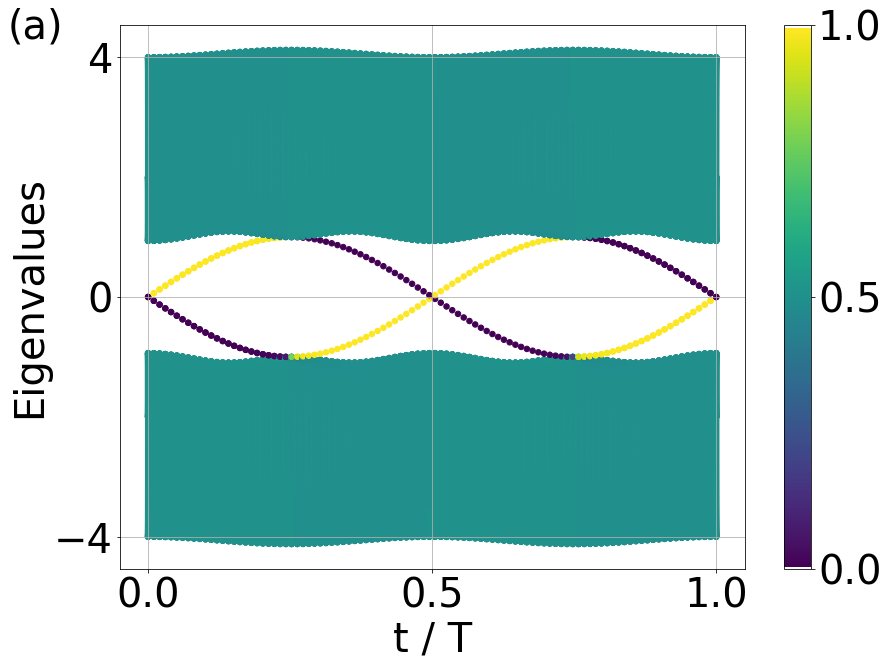}
    \includegraphics[width=0.23\textwidth]{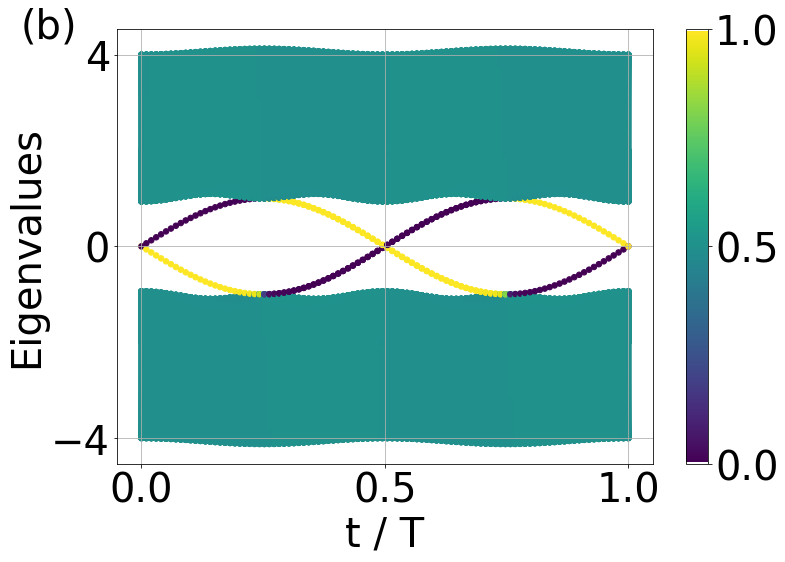}
    \includegraphics[width=0.23\textwidth]{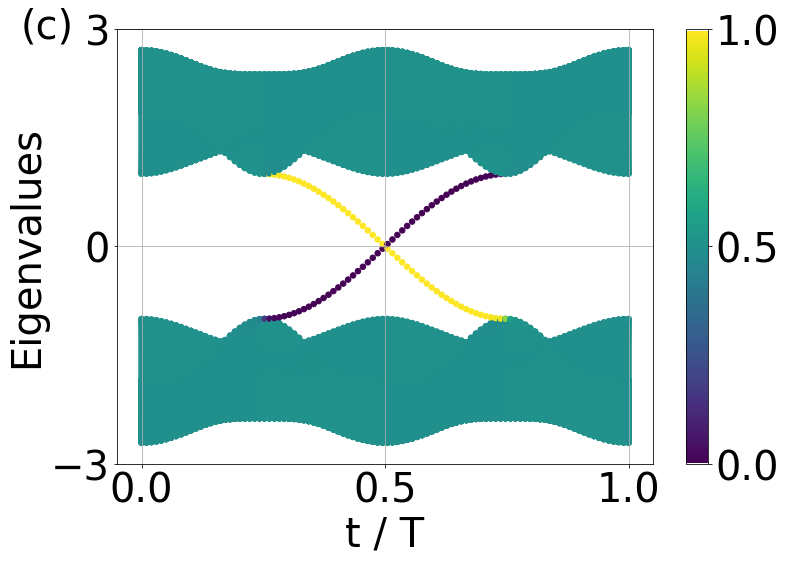}
    \includegraphics[width=0.23\textwidth]{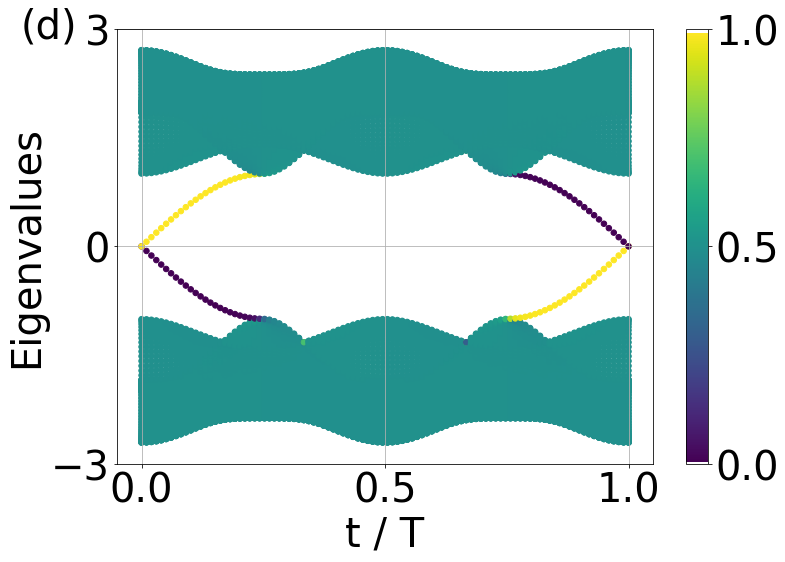}
    \caption{The variation of energy levels, computed from real space time-dependent SSHLR model Eq. (\ref{eq:sshlrt}), are shown as a function of time for $(t_0,d_0)=(1,1)$, $(1,-1)$, $(-0.5,1)$ and $(-0.5,-1)$ in (a), (b), (c) and (d), respectively, indicating the localization profile of the boundary modes along with their chirality. These choices of parameter are depicted by red dots in Fig. \ref{fig:chern_pd} (b). The initial, final, and mid points of quench path of the adiabatic drive are shown in Fig. \ref{fig:sshlr_rwinding}. }
\label{fig:sshlr_eng_t}    
\end{figure}

In order to investigate the Chern number further, we plot the quantity particle current $J_p(t)=\sum_{\rm n\in occ}\frac{1}{2\pi i} \int_{0}^{2\pi}  dk  \, F_{12,n}(k,t)$ and pumped charge $J(t)= \int_{0}^{t}  dt' J_p(t')$ in Figs. \ref{fig:current_t} (a,b) [(c,d)], respectively for driven SSH4 [SSHLR] model. We take a few representative points from the phase diagrams Figs. \ref{fig:chern_pd} (a,b) to demonstrate the $J_p(t)$ and $J(T)$. 
The time evolution of the particle current $J_p(t)$ shows an even nature with respect to the mid-point $t=T/2$  leading to the finite yet quantized value of pumped charge $J(T)$. The maximum value of $J_p(t)$ appears either around $t=T/2$ or $t=0$ suggesting to the fact that mid-gap states at zero-energy are visible at the above time instances, see Figs. \ref{fig:ssh4_eng_t} and \ref{fig:sshlr_eng_t}. The positive (negative) area under the $J_p(t)$ curve results in positive (negative) value of the pumped charge at the end of the cycle.
The final value of the pumped charge $J(T)$ is same as the Chern number $C$ while the monotonic behavior of $J(t)$ signifies the gradual accumulation of the pumped charge with time.
We do not find  $J_p(t)$ to be an odd function of time. In that case, 
the pumped charge vanishes leading to the trivial phase which does not show up for any of the driven models.

%Without including time, the observed Chern numbers are \( C = -1, 0, +1, +2 \). However, after the introduction of time, we no longer observe \( C = 0 \), and two new and higher topological phases emerge: \( C = -3 \) and \( C = +3 \) and phase boundary also changes.If we plot the energy spectrum versus time for all four cases of the Chern number, as highlighted in Figure~9, it becomes evident that the direction of charge motion is opposite for \( C = -3 \) and \( C = +3 \), as well as for \( C = -1 \) and \( C = +1 \). This is further illustrated in Figure~10.Additionally, we observe that the current flow is clearly visible and reaches its maximum at points where zero-energy modes are present, as shown in Figure~10(e). Furthermore, the charge pumped over one time period is quantized to an integer value, as depicted in Figure~10(f).

%---------------------------------------------------------------------------------------------------------------

%---------------------------------------------------------------------------------------------------------------

%---------------------------------------------------------------------------------------------------------------

%---------------------------------------------------------------------------------------------------------------

%------------------------------------------------------------------------------------------------------

Having examined the topological phases using Chern number, we now demonstrate the evolution of energy level, derived from real space Hamiltonian under OBC, as a function of time, see Figs. \ref{fig:ssh4_eng_t} and \ref{fig:sshlr_eng_t} for driven  SSH4 and SSHLR models, respectively. We first explain the Fig.  \ref{fig:ssh4_eng_t} (a). 
At $t=0$, the  zero-energy edge modes appear as the model resides in the topological phase.   As $t$ increases, the on-site energy $h_{\rm st}(t)$ lifts the degeneracy of the end mode pushing the right edge mode (yellow) to the positive band and the left edge mode (blue) to the negative band. 
At $t=T/2$, the two end states disappear as they have already evolved into the bulk states. When $t$ approaches $T$, edge modes start reappearing. To be precise, right (left) edge mode emerges from negative (positive) energy bands leading to a net flow of electron charge during the time evolution from $t=nT$ to $t=(n+1)T$, $n=0,1,2,...$. We can refer to yellow (energy increases with time) as positive chiral and blue (energy decreases with time) as negative chiral mode to understand the charge transfer. 
Interestingly, such chiral flow of a given edge mode is not observed for the finite energy mid-gap state. Therefore, there is no charge pumping associated with these mid-gap states.  Now going to Fig.  \ref{fig:ssh4_eng_t} (b), we find the charge transfer between $t=(n+1/2)T$ and $t=(n+3/2)T$. The two different chirality profiles shown in Figs.  \ref{fig:ssh4_eng_t} (a,b) cause a sign difference between the Chern numbers $C=\pm 1$.  One can identify the connection between charge transport and Chern number in the following way: $C=n_c(0) +n_c(T/2)$, where $n_c(t)$ denotes the number of chiral crossing at time instance $t$, $n_c=+1$ when blue (yellow) going up (down) and $n_c=-1$ when blue (yellow) going down (up).  The sign of the Chern number changes when the direction of motion of the edge states is reversed i.e., left $\to$ right transfer reverses to right $\to$ left.

Using the above analysis based on SSH4 model,  we now understand the connection between higher Chern numbers and the corresponding energy-time profile for SSHLR model, see Fig. \ref{fig:sshlr_eng_t}. We find one and two  pairs of zero-energy edge modes at $t=0$ and $t=T/2$, respectively,  
but  with opposite chiralitiy in Figs. \ref{fig:sshlr_eng_t} (a,b). This results in $C=-3$ [$3$] where the $n_c(0)$ yields $-1$ [$1$] and  $n_c(T/2)$ contribute $-2$ [$2$] during the complete evolution process.  The $+$ and $-$ signs in $n_c(t)$ are again determined whether the yellow band moves down and up, respectively, in energy-time plot.   On the other hand, Figs. \ref{fig:sshlr_eng_t} (c,d) represent the Chern number $C=1$ and $-1$ phases where  $n_c(0)=0,~n_c(T/2)=1$ and $n_c(0)=-1,~n_c(T/2)=0$, respectively. Therefore, the Chern number $C$ is connected to the chiral crossing of the energy levels at $t=0$ and $T/2$ in the similar way as described before for driven SSH4 model: $C=n_c(0) +n_c(T/2)$.  Therefore, the topology comes from the non-trivial winding of the edge modes along the time direction. However, it is important to have two distinct static SSH phases at $t=0$ and $t=T/2$.  One can think of this situation as the evolution of static phase diagram along the third time direction. Note that $t\ne nT, (n+1/2)T$ correspond to an on-site SSH model where the chiral symmetry is broken due to the presence of finite value of $h_{\rm st}$. During this adiabatic evolution  the gap is always maintained however, the time-dependent model traverses through a series of slices of different phase diagrams obtained from the above models.

%----------------------------------------------------------------------------------------------------

%----------------------------------------------------------------------------------------------------

%----------------------------------------------------------------------------------------------------
\subsection{Spatio-temporal Bott index}
\label{s5ss2}
%-------------------------------------------------------------------------------------------------------------------------------------------------
Having demonstrated the momentum space Chern number, we now shift our attention to verify the above phase diagram in terms of the spatio-temporal Bott index \cite{Huang18,Huang18b,Yoshii21,toniolo2022bott}.  Note that Bott index characterizes  disordered 2D Chern insulator phases, however, in the present context we use Bott index to characterize the Chern insulator phases of the driven SSH4 and SSHLR models. Importantly, we treat time as another dimension in addition to a single spatial dimension such that Bott index can be defined  on a 2D parametric space. To be precise,  the $x$-direction of the lattice represents spatial dimensions, and the $y$-direction corresponds to the temporal evolution; $(x,t) \to (x,y)$. Therefore, it is a spatio-temporal lattice where the already established framework of Bott index can be employed. We will examine the topological phases by adopting PBC and OBC both for the Bott index. 

\textit{Bott index from PBC or periodic-bulk Chern number:}
We now explore the topological phases using the periodic-bulk Chern number 
which is Bott index derived from a real space lattice with PBC \cite{Huang18,Zeng2020,Tang2020}. To begin with, one has to first construct  the projector operator for the occupied states, defined as:
\begin{equation}
P = \sum_i^{N_{\rm occ}} |\psi_i\rangle\langle\psi_i|
\end{equation}
where \(|\psi_i\rangle\) represents the wavefunction of the \(i\)-th state with eigenvalue \(\epsilon_i\) associated with the real space Hamiltonian under PBC.
Next, the periodic position operators are 
projected as follows:
\begin{eqnarray}
U = P e^{\frac{2i \pi x}{L_x}} P\\
V = P e^{\frac{2i \pi y}{L_y}} P
\label{eq:bott_pos}
\end{eqnarray}
where $z$ are $L_{z} \times L_z$ diagonal matrix with $z = x,y$. 
$x$ is the position operator matrix and $y$ represents the time operator matrix, however, they both take the form $z_{mn}=m \delta_{mn}$. Note that along the $x$-direction, $L_x$ denotes the system size while along the $y$-direction, $L_y$ represents the time period $T$ of the adiabatic drive. For simplicity without loss of generality, we call $U$ and $V$ both as the projected position operator. One can incorporate the complementary projectors \(Q = 1 - P\) into the definition of the projected $x,y$ position operators Eq. (\ref{eq:bott_pos}) such that the numerical stability of the algorithm increases without affecting the final results.  These new position operators are given by
\begin{eqnarray}
U = P e^{\frac{2i \pi x}{L_x}} P+ (1-P),\\
V = P e^{\frac{2i \pi x}{L_x}} P+ (1-P) 
\label{eq:bott_pos2}
\end{eqnarray}
Note that  \(UV U^\dagger V^\dagger\) is unitary leading to \(\det(UV U^\dagger V^\dagger) = 1\) and ensuring \(\log(U V U^\dagger V^\dagger)\) to become purely imaginary. The singular value decomposition of $V$ and $U$ can be further employed for even better convergence as well as preserve unitary nature of  $U, V$. We execute all the above steps for our calculation such that Bott index becomes purely real and well quantized inside a topological phase \cite{Huang18}. 
The Bott index, is then given by:
\begin{equation}
B = \frac{1}{2\pi} \textrm{Im}\{\textrm{Tr}[\log(U V U^\dagger V^\dagger)]\}.
\label{eq:bott_pbc}
\end{equation}

Investigating Eq. (\ref{eq:bott_pbc}) in details, one can connect it with the Wilson loop line element $W_{\Delta}(\bm {k})=\bra {n(\bm {k})} n(\bm {k+ \Delta })\rangle$ where $\ket{n(\bm {k})}$ represent the $n$-th occupied band for  the 2D parameter space ${\bm {k}}=(k_1,k_2)\equiv (k_x,k_y)\equiv(k,t)$ defined on a torus. The Berry curvature can be expressed in terms of a product of such line element such that a closed loop is formed $(k_1,k_2) \to (k_1+ \Delta_1, k_2 ) \to (k_1+ \Delta_1, k_2 +\Delta_2) \to (k_1, k_2 +\Delta_2) \to (k_1, k_2 )  $: $\Omega_{n,\bm k}=  \log(W_{\Delta_1}(\bm {k}) W_{\Delta_2}(\bm {k}+\Delta_1) W_{\Delta_1}(\bm {k}+\Delta_2)^{-1} W_{\Delta_2}(\bm {k})^{-1})$. This closed Wilson loop can be written down in terms of the link variable as followed in the Fukui formalism \cite{Fukui2005}. Now from the space of $(k_1,k_2)$, one can write the Berry curvature in the Fourier space where the line elements are replaced with projected position operators: $W_{\Delta_1}(\bm {k}) \to U$ and $W_{\Delta_2}(\bm {k}) \to V$, here, $\partial_k$ [$\ket{n(\bm {k})}\bra{n(\bm {k})}$] takes the form $\exp(2i \pi x)$ [$P$] in the real space. These result in the real space Berry curvature $U V U^\dagger V^\dagger$ leading to the Bott index as given in Eq. (\ref{eq:bott_pbc}).

%----------------------------------------------------------------------------------------------------
\begin{figure}
    \centering
    \includegraphics[width=0.235\textwidth]{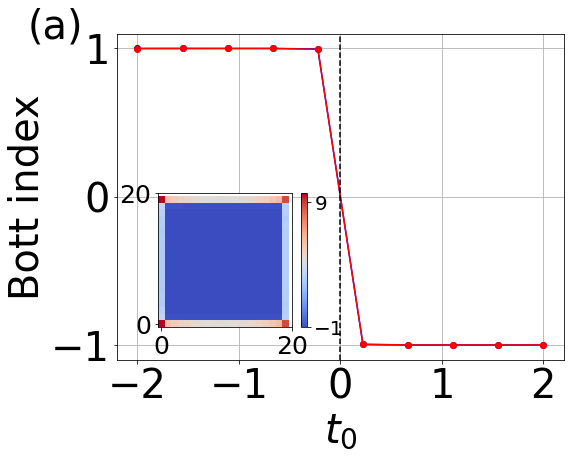}
    \includegraphics[width=0.235\textwidth]{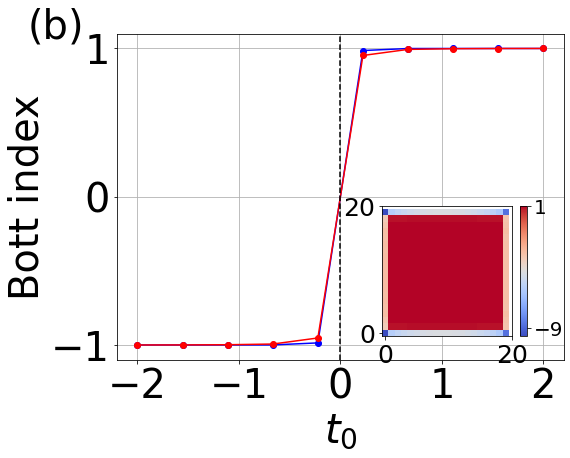}
    
    \caption{The variation of open-bulk and periodic-bulk Bott indices given in Eqs. (\ref{eq:bott_pbc}) and (\ref{eq:bott_obc}) with $t_0$, depicted by red and blue  points, respectively, for driven SSH4 model. We  consider a square lattice of size $L_x\times L_y = 20\times 20$ with $d_0=1$ in (a) and $d_0=-2$ in (b).  Insets in (a) and (b)  show the spatial distribution of {local Chern marker} under OBC for $(t_0, d_0)=(1,1)$ and  $(1,-1) $, respectively. } 
\label{fig:ssh4_bott_pbc}    
\end{figure}
%----------------------------------------------------------------------------------------------------

%----------------------------------------------------------------------------------------------------
\begin{figure}
    \centering
    \includegraphics[width=0.235\textwidth]{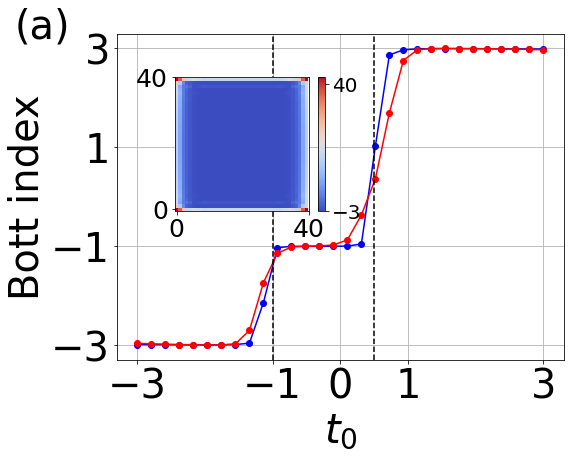}
    \includegraphics[width=0.235\textwidth]{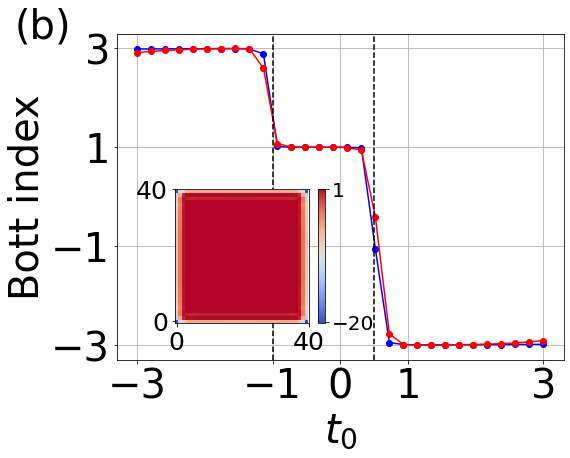}

    \caption{ The variation of open-bulk and periodic-bulk Bott indices
     given in Eqs. (\ref{eq:bott_pbc}) and (\ref{eq:bott_obc}) with $t_0$, depicted by red and blue  points, respectively, for driven SSH4 model with $d_0=-2$ in (a) and $d_0=1$ in (b).  We consider a square lattice of size $L_x\times L_y = 30\times 30$.  Insets in (a) and (b)  show the {local Chern marker} for $(t_0, d_0)=(1,1)$ and  $(1,-0.5) $, respectively in a square lattice of size $L_x\times L_x = 40\times 40$.  }
\label{fig:sshlr_bott_pbc}

\end{figure}
%----------------------------------------------------------------------------------------------------

\textit{Bott index from OBC or open-bulk Chern number:}
Having demonstrated the Bott index in terms of the unitary matrices $U$ and $V$ under PBC, we now extend the analysis to OBC. In this case, the Bott index is referred to as open-bulk Chern number. However, note that the Bott index under OBC yields primarily vanishing results. Starting from the Berry curvature in the momentum space  $F_{12}(k,t) = \partial_1 A_2(k,t) - \partial_2 A_1(k,t)= {\rm Tr}(P(k,t) [\partial_1 P(k,t),\partial_2 P(k,t)])$, where $P(k,t)$ represents the projector onto the occupied states, we can obtain the corresponding real space form. Here,  $\partial_1 P(k,t)= -i [x,P]$ and $\partial_2 P(k,t)= -i [y,P]$. Therefore, the real space Berry curvature takes the form $[P[x,P],P[y,P]]$ where $P$ represents the projector to many-body ground state obtained from the Hamiltonian under OBC. Here $x$, and $y$ represent the Fourier spaces associated with the $k$, and $t$, respectively.   After expanding the nested commutator, $[P[x,P],P[y,P]]$ reduces to $[PxP,PyP]$ resulting in the following expression of Bott index nder OBC
\begin{equation}
    B=\frac{2\pi}{\tilde{L}_x \tilde{L}_y} {\rm Im}\big({\rm Tr'}[PxP,PyP]\big)
    \label{eq:bott_obc}
\end{equation}
Where the term ${\rm Tr'}$ represent the partial trace discarding the boundaries. The physical dimension of the system is $L_x \times L_y$ with $L_{x,y}=2 l_{x,y} + \tilde{L}_{x,y} $ where $l_{x}$ ($l_{y}$) denotes the number of left and right (top and bottom) boundary sites along $x$ ($y$) directions while $L_{x,y}$ denotes the number of  bulk sites along $x$ and $y$ directions. The open-boundary Chern number is similar to the open-boundary winding number as discussed in Sec. \ref{s3ss2}  where the partial trace operation captures the absolute bulk contribution.

\begin{figure}
    \centering
    \includegraphics[width=0.235\textwidth]{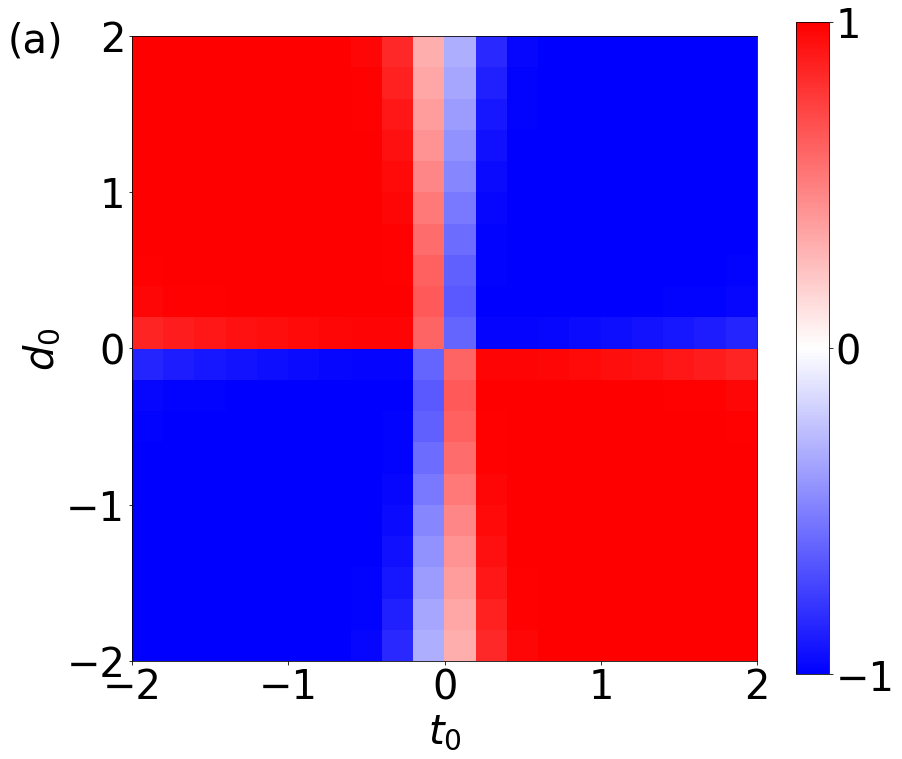}
    \includegraphics[width=0.235\textwidth]{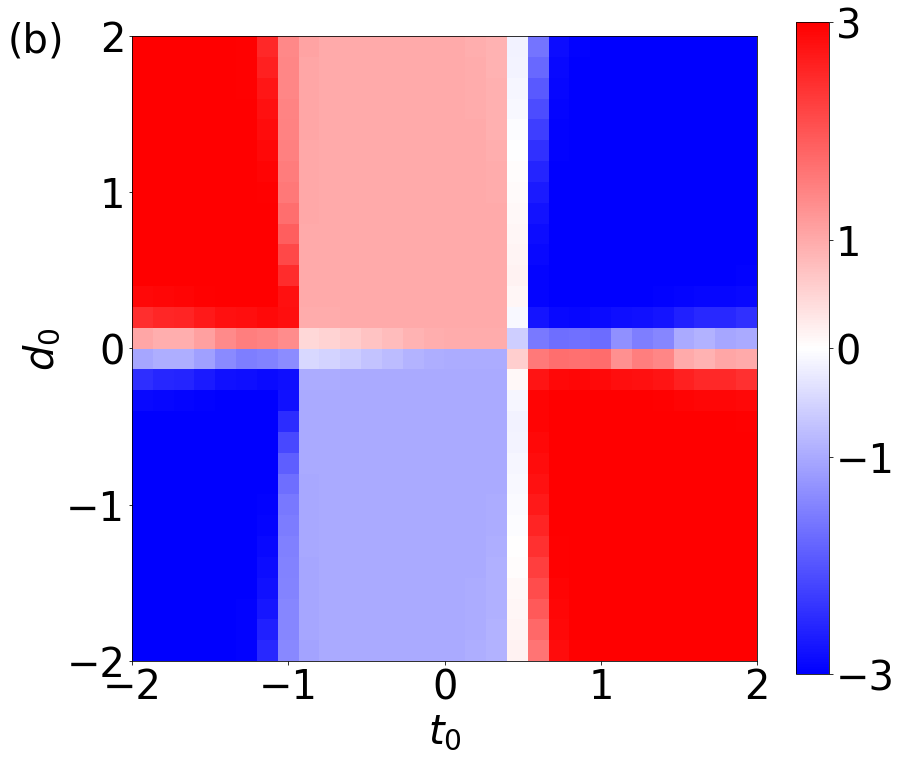}
    \caption{Topological phase diagram for driven SSH4 model and SSHLR model using open-bulk Bott index Eq. (\ref{eq:bott_obc}) are depicted in (a) and (b), respectively. Here $a=b=h_0=T=1$, and $L_x=L_y=50$. The color bar represents the values of  open-bulk Bott index.  }
\label{fig:ssh_bott_pbc}
\end{figure}

We also show the variation of spatio-temporal periodic (open)-bulk Bott index as a function of $t_0$ for the driven SSH4 model in Fig. \ref{fig:ssh4_bott_pbc} as depicted by blue (red) lines. We demonstrate different cases with 
$d_0=1$ and $-2$ in Fig. \ref{fig:ssh4_bott_pbc} (a,b), respectively, where $B$ changes from $1$ to $-1$ [$-1$ to $1$], crossing $0$ at $d_0=0$. This is consistent with the momentum space Chern number as shown in Fig. \ref{fig:chern_pd} (a).  A similar analysis on periodic (open)-bulk Bott index is carried out for the driven SSHLR model in  Fig. \ref{fig:ssh4_bott_pbc} (a,b) for $d_{0}=-2 $, and $1$, respectively, as depicted by blue (red) lines.  We find the jumps between $B=\pm 3$  passing through $B=-1$ ($1$) for $d_0=-2$, and $1$ for $d_0=1$.  In the insets of  Figs. \ref{fig:ssh4_bott_pbc} and \ref{fig:sshlr_bott_pbc}, we show the {local Chern marker} giving us the idea of how many lattice sites we have to consider for partial trace. {The local Chern marker can be thought of as the variation of Berry curvature over the real space lattice. The local Chern marker in the two-dimensional $(x,y)$-grid is computed from 
$2\pi ({\rm Im}[PxP,PyP]) $ without executing the partial trace in Eq. \ref{eq:bott_obc}, however, the contribution at each site in $(x,y)$-grid is obtained after summing over $4$ $[2]$ orbital degrees of freedom for SSH4 [SSHLR] model. } The transition of $B$ between two quantized plateau is more sharp for the periodic-bulk case as compared to the open-bulk case suggestion that the finite size effect is more in the open-bulk Bott index.  For sake of completeness, we show the topological phase diagram using open-bulk Bott index Eq. (\ref{eq:bott_obc}) for driven SSH4 and SSHLR models in Fig. \ref{fig:ssh_bott_pbc} (a,b), respectively.   One can clearly see
that these phase diagrams match quite  well with the Chern number phase diagram shown in Figs.  \ref{fig:chern_pd} (a,b) except for the fluctuations around the phase boundaries. This can be attributed to the finite size effect as discussed earlier for the OBC case.

{In this context, it is important to mention the connection with the Floquet theory \cite{ghosh2023generation, Ghosh22_dyn,Mondal23_dyn,Rodriguez-Vega_2018,Eckardt_2015,Nag15,Ghosh20_Flo}. The periodic drive can be analytically analyzed by various limits of Floquet theory that include high and low amplitude and frequency of the drive  as described by 
Floquet perturbation theory \cite{Nag15,Ghosh20_Flo}, low-frequency  \cite{Rodriguez-Vega_2018,Nag15} and 
high-frequency expansions \cite{Eckardt_2015,Mikami16}. The quasi-states and quasi-energies of a periodically driven system can be exactly obtained by numerically 
diagonalizing the stroboscopic Floquet Hamiltonian in the frequency space or the complete period evolution operator in the time space. Note that the matrix size of the stroboscopic Floquet Hamiltonian or the time step to compute the complete period evolution operator increases as frequency decreases. The topological gap closing and emergence of various topological phases by varying frequency is a major finding of Floquet theory. Adiabatic analysis is not able capture such variation as frequency is kept at an ultra-low value with time as a parameter and time-dependent Schr\"odinger equation no longer needs to be solved. However, the adiabatic dynamics leads to essential insight for the driven system in the ultra-low frequency limit. }

{Digging deep into the adiabatic analysis, one can find that  the quasi-states and  quasi-energies are obtained from the eigenstates of
$H(t=T)$ and $(1/T)\int_0^T E(t) dt$, respectively, where $E(t)$ is the eigenvalue of $H(t)$. 
Therefore, the adiabatic dynamics is a special case of Floquet theory. However, obtaining the quasi-energy and quasi-states  would become numerically difficult in the ultra-low frequency limit as number of photon sectors under consideration is enormously high leading to a shoot up in the size of the  stroboscopic Floquet Hamiltonian matrix. This problem is not there for the high-frequency limit where numerical and analytical solution both exist. Therefore, adiabatic theory can shed light on the ultra-low frequency limit which is very difficult to obtain using the Floquet theory exactly. Interestingly, for topological Floquet systems, boundary modes can exist at $0$ as well as $\pi$ quasi-energies \cite{Rudner13}. Adiabatic theory is not able to identify the $\pi$-quasi-energy states while $0$-quasi-energy states can only be identified by the above theory \cite{Rodriguez-Vega_2018,Nag15}. Therefore, adiabatic limit has its own limitation as well. Nonetheless, adiabatic analysis can qualitatively explain the Floquet result in the ultra-low frequency limit.}

\subsection{Quantum Metric}
\label{s5ss2}
We now focus on the quantum geometric tensor (QGT), defined as general covariant tensor in Hilbert space geometry ${Q^n_{\mu \nu }}= \langle \partial_\mu n(\bm {k})|(1-\ket{n(\bm {k})}\bra{n(\bm {k})})| \partial_\nu n(\bm {k})\rangle$ where $\ket{n(\bm{k})}$ denotes the $n$-th quantum state namely, $n$-th Bloch band. This provides a  measure of
 distance under the quantum adiabatic evolution path in parameter space from  $\ket{n(\bm{k})}$ to $\ket{n(\bm{k+\delta k})}$: $ds^2=\sum_{\mu,\nu} Q^n_{\mu \nu } dk_{\mu} dk_{\nu} $. The QGT can be decomposed into 
the symmetric part namely, Riemannian metric $g_{\mu \nu }$ and anti-symmetric part $F_{\mu \nu }$ such that 
$Q^n_{\mu \nu }= {g_{\mu \nu }} - \frac{i}{2}{F_{\mu \nu }}$. The anti-symmetric part vanishes after the summation leading to quantum distance 
$ds^2=\sum_{\mu,\nu} {\rm Re}[Q^n_{\mu \nu }] dk_{\mu} dk_{\nu}= g_{\mu \nu } dk_{\mu} dk_{\nu} $. The imaginary part $F_{\mu \nu }$ results in 
Berry curvature, encoding the winding of the Bloch wave-function over the Brillouin zone. 
So far we have been focusing on the topology associated with the Berry curvature part yielding quantized Chern number. Interestingly, the real part  ${g_{\mu \nu }}$ results in Euler characteristic number  after integration that characterizes the closed Bloch states manifold in the first Brillouin zone. The Euler characteristic number of all occupied bands is defined as 
\begin{equation}
    \chi  = \frac{1}{4\pi } \sum_n\int \int_{BZ} { R}^n {\sqrt{{\rm \det} (g_{\mu \nu })} \, dk_\mu \, dk_\nu}
\end{equation}
where ${ R}^n$ is the Ricci scalar curvature associated with $\ket{n(\bm{k})}$.
{The Ricci scalar is given by $R=g^{ij}R_{ij}$ and  Riemann tensor $R^i_{jkl}$ and Ricci tensor $R_{ij}$ are related by a contraction $R_{ij}=R^k_{ikj}$ with $g^{ij}=(g_{ij})^{-1}$ \cite{Ma2014}.  The relation between Ricci tensor $R_{jk}$ and quantum metric components $g_{ij}$ are given by the Christoffel symbols  $\Gamma_{ij}^k$ as follows: 
$R_{jk}=\partial_i \Gamma_{jk}^i- \partial_j \Gamma_{ki}^i + \Gamma_{ip}^i \Gamma_{jk}^p -\Gamma_{jp}^i\Gamma_{ik}^p$
and $\Gamma_{ij}^k=\frac{g^{kl}}{2}(\partial_i g_{jl} + \partial_j g_{il}- \partial_l g_{ij})$ \cite{do2016differential, lee2006riemannian}. For completeness, we mention the relation between the Riemann tensor and quantum metric  $R^i_{jkl}=\partial_k \Gamma_{jl}^i- \partial_l \Gamma_{jk}^i + \Gamma_{km}^i \Gamma_{jl}^m -\Gamma_{lm}^i\Gamma_{jk}^m$.
Interestingly, in two-dimensional Riemannian manifolds, the Gauss curvature $\kappa$ is half of the Ricci scalar curvature $R$, $R=2 \kappa$, resulting in the Euler number to acquire the form $\chi = (1/2\pi)\int \kappa dA= (1/4\pi)\int R \sqrt{{\rm det}[g_{ij}]} dk_{i} dk_{j}$ with  $\kappa=R_{ijij}/ {\rm det}[g_{ij}]$, and $dA=\sqrt{{\rm det}[g_{ij}]} dk_1 dk_2$, being the elemental surface area in $k_1$-$k_2$ plane. }

The quantum metric is related to fidelity susceptibility $\chi_F$ where $F= |\bra{n(\bm {k})} n(\bm {k+ \delta k})\rangle|^2=1-(\chi_F/2) \, dk_\mu \, dk_\nu$ \cite{gu2010fidelity}. 
This number may not always be quantized for topological phases, unlike the Chern number which is always quantized for the time reversal symmetry broken topological phases in 2D irrespective of the models \cite{Zeng_2024,Ozawa2021}. 
Interestingly, the phase boundaries can be captured by quantum metric where the Euler characteristics number  
is expected to change abruptly.

In the present case with driven extended-SSH models, one can identify $k_\mu,k_\nu \equiv 1,2 =k,t$ and quantum metric tensor is given by 
\begin{equation}
g =
\begin{bmatrix}
 g_{kk} & g_{kt} \\
 g_{tk} & g_{tt}    
\end{bmatrix}
\end{equation}
where
\begin{equation}
g_{\mu \nu} = \sum_{n} \frac{H_{\mu}^{0n}(k,t) H_{\nu}^{n0}(k,t)}{\left(E_n - E_0 \right)^2}
\end{equation}
where
\begin{equation}
H_{\mu(\nu)}^{mn}(k,t) = \langle m(k,t) | H_{\mu(\nu)}(k,t) | n(k,t) \rangle
\nonumber\\
\end{equation}
and the operator \(H_{\mu(\nu)}(k,t)\) is defined as
\begin{equation}
H_{\mu(\nu)}(k,t) = \frac{\partial H(k,t)}{\partial \mu(\nu)}.
\end{equation}
The  Euler characteristic number  is  given by
\begin{equation}
\chi  = \frac{2}{\pi }\int_{BZ} {\sqrt{\det g} \, dk \, dt}
\label{eq:qm_simplified}
\end{equation}
where Ricci scalar $R$ is considered as  a constant $R=8$ without loss of generality \cite{Zeng_2024,Ozawa2021,Ma2014}. %Note that Eq. (\ref{eq:qm_simplified}) also known as quantum volume \cite{Ozawa2021}. 

\begin{figure}
    \centering
    \includegraphics[width=0.5\textwidth]{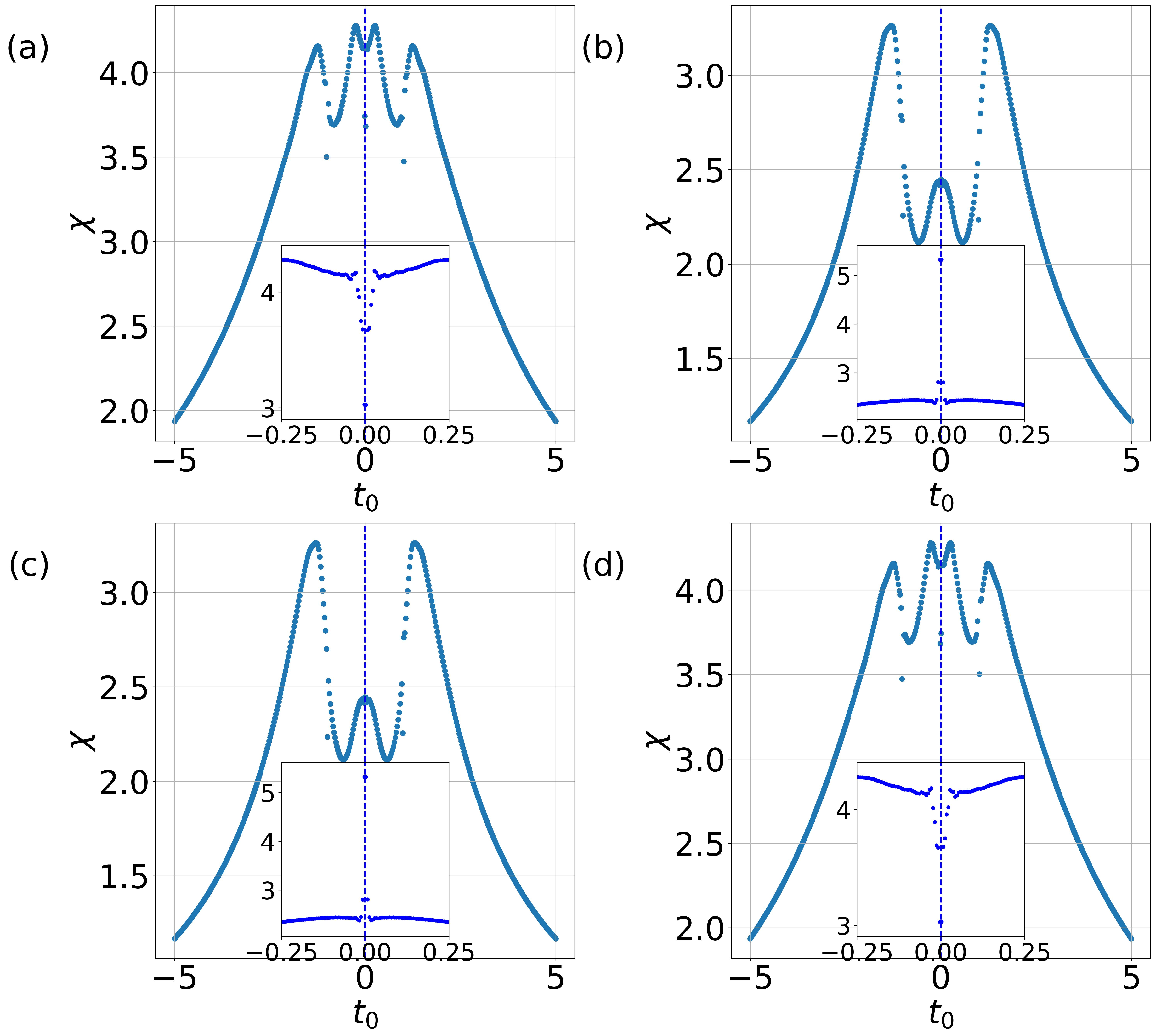}
    \caption{The variation of Euler characteristics number $\chi$ Eq. (\ref{eq:qm_simplified}) as a function of $t_0$ is shown in (a,b,c,d) with  $d_{0}=-2$, $-1$, $1$ and $2$, respectively, for driven SSH4 model.  {The fluctuations around $t_0=0$ are shown in the insets where the main plots are zoomed in within the window $<-0.25<t_0<0.25$.}}
\label{fig:SSH4_qm}    
\end{figure}

\begin{figure}
    \centering
    \includegraphics[width=0.45\textwidth]{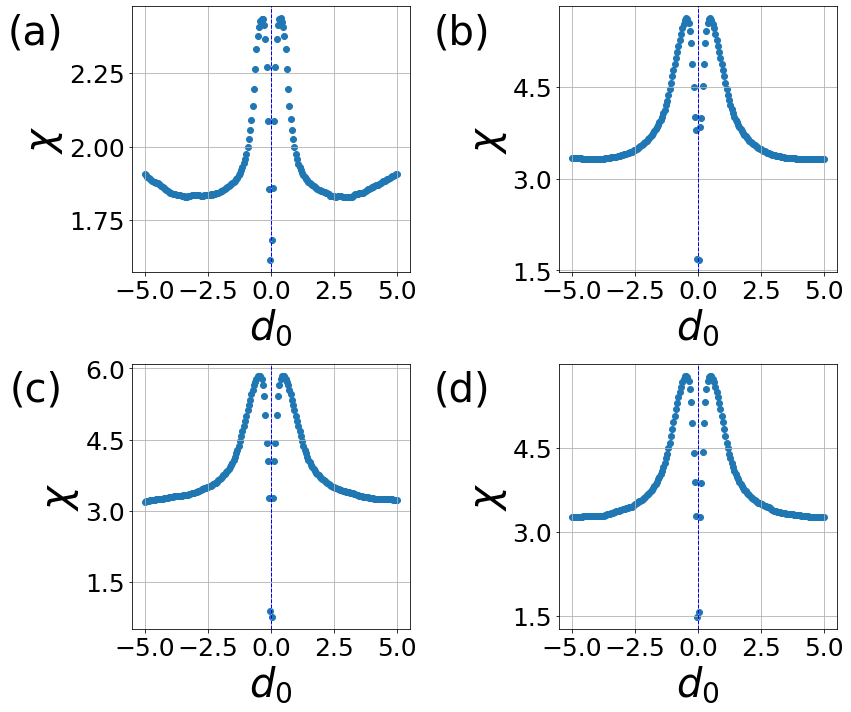}
    \includegraphics[width=0.45\textwidth]{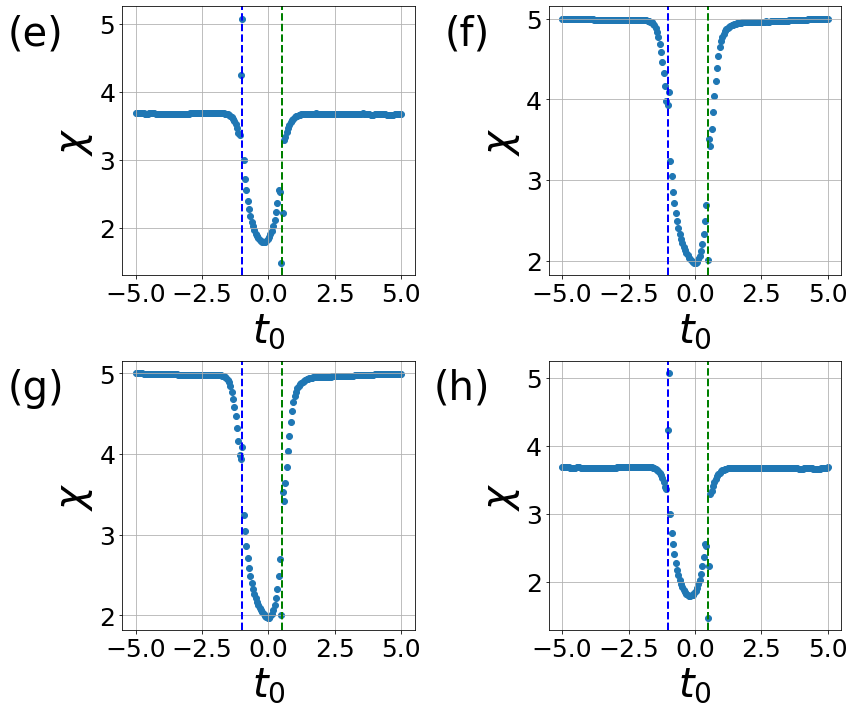}
    \caption{We examine the evolution of $\chi$ Eq. (\ref{eq:qm_simplified}) as a function of $d_0$ [$t_0$] in (a,b,c,d) and [(e,f,g,h)], respectively, for driven SSHLR model. We consider $t_0=0$, $1$, $2$, and $-1.5$ [$d_0=-2$, $-1$, $1$, and $2$] for (a,b,c,d) [(e,f,g,h)], respectively.}
\label{fig:SSH4lr_qm}        
\end{figure}

We examine the variation of $\chi$ with a parameter of the driven SSH4 model such that the phase transitions can be captured, see Fig.  \ref{fig:SSH4_qm} (a,b,c,d) for $d_0=-2$, $-1$, $1$ and $2$, respectively. Interestingly, $\chi$ behaves symmetrically around $t_0=0$ line while the Chern number changes its sign. When $t_0$ approaches $0$, $\chi$ shows fluctuations {as depicted in the insets of Fig.  \ref{fig:SSH4_qm} (a,b,c,d)}.  The amplitude of these fluctuations increases when $d_0$ approaches another critical point $d_0=0$  which can be understood from the fact that $\chi$ is more sensitive to the critical points. Interestingly, within a given phase, $\chi$ changes monotonically leading to the identical nature of the these phases. We find qualitatively similar behavior for driven SSHLR model when $d_0$ is varied keeping $t_0$ fixed at certain values, see Fig. \ref{fig:SSH4lr_qm} (e,f,g,h). The fluctuations around $d_0=0$ signify the existence of a  critical point at $d_0=0$. On the other hand, the fluctuations are clearly observed at $t=0.5$ and $-1$ when $\chi$ is plotted as a function of $t_0$. Interestingly, one can see the quantized behavior of $\chi$ for $t_0>1$ and $t_0<-2$, however, this quantization does not always correspond to an integer number.  Therefore, the $\chi$ is not expected to show integer quantization rather, its fluctuation around phase transition points refers to a change in phase. { We note that the quantum volume, computed using the integral of $\sqrt{{\rm det}[ g_{\mu \nu}]}$, shows (does not show) quantization in the two-band(four-band model) thus justifying our observation in  Figs. \ref{fig:SSH4_qm} and \ref{fig:SSH4lr_qm} for SSH4 and SSHLR models,    respectively \cite{Ozawa2021}.}

\section{CONCLUSION AND SUMMARY}
\label{s6}

In this work, we examine the extended SSH models namely, SSH4 and SSHLR models to investigate the variety of topological phases. Owing to the chiral symmetry, we compute winding number to characaterize these topological phases where SSHLR model exhibits higher values of winding number, associated with a larger number of zero-energy edge modes, than SSH4 model due to the presence of the long-range terms. We connect the momentum space invariant with the real space invariant to show the robustness of the open-bulk and periodic-bulk real space invariants with the different tuning parameters. Given the same system size, the  phase boundaries are more prominently observed for the case of periodic-bulk real space winding number indicating the system size effects.

Having understood the static phases, we introduce the periodic adiabatic dynamics where the charge pumping is examined after a full period. Due to the presence of another  periodic variable namely time in addition to the momentum, we can study the 2D Chern number to characterize the dynamic phases. Interestingly, we find that the adiabatically driven model always hosts topological phases such that there exists finite value of pumped charge throughout the parameter spaces. This is further supported by the time evolution of the open-boundary energy levels and time evolution of pumped charge, computed from real and momentum space versions of the time-dependent Hamiltonian, respectively. The underlying trivial phase of the static model is dynamically assembled with the topological phases leading eventually to a variety of topological phases. The intriguing windings of the edge modes along time direction are responsible for topological phases in the driven model. 
The adiabatic drive not only mediates topological phases in the trival part of the static phase diagram but also Chern number can acquire higher values than that of the static winding number. In order to analyze this phases further we compute the spatio-temporal Bott index where the time-direction is replaced by another lattice along the orthogonal direction to the original 1D lattice. This exercise validates topological phases  of the driven model using  the open-bulk and periodic-bulk spatio-temporal Bott index where the finite size effects are visible in the former one. We further extend our study to Euler characteristics number for the driven models that show fluctuations around the phase boundaries suggesting the fact that quantum metric is sensitive to different phases.

Now coming to the  experimental viability of our work, we note that topological phases in the SSH model have already been experimentally  reported using 1D  mechanical system  \cite{Thatcher_2022} where  elastic string with metallic masses emulate the  atomic sites. Importantly, ultracold atoms in  momentum lattices are found to extremely useful to experimentally realize the SSH and extended SSH models \cite{xie2019topological, meier2016observation}.  Using the platform of photonic lattices, SSH model is also reported experimentally   
\cite{Gabriel2022}. In the near future, the charge transport using the periodic drive can be implemented in this model using meta-material platforms \cite{peng2016experimental} where our findings could be of interest to the community. Apart from experimental connection, our study paves the way to understand the adiabatic charge transport from a different perspective. One can introduce disorder and examine its effect on the charge transport. Furthermore, the Floquet theory of charge transport can also be studied in which the adiabatic limit happens to be a limiting case. Therefore, our work has the potential to initiate multiple studies in the field of adiabatic charge transport.

\section{acknowledgement}
We thank Arnob K. Ghosh for useful discussions on real space winding numbers.   T.N. acknowledges the
NFSG ``NFSG/HYD/2023/H0911" from BITS Pilani.

%%%%%%%%%%%%%%%%%%%%%%%%%%%%%%%%%%%%%%%%%%%%%%%%%%%%%%%%%%%%%%%%%%%

%\bibliographystyle{apsrev4-2mod}
%\bibliography{bibfile}{}

\begin{thebibliography}{0}%
\makeatletter
\providecommand \@ifxundefined [1]{%
 \@ifx{#1\undefined}
}%
\providecommand \@ifnum [1]{%
 \ifnum #1\expandafter \@firstoftwo
 \else \expandafter \@secondoftwo
 \fi
}%
\providecommand \@ifx [1]{%
 \ifx #1\expandafter \@firstoftwo
 \else \expandafter \@secondoftwo
 \fi
}%
\providecommand \natexlab [1]{#1}%
\providecommand \enquote  [1]{``#1''}%
\providecommand \bibnamefont  [1]{#1}%
\providecommand \bibfnamefont [1]{#1}%
\providecommand \citenamefont [1]{#1}%
\providecommand \href@noop [0]{\@secondoftwo}%
\providecommand \href [0]{\begingroup \@sanitize@url \@href}%
\providecommand \@href[1]{\@@startlink{#1}\@@href}%
\providecommand \@@href[1]{\endgroup#1\@@endlink}%
\providecommand \@sanitize@url [0]{\catcode `\\12\catcode `\$12\catcode
  `\&12\catcode `\#12\catcode `\^12\catcode `\_12\catcode `\%12\relax}%
\providecommand \@@startlink[1]{}%
\providecommand \@@endlink[0]{}%
\providecommand \url  [0]{\begingroup\@sanitize@url \@url }%
\providecommand \@url [1]{\endgroup\@href {#1}{\urlprefix }}%
\providecommand \urlprefix  [0]{URL }%
\providecommand \Eprint [0]{\href }%
\providecommand \doibase [0]{http://dx.doi.org/}%
\providecommand \selectlanguage [0]{\@gobble}%
\providecommand \bibinfo  [0]{\@secondoftwo}%
\providecommand \bibfield  [0]{\@secondoftwo}%
\providecommand \translation [1]{[#1]}%
\providecommand \BibitemOpen [0]{}%
\providecommand \bibitemStop [0]{}%
\providecommand \bibitemNoStop [0]{.\EOS\space}%
\providecommand \EOS [0]{\spacefactor3000\relax}%
\providecommand \BibitemShut  [1]{\csname bibitem#1\endcsname}%
\let\auto@bib@innerbib\@empty
%</preamble>
\end{thebibliography}%


\begin{thebibliography}{107}%
\makeatletter
\providecommand \@ifxundefined [1]{%
 \@ifx{#1\undefined}
}%
\providecommand \@ifnum [1]{%
 \ifnum #1\expandafter \@firstoftwo
 \else \expandafter \@secondoftwo
 \fi
}%
\providecommand \@ifx [1]{%
 \ifx #1\expandafter \@firstoftwo
 \else \expandafter \@secondoftwo
 \fi
}%
\providecommand \natexlab [1]{#1}%
\providecommand \enquote  [1]{``#1''}%
\providecommand \bibnamefont  [1]{#1}%
\providecommand \bibfnamefont [1]{#1}%
\providecommand \citenamefont [1]{#1}%
\providecommand \href@noop [0]{\@secondoftwo}%
\providecommand \href [0]{\begingroup \@sanitize@url \@href}%
\providecommand \@href[1]{\@@startlink{#1}\@@href}%
\providecommand \@@href[1]{\endgroup#1\@@endlink}%
\providecommand \@sanitize@url [0]{\catcode `\\12\catcode `\$12\catcode
  `\&12\catcode `\#12\catcode `\^12\catcode `\_12\catcode `\%12\relax}%
\providecommand \@@startlink[1]{}%
\providecommand \@@endlink[0]{}%
\providecommand \url  [0]{\begingroup\@sanitize@url \@url }%
\providecommand \@url [1]{\endgroup\@href {#1}{\urlprefix }}%
\providecommand \urlprefix  [0]{URL }%
\providecommand \Eprint [0]{\href }%
\providecommand \doibase [0]{http://dx.doi.org/}%
\providecommand \selectlanguage [0]{\@gobble}%
\providecommand \bibinfo  [0]{\@secondoftwo}%
\providecommand \bibfield  [0]{\@secondoftwo}%
\providecommand \translation [1]{[#1]}%
\providecommand \BibitemOpen [0]{}%
\providecommand \bibitemStop [0]{}%
\providecommand \bibitemNoStop [0]{.\EOS\space}%
\providecommand \EOS [0]{\spacefactor3000\relax}%
\providecommand \BibitemShut  [1]{\csname bibitem#1\endcsname}%
\let\auto@bib@innerbib\@empty
%</preamble>
\bibitem [{\citenamefont {Nakahara}(2018)}]{nakahara2018geometry}%
  \BibitemOpen
  \bibfield  {author} {\bibinfo {author} {\bibfnamefont {Mikio}\ \bibnamefont
  {Nakahara}},\ }\href@noop {} {\emph {\bibinfo {title} {Geometry, topology and
  physics}}}\ (\bibinfo  {publisher} {CRC press},\ \bibinfo {year}
  {2018})\BibitemShut {NoStop}%
\bibitem [{\citenamefont {Mukhi}\ and\ \citenamefont
  {Mukunda}(1990)}]{mukhi1990introduction}%
  \BibitemOpen
  \bibfield  {author} {\bibinfo {author} {\bibfnamefont {Sunil}\ \bibnamefont
  {Mukhi}}\ and\ \bibinfo {author} {\bibfnamefont {N}~\bibnamefont {Mukunda}},\
  }\href@noop {} {\emph {\bibinfo {title} {Introduction to topology,
  differential geometry and group theory for physicists}}}\ (\bibinfo
  {publisher} {Wiley Eastern},\ \bibinfo {year} {1990})\BibitemShut {NoStop}%
\bibitem [{\citenamefont {Qi}\ and\ \citenamefont
  {Zhang}(2011{\natexlab{a}})}]{Shou-Cheng11}%
  \BibitemOpen
  \bibfield  {author} {\bibinfo {author} {\bibfnamefont {Xiao-Liang}\
  \bibnamefont {Qi}}\ and\ \bibinfo {author} {\bibfnamefont {Shou-Cheng}\
  \bibnamefont {Zhang}},\ }\bibfield  {title} {\enquote {\bibinfo {title}
  {Topological insulators and superconductors},}\ }\href {\doibase
  10.1103/RevModPhys.83.1057} {\bibfield  {journal} {\bibinfo  {journal} {Rev.
  Mod. Phys.}\ }\textbf {\bibinfo {volume} {83}},\ \bibinfo {pages}
  {1057--1110} (\bibinfo {year} {2011}{\natexlab{a}})}\BibitemShut {NoStop}%
\bibitem [{\citenamefont {Bernevig}\ and\ \citenamefont
  {Hughes}(2013)}]{bernevig2013topological}%
  \BibitemOpen
  \bibfield  {author} {\bibinfo {author} {\bibfnamefont {BA}~\bibnamefont
  {Bernevig}}\ and\ \bibinfo {author} {\bibfnamefont {TL}~\bibnamefont
  {Hughes}},\ }\bibfield  {title} {\enquote {\bibinfo {title} {Topological
  insulators and topological superconductors princeton university press},}\
  }\href@noop {} {\bibfield  {journal} {\bibinfo  {journal} {New Jersey}\ }
  (\bibinfo {year} {2013})}\BibitemShut {NoStop}%
\bibitem [{\citenamefont {Halperin}(2019)}]{Halperin2019}%
  \BibitemOpen
  \bibfield  {author} {\bibinfo {author} {\bibfnamefont {Bertrand~I.}\
  \bibnamefont {Halperin}},\ }\bibfield  {title} {\enquote {\bibinfo {title}
  {On the hohenberg--mermin--wagner theorem and its limitations},}\ }\href
  {\doibase 10.1007/s10955-018-2202-y} {\bibfield  {journal} {\bibinfo
  {journal} {Journal of Statistical Physics}\ }\textbf {\bibinfo {volume}
  {175}},\ \bibinfo {pages} {521--529} (\bibinfo {year} {2019})}\BibitemShut
  {NoStop}%
\bibitem [{\citenamefont {Guo}(2016)}]{Guo2016}%
  \BibitemOpen
  \bibfield  {author} {\bibinfo {author} {\bibfnamefont {Huai-Ming}\
  \bibnamefont {Guo}},\ }\bibfield  {title} {\enquote {\bibinfo {title} {A
  brief review on one-dimensional topological insulators and
  superconductors},}\ }\href {\doibase 10.1007/s11433-015-5773-5} {\bibfield
  {journal} {\bibinfo  {journal} {Science China Physics, Mechanics {\&}
  Astronomy}\ }\textbf {\bibinfo {volume} {59}},\ \bibinfo {pages} {637401}
  (\bibinfo {year} {2016})}\BibitemShut {NoStop}%
\bibitem [{\citenamefont {Altland}\ and\ \citenamefont
  {Zirnbauer}(1997)}]{Altland97}%
  \BibitemOpen
  \bibfield  {author} {\bibinfo {author} {\bibfnamefont {Alexander}\
  \bibnamefont {Altland}}\ and\ \bibinfo {author} {\bibfnamefont {Martin~R.}\
  \bibnamefont {Zirnbauer}},\ }\bibfield  {title} {\enquote {\bibinfo {title}
  {Nonstandard symmetry classes in mesoscopic normal-superconducting hybrid
  structures},}\ }\href {\doibase 10.1103/PhysRevB.55.1142} {\bibfield
  {journal} {\bibinfo  {journal} {Phys. Rev. B}\ }\textbf {\bibinfo {volume}
  {55}},\ \bibinfo {pages} {1142--1161} (\bibinfo {year} {1997})}\BibitemShut
  {NoStop}%
\bibitem [{\citenamefont {Chiu}\ \emph {et~al.}(2016)\citenamefont {Chiu},
  \citenamefont {Teo}, \citenamefont {Schnyder},\ and\ \citenamefont
  {Ryu}}]{Chiu16}%
  \BibitemOpen
  \bibfield  {author} {\bibinfo {author} {\bibfnamefont {Ching-Kai}\
  \bibnamefont {Chiu}}, \bibinfo {author} {\bibfnamefont {Jeffrey C.~Y.}\
  \bibnamefont {Teo}}, \bibinfo {author} {\bibfnamefont {Andreas~P.}\
  \bibnamefont {Schnyder}}, \ and\ \bibinfo {author} {\bibfnamefont {Shinsei}\
  \bibnamefont {Ryu}},\ }\bibfield  {title} {\enquote {\bibinfo {title}
  {Classification of topological quantum matter with symmetries},}\ }\href
  {\doibase 10.1103/RevModPhys.88.035005} {\bibfield  {journal} {\bibinfo
  {journal} {Rev. Mod. Phys.}\ }\textbf {\bibinfo {volume} {88}},\ \bibinfo
  {pages} {035005} (\bibinfo {year} {2016})}\BibitemShut {NoStop}%
\bibitem [{\citenamefont {Ryu}\ \emph {et~al.}(2010)\citenamefont {Ryu},
  \citenamefont {Schnyder}, \citenamefont {Furusaki},\ and\ \citenamefont
  {Ludwig}}]{Ryu_2010}%
  \BibitemOpen
  \bibfield  {author} {\bibinfo {author} {\bibfnamefont {Shinsei}\ \bibnamefont
  {Ryu}}, \bibinfo {author} {\bibfnamefont {Andreas~P}\ \bibnamefont
  {Schnyder}}, \bibinfo {author} {\bibfnamefont {Akira}\ \bibnamefont
  {Furusaki}}, \ and\ \bibinfo {author} {\bibfnamefont {Andreas W~W}\
  \bibnamefont {Ludwig}},\ }\bibfield  {title} {\enquote {\bibinfo {title}
  {Topological insulators and superconductors: tenfold way and dimensional
  hierarchy},}\ }\href {\doibase 10.1088/1367-2630/12/6/065010} {\bibfield
  {journal} {\bibinfo  {journal} {New Journal of Physics}\ }\textbf {\bibinfo
  {volume} {12}},\ \bibinfo {pages} {065010} (\bibinfo {year}
  {2010})}\BibitemShut {NoStop}%
\bibitem [{\citenamefont {von Klitzing}(1986)}]{Klitzing86}%
  \BibitemOpen
  \bibfield  {author} {\bibinfo {author} {\bibfnamefont {Klaus}\ \bibnamefont
  {von Klitzing}},\ }\bibfield  {title} {\enquote {\bibinfo {title} {The
  quantized hall effect},}\ }\href {\doibase 10.1103/RevModPhys.58.519}
  {\bibfield  {journal} {\bibinfo  {journal} {Rev. Mod. Phys.}\ }\textbf
  {\bibinfo {volume} {58}},\ \bibinfo {pages} {519--531} (\bibinfo {year}
  {1986})}\BibitemShut {NoStop}%
\bibitem [{\citenamefont {Thouless}\ \emph {et~al.}(1982)\citenamefont
  {Thouless}, \citenamefont {Kohmoto}, \citenamefont {Nightingale},\ and\
  \citenamefont {den Nijs}}]{Thouless82}%
  \BibitemOpen
  \bibfield  {author} {\bibinfo {author} {\bibfnamefont {D.~J.}\ \bibnamefont
  {Thouless}}, \bibinfo {author} {\bibfnamefont {M.}~\bibnamefont {Kohmoto}},
  \bibinfo {author} {\bibfnamefont {M.~P.}\ \bibnamefont {Nightingale}}, \ and\
  \bibinfo {author} {\bibfnamefont {M.}~\bibnamefont {den Nijs}},\ }\bibfield
  {title} {\enquote {\bibinfo {title} {Quantized hall conductance in a
  two-dimensional periodic potential},}\ }\href {\doibase
  10.1103/PhysRevLett.49.405} {\bibfield  {journal} {\bibinfo  {journal} {Phys.
  Rev. Lett.}\ }\textbf {\bibinfo {volume} {49}},\ \bibinfo {pages} {405--408}
  (\bibinfo {year} {1982})}\BibitemShut {NoStop}%
\bibitem [{\citenamefont {Jotzu}\ \emph {et~al.}(2014)\citenamefont {Jotzu},
  \citenamefont {Messer}, \citenamefont {Desbuquois}, \citenamefont {Lebrat},
  \citenamefont {Uehlinger}, \citenamefont {Greif},\ and\ \citenamefont
  {Esslinger}}]{jotzu2014chern}%
  \BibitemOpen
  \bibfield  {author} {\bibinfo {author} {\bibfnamefont {G}~\bibnamefont
  {Jotzu}}, \bibinfo {author} {\bibfnamefont {M}~\bibnamefont {Messer}},
  \bibinfo {author} {\bibfnamefont {R}~\bibnamefont {Desbuquois}}, \bibinfo
  {author} {\bibfnamefont {M}~\bibnamefont {Lebrat}}, \bibinfo {author}
  {\bibfnamefont {T}~\bibnamefont {Uehlinger}}, \bibinfo {author}
  {\bibfnamefont {D}~\bibnamefont {Greif}}, \ and\ \bibinfo {author}
  {\bibfnamefont {T}~\bibnamefont {Esslinger}},\ }\bibfield  {title} {\enquote
  {\bibinfo {title} {From chern insulators to 3d topological insulators},}\
  }\href@noop {} {\bibfield  {journal} {\bibinfo  {journal} {Nature}\ }\textbf
  {\bibinfo {volume} {515}},\ \bibinfo {pages} {237} (\bibinfo {year}
  {2014})}\BibitemShut {NoStop}%
\bibitem [{\citenamefont {Haldane}(1988)}]{Haldane88}%
  \BibitemOpen
  \bibfield  {author} {\bibinfo {author} {\bibfnamefont {F.~D.~M.}\
  \bibnamefont {Haldane}},\ }\bibfield  {title} {\enquote {\bibinfo {title}
  {Model for a quantum hall effect without landau levels: Condensed-matter
  realization of the "parity anomaly"},}\ }\href {\doibase
  10.1103/PhysRevLett.61.2015} {\bibfield  {journal} {\bibinfo  {journal}
  {Phys. Rev. Lett.}\ }\textbf {\bibinfo {volume} {61}},\ \bibinfo {pages}
  {2015--2018} (\bibinfo {year} {1988})}\BibitemShut {NoStop}%
\bibitem [{\citenamefont {Kane}\ and\ \citenamefont {Mele}(2005)}]{Kane05}%
  \BibitemOpen
  \bibfield  {author} {\bibinfo {author} {\bibfnamefont {C.~L.}\ \bibnamefont
  {Kane}}\ and\ \bibinfo {author} {\bibfnamefont {E.~J.}\ \bibnamefont
  {Mele}},\ }\bibfield  {title} {\enquote {\bibinfo {title} {Quantum spin hall
  effect in graphene},}\ }\href {\doibase 10.1103/PhysRevLett.95.226801}
  {\bibfield  {journal} {\bibinfo  {journal} {Phys. Rev. Lett.}\ }\textbf
  {\bibinfo {volume} {95}},\ \bibinfo {pages} {226801} (\bibinfo {year}
  {2005})}\BibitemShut {NoStop}%
\bibitem [{\citenamefont {Br{\"u}ne}\ \emph {et~al.}(2012)\citenamefont
  {Br{\"u}ne}, \citenamefont {Roth}, \citenamefont {Buhmann}, \citenamefont
  {Hankiewicz}, \citenamefont {Molenkamp}, \citenamefont {Maciejko},
  \citenamefont {Qi},\ and\ \citenamefont {Zhang}}]{brune2012spin}%
  \BibitemOpen
  \bibfield  {author} {\bibinfo {author} {\bibfnamefont {Christoph}\
  \bibnamefont {Br{\"u}ne}}, \bibinfo {author} {\bibfnamefont {Andreas}\
  \bibnamefont {Roth}}, \bibinfo {author} {\bibfnamefont {Hartmut}\
  \bibnamefont {Buhmann}}, \bibinfo {author} {\bibfnamefont {Ewelina~M}\
  \bibnamefont {Hankiewicz}}, \bibinfo {author} {\bibfnamefont {Laurens~W}\
  \bibnamefont {Molenkamp}}, \bibinfo {author} {\bibfnamefont {Joseph}\
  \bibnamefont {Maciejko}}, \bibinfo {author} {\bibfnamefont {Xiao-Liang}\
  \bibnamefont {Qi}}, \ and\ \bibinfo {author} {\bibfnamefont {Shou-Cheng}\
  \bibnamefont {Zhang}},\ }\bibfield  {title} {\enquote {\bibinfo {title} {Spin
  polarization of the quantum spin hall edge states},}\ }\href@noop {}
  {\bibfield  {journal} {\bibinfo  {journal} {Nature Physics}\ }\textbf
  {\bibinfo {volume} {8}},\ \bibinfo {pages} {485--490} (\bibinfo {year}
  {2012})}\BibitemShut {NoStop}%
\bibitem [{\citenamefont {Benalcazar}\ \emph {et~al.}(2017)\citenamefont
  {Benalcazar}, \citenamefont {Bernevig},\ and\ \citenamefont
  {Hughes}}]{Benalcazar17}%
  \BibitemOpen
  \bibfield  {author} {\bibinfo {author} {\bibfnamefont {Wladimir~A.}\
  \bibnamefont {Benalcazar}}, \bibinfo {author} {\bibfnamefont {B.~Andrei}\
  \bibnamefont {Bernevig}}, \ and\ \bibinfo {author} {\bibfnamefont
  {Taylor~L.}\ \bibnamefont {Hughes}},\ }\bibfield  {title} {\enquote {\bibinfo
  {title} {Electric multipole moments, topological multipole moment pumping,
  and chiral hinge states in crystalline insulators},}\ }\href {\doibase
  10.1103/PhysRevB.96.245115} {\bibfield  {journal} {\bibinfo  {journal} {Phys.
  Rev. B}\ }\textbf {\bibinfo {volume} {96}},\ \bibinfo {pages} {245115}
  (\bibinfo {year} {2017})}\BibitemShut {NoStop}%
\bibitem [{\citenamefont {Schindler}\ \emph {et~al.}(2018)\citenamefont
  {Schindler}, \citenamefont {Wang}, \citenamefont {Vergniory}, \citenamefont
  {Cook}, \citenamefont {Murani}, \citenamefont {Sengupta}, \citenamefont
  {Kasumov}, \citenamefont {Deblock}, \citenamefont {Jeon}, \citenamefont
  {Drozdov} \emph {et~al.}}]{schindler2018higher}%
  \BibitemOpen
  \bibfield  {author} {\bibinfo {author} {\bibfnamefont {Frank}\ \bibnamefont
  {Schindler}}, \bibinfo {author} {\bibfnamefont {Zhijun}\ \bibnamefont
  {Wang}}, \bibinfo {author} {\bibfnamefont {Maia~G}\ \bibnamefont
  {Vergniory}}, \bibinfo {author} {\bibfnamefont {Ashley~M}\ \bibnamefont
  {Cook}}, \bibinfo {author} {\bibfnamefont {Anil}\ \bibnamefont {Murani}},
  \bibinfo {author} {\bibfnamefont {Shamashis}\ \bibnamefont {Sengupta}},
  \bibinfo {author} {\bibfnamefont {Alik~Yu}\ \bibnamefont {Kasumov}}, \bibinfo
  {author} {\bibfnamefont {Richard}\ \bibnamefont {Deblock}}, \bibinfo {author}
  {\bibfnamefont {Sangjun}\ \bibnamefont {Jeon}}, \bibinfo {author}
  {\bibfnamefont {Ilya}\ \bibnamefont {Drozdov}},  \emph {et~al.},\ }\bibfield
  {title} {\enquote {\bibinfo {title} {Higher-order topology in bismuth},}\
  }\href@noop {} {\bibfield  {journal} {\bibinfo  {journal} {Nature physics}\
  }\textbf {\bibinfo {volume} {14}},\ \bibinfo {pages} {918--924} (\bibinfo
  {year} {2018})}\BibitemShut {NoStop}%
\bibitem [{\citenamefont {Ghosh}\ \emph {et~al.}(2023)\citenamefont {Ghosh},
  \citenamefont {Nag},\ and\ \citenamefont {Saha}}]{ghosh2023generation}%
  \BibitemOpen
  \bibfield  {author} {\bibinfo {author} {\bibfnamefont {Arnob~Kumar}\
  \bibnamefont {Ghosh}}, \bibinfo {author} {\bibfnamefont {Tanay}\ \bibnamefont
  {Nag}}, \ and\ \bibinfo {author} {\bibfnamefont {Arijit}\ \bibnamefont
  {Saha}},\ }\bibfield  {title} {\enquote {\bibinfo {title} {Generation of
  higher-order topological insulators using periodic driving},}\ }\href@noop {}
  {\bibfield  {journal} {\bibinfo  {journal} {Journal of Physics: Condensed
  Matter}\ }\textbf {\bibinfo {volume} {36}},\ \bibinfo {pages} {093001}
  (\bibinfo {year} {2023})}\BibitemShut {NoStop}%
\bibitem [{\citenamefont {Ghosh}\ \emph {et~al.}(2021)\citenamefont {Ghosh},
  \citenamefont {Nag},\ and\ \citenamefont {Saha}}]{Ghosh21}%
  \BibitemOpen
  \bibfield  {author} {\bibinfo {author} {\bibfnamefont {Arnob~Kumar}\
  \bibnamefont {Ghosh}}, \bibinfo {author} {\bibfnamefont {Tanay}\ \bibnamefont
  {Nag}}, \ and\ \bibinfo {author} {\bibfnamefont {Arijit}\ \bibnamefont
  {Saha}},\ }\bibfield  {title} {\enquote {\bibinfo {title} {Hierarchy of
  higher-order topological superconductors in three dimensions},}\ }\href
  {\doibase 10.1103/PhysRevB.104.134508} {\bibfield  {journal} {\bibinfo
  {journal} {Phys. Rev. B}\ }\textbf {\bibinfo {volume} {104}},\ \bibinfo
  {pages} {134508} (\bibinfo {year} {2021})}\BibitemShut {NoStop}%
\bibitem [{\citenamefont {Arouca}\ \emph {et~al.}(2024)\citenamefont {Arouca},
  \citenamefont {Nag},\ and\ \citenamefont {Black-Schaffer}}]{Arouca24}%
  \BibitemOpen
  \bibfield  {author} {\bibinfo {author} {\bibfnamefont {Rodrigo}\ \bibnamefont
  {Arouca}}, \bibinfo {author} {\bibfnamefont {Tanay}\ \bibnamefont {Nag}}, \
  and\ \bibinfo {author} {\bibfnamefont {Annica~M.}\ \bibnamefont
  {Black-Schaffer}},\ }\bibfield  {title} {\enquote {\bibinfo {title} {Mixed
  higher-order topology, and nodal and nodeless flat band topological phases in
  a superconducting multiorbital model},}\ }\href {\doibase
  10.1103/PhysRevB.110.064520} {\bibfield  {journal} {\bibinfo  {journal}
  {Phys. Rev. B}\ }\textbf {\bibinfo {volume} {110}},\ \bibinfo {pages}
  {064520} (\bibinfo {year} {2024})}\BibitemShut {NoStop}%
\bibitem [{\citenamefont {Stuhl}\ \emph {et~al.}(2015)\citenamefont {Stuhl},
  \citenamefont {Lu}, \citenamefont {Aycock}, \citenamefont {Genkina},\ and\
  \citenamefont {Spielman}}]{Stuhl15}%
  \BibitemOpen
  \bibfield  {author} {\bibinfo {author} {\bibfnamefont {B.~K.}\ \bibnamefont
  {Stuhl}}, \bibinfo {author} {\bibfnamefont {H.-I.}\ \bibnamefont {Lu}},
  \bibinfo {author} {\bibfnamefont {L.~M.}\ \bibnamefont {Aycock}}, \bibinfo
  {author} {\bibfnamefont {D.}~\bibnamefont {Genkina}}, \ and\ \bibinfo
  {author} {\bibfnamefont {I.~B.}\ \bibnamefont {Spielman}},\ }\bibfield
  {title} {\enquote {\bibinfo {title} {Visualizing edge states with an atomic
  bose gas in the quantum hall regime},}\ }\href {\doibase
  10.1126/science.aaa8515} {\bibfield  {journal} {\bibinfo  {journal}
  {Science}\ }\textbf {\bibinfo {volume} {349}},\ \bibinfo {pages} {1514--1518}
  (\bibinfo {year} {2015})},\ \Eprint
  {http://arxiv.org/abs/https://www.science.org/doi/pdf/10.1126/science.aaa8515}
  {https://www.science.org/doi/pdf/10.1126/science.aaa8515} \BibitemShut
  {NoStop}%
\bibitem [{\citenamefont {Meier}\ \emph
  {et~al.}(2016{\natexlab{a}})\citenamefont {Meier}, \citenamefont {An},\ and\
  \citenamefont {Gadway}}]{Meier2016}%
  \BibitemOpen
  \bibfield  {author} {\bibinfo {author} {\bibfnamefont {Eric~J.}\ \bibnamefont
  {Meier}}, \bibinfo {author} {\bibfnamefont {Fangzhao~Alex}\ \bibnamefont
  {An}}, \ and\ \bibinfo {author} {\bibfnamefont {Bryce}\ \bibnamefont
  {Gadway}},\ }\bibfield  {title} {\enquote {\bibinfo {title} {Observation of
  the topological soliton state in the su--schrieffer--heeger model},}\ }\href
  {\doibase 10.1038/ncomms13986} {\bibfield  {journal} {\bibinfo  {journal}
  {Nature Communications}\ }\textbf {\bibinfo {volume} {7}},\ \bibinfo {pages}
  {13986} (\bibinfo {year} {2016}{\natexlab{a}})}\BibitemShut {NoStop}%
\bibitem [{\citenamefont {Rechtsman}\ \emph {et~al.}(2013)\citenamefont
  {Rechtsman}, \citenamefont {Zeuner}, \citenamefont {Plotnik}, \citenamefont
  {Lumer}, \citenamefont {Podolsky}, \citenamefont {Dreisow}, \citenamefont
  {Nolte}, \citenamefont {Segev},\ and\ \citenamefont
  {Szameit}}]{Rechtsman2013}%
  \BibitemOpen
  \bibfield  {author} {\bibinfo {author} {\bibfnamefont {M.~C.}\ \bibnamefont
  {Rechtsman}}, \bibinfo {author} {\bibfnamefont {J.~M.}\ \bibnamefont
  {Zeuner}}, \bibinfo {author} {\bibfnamefont {Y.}~\bibnamefont {Plotnik}},
  \bibinfo {author} {\bibfnamefont {Y.}~\bibnamefont {Lumer}}, \bibinfo
  {author} {\bibfnamefont {D.}~\bibnamefont {Podolsky}}, \bibinfo {author}
  {\bibfnamefont {F.}~\bibnamefont {Dreisow}}, \bibinfo {author} {\bibfnamefont
  {S.}~\bibnamefont {Nolte}}, \bibinfo {author} {\bibfnamefont
  {M.}~\bibnamefont {Segev}}, \ and\ \bibinfo {author} {\bibfnamefont
  {A.}~\bibnamefont {Szameit}},\ }\bibfield  {title} {\enquote {\bibinfo
  {title} {Photonic floquet topological insulators},}\ }\href
  {https://doi.org/10.1038/nature12066} {\bibfield  {journal} {\bibinfo
  {journal} {Nature}\ }\textbf {\bibinfo {volume} {496}},\ \bibinfo {pages}
  {196--200} (\bibinfo {year} {2013})}\BibitemShut {NoStop}%
\bibitem [{\citenamefont {Peano}\ \emph {et~al.}(2015)\citenamefont {Peano},
  \citenamefont {Brendel}, \citenamefont {Schmidt},\ and\ \citenamefont
  {Marquardt}}]{Peano15}%
  \BibitemOpen
  \bibfield  {author} {\bibinfo {author} {\bibfnamefont {V.}~\bibnamefont
  {Peano}}, \bibinfo {author} {\bibfnamefont {C.}~\bibnamefont {Brendel}},
  \bibinfo {author} {\bibfnamefont {M.}~\bibnamefont {Schmidt}}, \ and\
  \bibinfo {author} {\bibfnamefont {F.}~\bibnamefont {Marquardt}},\ }\bibfield
  {title} {\enquote {\bibinfo {title} {Topological phases of sound and
  light},}\ }\href {\doibase 10.1103/PhysRevX.5.031011} {\bibfield  {journal}
  {\bibinfo  {journal} {Phys. Rev. X}\ }\textbf {\bibinfo {volume} {5}},\
  \bibinfo {pages} {031011} (\bibinfo {year} {2015})}\BibitemShut {NoStop}%
\bibitem [{\citenamefont {Lu}\ \emph {et~al.}(2016)\citenamefont {Lu},
  \citenamefont {Joannopoulos},\ and\ \citenamefont
  {Solja{\v{c}}i{\'{c}}}}]{Lu2016}%
  \BibitemOpen
  \bibfield  {author} {\bibinfo {author} {\bibfnamefont {Ling}\ \bibnamefont
  {Lu}}, \bibinfo {author} {\bibfnamefont {John~D.}\ \bibnamefont
  {Joannopoulos}}, \ and\ \bibinfo {author} {\bibfnamefont {Marin}\
  \bibnamefont {Solja{\v{c}}i{\'{c}}}},\ }\bibfield  {title} {\enquote
  {\bibinfo {title} {Topological states in photonic systems},}\ }\href
  {\doibase 10.1038/nphys3796} {\bibfield  {journal} {\bibinfo  {journal}
  {Nature Physics}\ }\textbf {\bibinfo {volume} {12}},\ \bibinfo {pages}
  {626--629} (\bibinfo {year} {2016})}\BibitemShut {NoStop}%
\bibitem [{\citenamefont {Chen}\ \emph {et~al.}(2019)\citenamefont {Chen},
  \citenamefont {Deng}, \citenamefont {Shi}, \citenamefont {Zhao},
  \citenamefont {Chen},\ and\ \citenamefont {Dong}}]{PhotonicChen}%
  \BibitemOpen
  \bibfield  {author} {\bibinfo {author} {\bibfnamefont {Xiao-Dong}\
  \bibnamefont {Chen}}, \bibinfo {author} {\bibfnamefont {Wei-Min}\
  \bibnamefont {Deng}}, \bibinfo {author} {\bibfnamefont {Fu-Long}\
  \bibnamefont {Shi}}, \bibinfo {author} {\bibfnamefont {Fu-Li}\ \bibnamefont
  {Zhao}}, \bibinfo {author} {\bibfnamefont {Min}\ \bibnamefont {Chen}}, \ and\
  \bibinfo {author} {\bibfnamefont {Jian-Wen}\ \bibnamefont {Dong}},\
  }\bibfield  {title} {\enquote {\bibinfo {title} {Direct observation of corner
  states in second-order topological photonic crystal slabs},}\ }\href
  {\doibase 10.1103/PhysRevLett.122.233902} {\bibfield  {journal} {\bibinfo
  {journal} {Phys. Rev. Lett.}\ }\textbf {\bibinfo {volume} {122}},\ \bibinfo
  {pages} {233902} (\bibinfo {year} {2019})}\BibitemShut {NoStop}%
\bibitem [{\citenamefont {Xue}\ \emph {et~al.}(2019)\citenamefont {Xue},
  \citenamefont {Yang}, \citenamefont {Gao}, \citenamefont {Chong},\ and\
  \citenamefont {Zhang}}]{XueAcousticKagome}%
  \BibitemOpen
  \bibfield  {author} {\bibinfo {author} {\bibfnamefont {Haoran}\ \bibnamefont
  {Xue}}, \bibinfo {author} {\bibfnamefont {Yahui}\ \bibnamefont {Yang}},
  \bibinfo {author} {\bibfnamefont {Fei}\ \bibnamefont {Gao}}, \bibinfo
  {author} {\bibfnamefont {Yidong}\ \bibnamefont {Chong}}, \ and\ \bibinfo
  {author} {\bibfnamefont {Baile}\ \bibnamefont {Zhang}},\ }\bibfield  {title}
  {\enquote {\bibinfo {title} {Acoustic higher-order topological insulator on a
  kagome lattice},}\ }\href {\doibase 10.1038/s41563-018-0251-x} {\bibfield
  {journal} {\bibinfo  {journal} {Nature Materials}\ }\textbf {\bibinfo
  {volume} {18}},\ \bibinfo {pages} {108--112} (\bibinfo {year}
  {2019})}\BibitemShut {NoStop}%
\bibitem [{\citenamefont {Xiao}\ \emph {et~al.}(2015)\citenamefont {Xiao},
  \citenamefont {Ma}, \citenamefont {Yang}, \citenamefont {Sheng},
  \citenamefont {Zhang},\ and\ \citenamefont {Chan}}]{Xiao2015}%
  \BibitemOpen
  \bibfield  {author} {\bibinfo {author} {\bibfnamefont {Meng}\ \bibnamefont
  {Xiao}}, \bibinfo {author} {\bibfnamefont {Guancong}\ \bibnamefont {Ma}},
  \bibinfo {author} {\bibfnamefont {Zhiyu}\ \bibnamefont {Yang}}, \bibinfo
  {author} {\bibfnamefont {Ping}\ \bibnamefont {Sheng}}, \bibinfo {author}
  {\bibfnamefont {Z.~Q.}\ \bibnamefont {Zhang}}, \ and\ \bibinfo {author}
  {\bibfnamefont {C.~T.}\ \bibnamefont {Chan}},\ }\bibfield  {title} {\enquote
  {\bibinfo {title} {Geometric phase and band inversion in periodic acoustic
  systems},}\ }\href {\doibase 10.1038/nphys3228} {\bibfield  {journal}
  {\bibinfo  {journal} {Nature Physics}\ }\textbf {\bibinfo {volume} {11}},\
  \bibinfo {pages} {240--244} (\bibinfo {year} {2015})}\BibitemShut {NoStop}%
\bibitem [{\citenamefont {Liu}\ \emph {et~al.}(2023)\citenamefont {Liu},
  \citenamefont {Matveeva}, \citenamefont {Gutman},\ and\ \citenamefont
  {Carr}}]{Liu2023}%
  \BibitemOpen
  \bibfield  {author} {\bibinfo {author} {\bibfnamefont {Donghao}\ \bibnamefont
  {Liu}}, \bibinfo {author} {\bibfnamefont {Polina}\ \bibnamefont {Matveeva}},
  \bibinfo {author} {\bibfnamefont {Dmitri}\ \bibnamefont {Gutman}}, \ and\
  \bibinfo {author} {\bibfnamefont {Sam~T.}\ \bibnamefont {Carr}},\ }\bibfield
  {title} {\enquote {\bibinfo {title} {Elementary models of three-dimensional
  topological insulators with chiral symmetry},}\ }\href {\doibase
  10.1103/PhysRevB.108.035418} {\bibfield  {journal} {\bibinfo  {journal}
  {Phys. Rev. B}\ }\textbf {\bibinfo {volume} {108}},\ \bibinfo {pages}
  {035418} (\bibinfo {year} {2023})}\BibitemShut {NoStop}%
\bibitem [{\citenamefont {Matveeva}\ \emph {et~al.}(2023)\citenamefont
  {Matveeva}, \citenamefont {Hewitt}, \citenamefont {Liu}, \citenamefont
  {Reddy}, \citenamefont {Gutman},\ and\ \citenamefont {Carr}}]{Matveeva2023}%
  \BibitemOpen
  \bibfield  {author} {\bibinfo {author} {\bibfnamefont {Polina}\ \bibnamefont
  {Matveeva}}, \bibinfo {author} {\bibfnamefont {Tyler}\ \bibnamefont
  {Hewitt}}, \bibinfo {author} {\bibfnamefont {Donghao}\ \bibnamefont {Liu}},
  \bibinfo {author} {\bibfnamefont {Kethan}\ \bibnamefont {Reddy}}, \bibinfo
  {author} {\bibfnamefont {Dmitri}\ \bibnamefont {Gutman}}, \ and\ \bibinfo
  {author} {\bibfnamefont {Sam~T.}\ \bibnamefont {Carr}},\ }\bibfield  {title}
  {\enquote {\bibinfo {title} {One-dimensional noninteracting topological
  insulators with chiral symmetry},}\ }\href {\doibase
  10.1103/PhysRevB.107.075422} {\bibfield  {journal} {\bibinfo  {journal}
  {Phys. Rev. B}\ }\textbf {\bibinfo {volume} {107}},\ \bibinfo {pages}
  {075422} (\bibinfo {year} {2023})}\BibitemShut {NoStop}%
\bibitem [{\citenamefont {Su}\ \emph {et~al.}(1979)\citenamefont {Su},
  \citenamefont {Schrieffer},\ and\ \citenamefont {Heeger}}]{SSH79}%
  \BibitemOpen
  \bibfield  {author} {\bibinfo {author} {\bibfnamefont {W.~P.}\ \bibnamefont
  {Su}}, \bibinfo {author} {\bibfnamefont {J.~R.}\ \bibnamefont {Schrieffer}},
  \ and\ \bibinfo {author} {\bibfnamefont {A.~J.}\ \bibnamefont {Heeger}},\
  }\bibfield  {title} {\enquote {\bibinfo {title} {Solitons in
  polyacetylene},}\ }\href {\doibase 10.1103/PhysRevLett.42.1698} {\bibfield
  {journal} {\bibinfo  {journal} {Phys. Rev. Lett.}\ }\textbf {\bibinfo
  {volume} {42}},\ \bibinfo {pages} {1698--1701} (\bibinfo {year}
  {1979})}\BibitemShut {NoStop}%
\bibitem [{\citenamefont {Hasan}\ and\ \citenamefont
  {Kane}(2010)}]{hasan2010colloquium}%
  \BibitemOpen
  \bibfield  {author} {\bibinfo {author} {\bibfnamefont {M~Zahid}\ \bibnamefont
  {Hasan}}\ and\ \bibinfo {author} {\bibfnamefont {Charles~L}\ \bibnamefont
  {Kane}},\ }\bibfield  {title} {\enquote {\bibinfo {title} {Colloquium:
  topological insulators},}\ }\href {\doibase 10.1103/RevModPhys.82.3045}
  {\bibfield  {journal} {\bibinfo  {journal} {Rev. Mod. Phys.}\ }\textbf
  {\bibinfo {volume} {82}},\ \bibinfo {pages} {3045} (\bibinfo {year}
  {2010})}\BibitemShut {NoStop}%
\bibitem [{\citenamefont {Qi}\ and\ \citenamefont
  {Zhang}(2011{\natexlab{b}})}]{qi2011topological}%
  \BibitemOpen
  \bibfield  {author} {\bibinfo {author} {\bibfnamefont {Xiao-Liang}\
  \bibnamefont {Qi}}\ and\ \bibinfo {author} {\bibfnamefont {Shou-Cheng}\
  \bibnamefont {Zhang}},\ }\bibfield  {title} {\enquote {\bibinfo {title}
  {Topological insulators and superconductors},}\ }\href {\doibase
  10.1103/RevModPhys.83.1057} {\bibfield  {journal} {\bibinfo  {journal} {Rev.
  Mod. Phys.}\ }\textbf {\bibinfo {volume} {83}},\ \bibinfo {pages} {1057}
  (\bibinfo {year} {2011}{\natexlab{b}})}\BibitemShut {NoStop}%
\bibitem [{\citenamefont {Asbóth}\ \emph {et~al.}(2016)\citenamefont
  {Asbóth}, \citenamefont {Oroszlány},\ and\ \citenamefont
  {Pályi}}]{Asb_th_2016}%
  \BibitemOpen
  \bibfield  {author} {\bibinfo {author} {\bibfnamefont {János~K.}\
  \bibnamefont {Asbóth}}, \bibinfo {author} {\bibfnamefont {László}\
  \bibnamefont {Oroszlány}}, \ and\ \bibinfo {author} {\bibfnamefont
  {András}\ \bibnamefont {Pályi}},\ }\href {\doibase
  10.1007/978-3-319-25607-8} {\emph {\bibinfo {title} {A Short Course on
  Topological Insulators}}}\ (\bibinfo  {publisher} {Springer International
  Publishing},\ \bibinfo {year} {2016})\BibitemShut {NoStop}%
\bibitem [{\citenamefont {Jin}\ \emph {et~al.}(2024)\citenamefont {Jin},
  \citenamefont {Oriekhov}, \citenamefont {Splitthoff},\ and\ \citenamefont
  {Greplova}}]{jin2024topological}%
  \BibitemOpen
  \bibfield  {author} {\bibinfo {author} {\bibfnamefont {Guliuxin}\
  \bibnamefont {Jin}}, \bibinfo {author} {\bibfnamefont {DO}~\bibnamefont
  {Oriekhov}}, \bibinfo {author} {\bibfnamefont {Lukas~Johannes}\ \bibnamefont
  {Splitthoff}}, \ and\ \bibinfo {author} {\bibfnamefont {Eliska}\ \bibnamefont
  {Greplova}},\ }\bibfield  {title} {\enquote {\bibinfo {title} {Topological
  finite size effect in one-dimensional chiral symmetric systems},}\
  }\href@noop {} {\bibfield  {journal} {\bibinfo  {journal} {arXiv preprint
  arXiv:2411.17822}\ } (\bibinfo {year} {2024})}\BibitemShut {NoStop}%
\bibitem [{\citenamefont {Nava}\ \emph {et~al.}(2023)\citenamefont {Nava},
  \citenamefont {Campagnano}, \citenamefont {Sodano},\ and\ \citenamefont
  {Giuliano}}]{Nava23}%
  \BibitemOpen
  \bibfield  {author} {\bibinfo {author} {\bibfnamefont {Andrea}\ \bibnamefont
  {Nava}}, \bibinfo {author} {\bibfnamefont {Gabriele}\ \bibnamefont
  {Campagnano}}, \bibinfo {author} {\bibfnamefont {Pasquale}\ \bibnamefont
  {Sodano}}, \ and\ \bibinfo {author} {\bibfnamefont {Domenico}\ \bibnamefont
  {Giuliano}},\ }\bibfield  {title} {\enquote {\bibinfo {title} {Lindblad
  master equation approach to the topological phase transition in the
  disordered su-schrieffer-heeger model},}\ }\href {\doibase
  10.1103/PhysRevB.107.035113} {\bibfield  {journal} {\bibinfo  {journal}
  {Phys. Rev. B}\ }\textbf {\bibinfo {volume} {107}},\ \bibinfo {pages}
  {035113} (\bibinfo {year} {2023})}\BibitemShut {NoStop}%
\bibitem [{\citenamefont {Cinnirella}\ \emph {et~al.}(2024)\citenamefont
  {Cinnirella}, \citenamefont {Nava}, \citenamefont {Campagnano},\ and\
  \citenamefont {Giuliano}}]{Cinnirella24}%
  \BibitemOpen
  \bibfield  {author} {\bibinfo {author} {\bibfnamefont {Emmanuele~G.}\
  \bibnamefont {Cinnirella}}, \bibinfo {author} {\bibfnamefont {Andrea}\
  \bibnamefont {Nava}}, \bibinfo {author} {\bibfnamefont {Gabriele}\
  \bibnamefont {Campagnano}}, \ and\ \bibinfo {author} {\bibfnamefont
  {Domenico}\ \bibnamefont {Giuliano}},\ }\bibfield  {title} {\enquote
  {\bibinfo {title} {Fate of high winding number topological phases in the
  disordered extended su-schrieffer-heeger model},}\ }\href {\doibase
  10.1103/PhysRevB.109.035114} {\bibfield  {journal} {\bibinfo  {journal}
  {Phys. Rev. B}\ }\textbf {\bibinfo {volume} {109}},\ \bibinfo {pages}
  {035114} (\bibinfo {year} {2024})}\BibitemShut {NoStop}%
\bibitem [{\citenamefont {Kitaev}(2001)}]{Kitaev_2001}%
  \BibitemOpen
  \bibfield  {author} {\bibinfo {author} {\bibfnamefont {A~Yu}\ \bibnamefont
  {Kitaev}},\ }\bibfield  {title} {\enquote {\bibinfo {title} {Unpaired
  majorana fermions in quantum wires},}\ }\href {\doibase
  10.1070/1063-7869/44/10S/S29} {\bibfield  {journal} {\bibinfo  {journal}
  {Physics-Uspekhi}\ }\textbf {\bibinfo {volume} {44}},\ \bibinfo {pages} {131}
  (\bibinfo {year} {2001})}\BibitemShut {NoStop}%
\bibitem [{\citenamefont {DeGottardi}\ \emph
  {et~al.}(2013{\natexlab{a}})\citenamefont {DeGottardi}, \citenamefont {Sen},\
  and\ \citenamefont {Vishveshwara}}]{DeGottardi13}%
  \BibitemOpen
  \bibfield  {author} {\bibinfo {author} {\bibfnamefont {Wade}\ \bibnamefont
  {DeGottardi}}, \bibinfo {author} {\bibfnamefont {Diptiman}\ \bibnamefont
  {Sen}}, \ and\ \bibinfo {author} {\bibfnamefont {Smitha}\ \bibnamefont
  {Vishveshwara}},\ }\bibfield  {title} {\enquote {\bibinfo {title} {Majorana
  fermions in superconducting 1d systems having periodic, quasiperiodic, and
  disordered potentials},}\ }\href {\doibase 10.1103/PhysRevLett.110.146404}
  {\bibfield  {journal} {\bibinfo  {journal} {Phys. Rev. Lett.}\ }\textbf
  {\bibinfo {volume} {110}},\ \bibinfo {pages} {146404} (\bibinfo {year}
  {2013}{\natexlab{a}})}\BibitemShut {NoStop}%
\bibitem [{\citenamefont {DeGottardi}\ \emph
  {et~al.}(2013{\natexlab{b}})\citenamefont {DeGottardi}, \citenamefont
  {Thakurathi}, \citenamefont {Vishveshwara},\ and\ \citenamefont
  {Sen}}]{DeGottardi13b}%
  \BibitemOpen
  \bibfield  {author} {\bibinfo {author} {\bibfnamefont {Wade}\ \bibnamefont
  {DeGottardi}}, \bibinfo {author} {\bibfnamefont {Manisha}\ \bibnamefont
  {Thakurathi}}, \bibinfo {author} {\bibfnamefont {Smitha}\ \bibnamefont
  {Vishveshwara}}, \ and\ \bibinfo {author} {\bibfnamefont {Diptiman}\
  \bibnamefont {Sen}},\ }\bibfield  {title} {\enquote {\bibinfo {title}
  {Majorana fermions in superconducting wires: Effects of long-range hopping,
  broken time-reversal symmetry, and potential landscapes},}\ }\href {\doibase
  10.1103/PhysRevB.88.165111} {\bibfield  {journal} {\bibinfo  {journal} {Phys.
  Rev. B}\ }\textbf {\bibinfo {volume} {88}},\ \bibinfo {pages} {165111}
  (\bibinfo {year} {2013}{\natexlab{b}})}\BibitemShut {NoStop}%
\bibitem [{\citenamefont {Rajak}\ \emph {et~al.}(2014)\citenamefont {Rajak},
  \citenamefont {Nag},\ and\ \citenamefont {Dutta}}]{Rajak14}%
  \BibitemOpen
  \bibfield  {author} {\bibinfo {author} {\bibfnamefont {Atanu}\ \bibnamefont
  {Rajak}}, \bibinfo {author} {\bibfnamefont {Tanay}\ \bibnamefont {Nag}}, \
  and\ \bibinfo {author} {\bibfnamefont {Amit}\ \bibnamefont {Dutta}},\
  }\bibfield  {title} {\enquote {\bibinfo {title} {Possibility of adiabatic
  transport of a majorana edge state through an extended gapless region},}\
  }\href {\doibase 10.1103/PhysRevE.90.042107} {\bibfield  {journal} {\bibinfo
  {journal} {Phys. Rev. E}\ }\textbf {\bibinfo {volume} {90}},\ \bibinfo
  {pages} {042107} (\bibinfo {year} {2014})}\BibitemShut {NoStop}%
\bibitem [{\citenamefont {Jackiw}\ and\ \citenamefont
  {Rebbi}(1976)}]{jackiw1976}%
  \BibitemOpen
  \bibfield  {author} {\bibinfo {author} {\bibfnamefont {Roman}\ \bibnamefont
  {Jackiw}}\ and\ \bibinfo {author} {\bibfnamefont {Cl{\'a}udio}\ \bibnamefont
  {Rebbi}},\ }\bibfield  {title} {\enquote {\bibinfo {title} {Solitons with
  fermion number $1/2$},}\ }\href {\doibase 10.1103/PhysRevD.13.3398}
  {\bibfield  {journal} {\bibinfo  {journal} {Phys. Rev. D}\ }\textbf {\bibinfo
  {volume} {13}},\ \bibinfo {pages} {3398} (\bibinfo {year}
  {1976})}\BibitemShut {NoStop}%
\bibitem [{\citenamefont {Rhim}\ \emph {et~al.}(2018)\citenamefont {Rhim},
  \citenamefont {Bardarson},\ and\ \citenamefont {Slager}}]{Rhim18}%
  \BibitemOpen
  \bibfield  {author} {\bibinfo {author} {\bibfnamefont {Jun-Won}\ \bibnamefont
  {Rhim}}, \bibinfo {author} {\bibfnamefont {Jens~H.}\ \bibnamefont
  {Bardarson}}, \ and\ \bibinfo {author} {\bibfnamefont {Robert-Jan}\
  \bibnamefont {Slager}},\ }\bibfield  {title} {\enquote {\bibinfo {title}
  {Unified bulk-boundary correspondence for band insulators},}\ }\href
  {\doibase 10.1103/PhysRevB.97.115143} {\bibfield  {journal} {\bibinfo
  {journal} {Phys. Rev. B}\ }\textbf {\bibinfo {volume} {97}},\ \bibinfo
  {pages} {115143} (\bibinfo {year} {2018})}\BibitemShut {NoStop}%
\bibitem [{\citenamefont {Atala}\ \emph {et~al.}(2013)\citenamefont {Atala},
  \citenamefont {Aidelsburger}, \citenamefont {Barreiro}, \citenamefont
  {Abanin}, \citenamefont {Kitagawa}, \citenamefont {Demler},\ and\
  \citenamefont {Bloch}}]{atala2013direct}%
  \BibitemOpen
  \bibfield  {author} {\bibinfo {author} {\bibfnamefont {Marcos}\ \bibnamefont
  {Atala}}, \bibinfo {author} {\bibfnamefont {Monika}\ \bibnamefont
  {Aidelsburger}}, \bibinfo {author} {\bibfnamefont {Julio~T}\ \bibnamefont
  {Barreiro}}, \bibinfo {author} {\bibfnamefont {Dmitry}\ \bibnamefont
  {Abanin}}, \bibinfo {author} {\bibfnamefont {Takuya}\ \bibnamefont
  {Kitagawa}}, \bibinfo {author} {\bibfnamefont {Eugene}\ \bibnamefont
  {Demler}}, \ and\ \bibinfo {author} {\bibfnamefont {Immanuel}\ \bibnamefont
  {Bloch}},\ }\bibfield  {title} {\enquote {\bibinfo {title} {Direct
  measurement of the zak phase in topological bloch bands},}\ }\href@noop {}
  {\bibfield  {journal} {\bibinfo  {journal} {Nature Physics}\ }\textbf
  {\bibinfo {volume} {9}},\ \bibinfo {pages} {795--800} (\bibinfo {year}
  {2013})}\BibitemShut {NoStop}%
\bibitem [{\citenamefont {Lin}\ \emph {et~al.}(2021)\citenamefont {Lin},
  \citenamefont {Ke},\ and\ \citenamefont {Lee}}]{Lin2021}%
  \BibitemOpen
  \bibfield  {author} {\bibinfo {author} {\bibfnamefont {Ling}\ \bibnamefont
  {Lin}}, \bibinfo {author} {\bibfnamefont {Yongguan}\ \bibnamefont {Ke}}, \
  and\ \bibinfo {author} {\bibfnamefont {Chaohong}\ \bibnamefont {Lee}},\
  }\bibfield  {title} {\enquote {\bibinfo {title} {Real-space representation of
  the winding number for a one-dimensional chiral-symmetric topological
  insulator},}\ }\href {\doibase 10.1103/PhysRevB.103.224208} {\bibfield
  {journal} {\bibinfo  {journal} {Phys. Rev. B}\ }\textbf {\bibinfo {volume}
  {103}},\ \bibinfo {pages} {224208} (\bibinfo {year} {2021})}\BibitemShut
  {NoStop}%
\bibitem [{\citenamefont {Ghosh}\ \emph {et~al.}(2024)\citenamefont {Ghosh},
  \citenamefont {Saha},\ and\ \citenamefont {Nag}}]{Ghosh2024}%
  \BibitemOpen
  \bibfield  {author} {\bibinfo {author} {\bibfnamefont {Arnob~Kumar}\
  \bibnamefont {Ghosh}}, \bibinfo {author} {\bibfnamefont {Arijit}\
  \bibnamefont {Saha}}, \ and\ \bibinfo {author} {\bibfnamefont {Tanay}\
  \bibnamefont {Nag}},\ }\bibfield  {title} {\enquote {\bibinfo {title} {Corner
  modes in non-hermitian next-nearest-neighbor hopping model},}\ }\href
  {\doibase 10.1103/PhysRevB.110.115403} {\bibfield  {journal} {\bibinfo
  {journal} {Phys. Rev. B}\ }\textbf {\bibinfo {volume} {110}},\ \bibinfo
  {pages} {115403} (\bibinfo {year} {2024})}\BibitemShut {NoStop}%
\bibitem [{\citenamefont {Zhang}\ \emph {et~al.}(2022)\citenamefont {Zhang},
  \citenamefont {Wu}, \citenamefont {Tang},\ and\ \citenamefont
  {Zhang}}]{Zhang_2022}%
  \BibitemOpen
  \bibfield  {author} {\bibinfo {author} {\bibfnamefont {Wei-Jie}\ \bibnamefont
  {Zhang}}, \bibinfo {author} {\bibfnamefont {Yi-Piao}\ \bibnamefont {Wu}},
  \bibinfo {author} {\bibfnamefont {Ling-Zhi}\ \bibnamefont {Tang}}, \ and\
  \bibinfo {author} {\bibfnamefont {Guo-Qing}\ \bibnamefont {Zhang}},\
  }\bibfield  {title} {\enquote {\bibinfo {title} {Topological and dynamical
  phase transitions in the su–schrieffer–heeger model with quasiperiodic
  and long-range hoppings},}\ }\href {\doibase 10.1088/1572-9494/ac75db}
  {\bibfield  {journal} {\bibinfo  {journal} {Communications in Theoretical
  Physics}\ }\textbf {\bibinfo {volume} {74}},\ \bibinfo {pages} {075702}
  (\bibinfo {year} {2022})}\BibitemShut {NoStop}%
\bibitem [{\citenamefont {Tang}\ \emph {et~al.}(2020)\citenamefont {Tang},
  \citenamefont {Zhang}, \citenamefont {Zhang},\ and\ \citenamefont
  {Zhang}}]{Tang2020}%
  \BibitemOpen
  \bibfield  {author} {\bibinfo {author} {\bibfnamefont {Ling-Zhi}\
  \bibnamefont {Tang}}, \bibinfo {author} {\bibfnamefont {Ling-Feng}\
  \bibnamefont {Zhang}}, \bibinfo {author} {\bibfnamefont {Guo-Qing}\
  \bibnamefont {Zhang}}, \ and\ \bibinfo {author} {\bibfnamefont {Dan-Wei}\
  \bibnamefont {Zhang}},\ }\bibfield  {title} {\enquote {\bibinfo {title}
  {Topological anderson insulators in two-dimensional non-hermitian disordered
  systems},}\ }\href {\doibase 10.1103/PhysRevA.101.063612} {\bibfield
  {journal} {\bibinfo  {journal} {Phys. Rev. A}\ }\textbf {\bibinfo {volume}
  {101}},\ \bibinfo {pages} {063612} (\bibinfo {year} {2020})}\BibitemShut
  {NoStop}%
\bibitem [{\citenamefont {Song}\ \emph {et~al.}(2019)\citenamefont {Song},
  \citenamefont {Yao},\ and\ \citenamefont {Wang}}]{Song2019}%
  \BibitemOpen
  \bibfield  {author} {\bibinfo {author} {\bibfnamefont {Fei}\ \bibnamefont
  {Song}}, \bibinfo {author} {\bibfnamefont {Shunyu}\ \bibnamefont {Yao}}, \
  and\ \bibinfo {author} {\bibfnamefont {Zhong}\ \bibnamefont {Wang}},\
  }\bibfield  {title} {\enquote {\bibinfo {title} {Non-hermitian topological
  invariants in real space},}\ }\href {\doibase 10.1103/PhysRevLett.123.246801}
  {\bibfield  {journal} {\bibinfo  {journal} {Phys. Rev. Lett.}\ }\textbf
  {\bibinfo {volume} {123}},\ \bibinfo {pages} {246801} (\bibinfo {year}
  {2019})}\BibitemShut {NoStop}%
\bibitem [{\citenamefont {He}\ and\ \citenamefont {Chien}(2020)}]{He2020}%
  \BibitemOpen
  \bibfield  {author} {\bibinfo {author} {\bibfnamefont {Yan}\ \bibnamefont
  {He}}\ and\ \bibinfo {author} {\bibfnamefont {Chih-Chun}\ \bibnamefont
  {Chien}},\ }\bibfield  {title} {\enquote {\bibinfo {title} {Non-hermitian
  generalizations of extended su-schrieffer-heeger models},}\ }\href {\doibase
  10.1088/1361-648X/abc974} {\bibfield  {journal} {\bibinfo  {journal} {Journal
  of physics. Condensed matter : an Institute of Physics journal}\ }\textbf
  {\bibinfo {volume} {33}} (\bibinfo {year} {2020}),\
  10.1088/1361-648X/abc974}\BibitemShut {NoStop}%
\bibitem [{\citenamefont {Kivelson}(1982)}]{Kivelson82}%
  \BibitemOpen
  \bibfield  {author} {\bibinfo {author} {\bibfnamefont {S.}~\bibnamefont
  {Kivelson}},\ }\bibfield  {title} {\enquote {\bibinfo {title} {Wannier
  functions in one-dimensional disordered systems: Application to fractionally
  charged solitons},}\ }\href {\doibase 10.1103/PhysRevB.26.4269} {\bibfield
  {journal} {\bibinfo  {journal} {Phys. Rev. B}\ }\textbf {\bibinfo {volume}
  {26}},\ \bibinfo {pages} {4269--4277} (\bibinfo {year} {1982})}\BibitemShut
  {NoStop}%
\bibitem [{\citenamefont {Mondragon-Shem}\ \emph {et~al.}(2014)\citenamefont
  {Mondragon-Shem}, \citenamefont {Hughes}, \citenamefont {Song},\ and\
  \citenamefont {Prodan}}]{Mondragon-Shem14}%
  \BibitemOpen
  \bibfield  {author} {\bibinfo {author} {\bibfnamefont {Ian}\ \bibnamefont
  {Mondragon-Shem}}, \bibinfo {author} {\bibfnamefont {Taylor~L.}\ \bibnamefont
  {Hughes}}, \bibinfo {author} {\bibfnamefont {Juntao}\ \bibnamefont {Song}}, \
  and\ \bibinfo {author} {\bibfnamefont {Emil}\ \bibnamefont {Prodan}},\
  }\bibfield  {title} {\enquote {\bibinfo {title} {Topological criticality in
  the chiral-symmetric aiii class at strong disorder},}\ }\href {\doibase
  10.1103/PhysRevLett.113.046802} {\bibfield  {journal} {\bibinfo  {journal}
  {Phys. Rev. Lett.}\ }\textbf {\bibinfo {volume} {113}},\ \bibinfo {pages}
  {046802} (\bibinfo {year} {2014})}\BibitemShut {NoStop}%
\bibitem [{\citenamefont {Marzari}\ \emph {et~al.}(2012)\citenamefont
  {Marzari}, \citenamefont {Mostofi}, \citenamefont {Yates}, \citenamefont
  {Souza},\ and\ \citenamefont {Vanderbilt}}]{Marzari12}%
  \BibitemOpen
  \bibfield  {author} {\bibinfo {author} {\bibfnamefont {Nicola}\ \bibnamefont
  {Marzari}}, \bibinfo {author} {\bibfnamefont {Arash~A.}\ \bibnamefont
  {Mostofi}}, \bibinfo {author} {\bibfnamefont {Jonathan~R.}\ \bibnamefont
  {Yates}}, \bibinfo {author} {\bibfnamefont {Ivo}\ \bibnamefont {Souza}}, \
  and\ \bibinfo {author} {\bibfnamefont {David}\ \bibnamefont {Vanderbilt}},\
  }\bibfield  {title} {\enquote {\bibinfo {title} {Maximally localized wannier
  functions: Theory and applications},}\ }\href {\doibase
  10.1103/RevModPhys.84.1419} {\bibfield  {journal} {\bibinfo  {journal} {Rev.
  Mod. Phys.}\ }\textbf {\bibinfo {volume} {84}},\ \bibinfo {pages}
  {1419--1475} (\bibinfo {year} {2012})}\BibitemShut {NoStop}%
\bibitem [{\citenamefont {Flurin}\ \emph {et~al.}(2017)\citenamefont {Flurin},
  \citenamefont {Ramasesh}, \citenamefont {Hacohen-Gourgy}, \citenamefont
  {Martin}, \citenamefont {Yao},\ and\ \citenamefont {Siddiqi}}]{Flurin17}%
  \BibitemOpen
  \bibfield  {author} {\bibinfo {author} {\bibfnamefont {E.}~\bibnamefont
  {Flurin}}, \bibinfo {author} {\bibfnamefont {V.~V.}\ \bibnamefont
  {Ramasesh}}, \bibinfo {author} {\bibfnamefont {S.}~\bibnamefont
  {Hacohen-Gourgy}}, \bibinfo {author} {\bibfnamefont {L.~S.}\ \bibnamefont
  {Martin}}, \bibinfo {author} {\bibfnamefont {N.~Y.}\ \bibnamefont {Yao}}, \
  and\ \bibinfo {author} {\bibfnamefont {I.}~\bibnamefont {Siddiqi}},\
  }\bibfield  {title} {\enquote {\bibinfo {title} {Observing topological
  invariants using quantum walks in superconducting circuits},}\ }\href
  {\doibase 10.1103/PhysRevX.7.031023} {\bibfield  {journal} {\bibinfo
  {journal} {Phys. Rev. X}\ }\textbf {\bibinfo {volume} {7}},\ \bibinfo {pages}
  {031023} (\bibinfo {year} {2017})}\BibitemShut {NoStop}%
\bibitem [{\citenamefont {Barkhofen}\ \emph {et~al.}(2017)\citenamefont
  {Barkhofen}, \citenamefont {Nitsche}, \citenamefont {Elster}, \citenamefont
  {Lorz}, \citenamefont {G\'abris}, \citenamefont {Jex},\ and\ \citenamefont
  {Silberhorn}}]{Barkhofen17}%
  \BibitemOpen
  \bibfield  {author} {\bibinfo {author} {\bibfnamefont {Sonja}\ \bibnamefont
  {Barkhofen}}, \bibinfo {author} {\bibfnamefont {Thomas}\ \bibnamefont
  {Nitsche}}, \bibinfo {author} {\bibfnamefont {Fabian}\ \bibnamefont
  {Elster}}, \bibinfo {author} {\bibfnamefont {Lennart}\ \bibnamefont {Lorz}},
  \bibinfo {author} {\bibfnamefont {Aur\'el}\ \bibnamefont {G\'abris}},
  \bibinfo {author} {\bibfnamefont {Igor}\ \bibnamefont {Jex}}, \ and\ \bibinfo
  {author} {\bibfnamefont {Christine}\ \bibnamefont {Silberhorn}},\ }\bibfield
  {title} {\enquote {\bibinfo {title} {Measuring topological invariants in
  disordered discrete-time quantum walks},}\ }\href {\doibase
  10.1103/PhysRevA.96.033846} {\bibfield  {journal} {\bibinfo  {journal} {Phys.
  Rev. A}\ }\textbf {\bibinfo {volume} {96}},\ \bibinfo {pages} {033846}
  (\bibinfo {year} {2017})}\BibitemShut {NoStop}%
\bibitem [{\citenamefont {Meier}\ \emph {et~al.}(2018)\citenamefont {Meier},
  \citenamefont {An}, \citenamefont {Dauphin}, \citenamefont {Maffei},
  \citenamefont {Massignan}, \citenamefont {Hughes},\ and\ \citenamefont
  {Gadway}}]{Eric18}%
  \BibitemOpen
  \bibfield  {author} {\bibinfo {author} {\bibfnamefont {Eric~J.}\ \bibnamefont
  {Meier}}, \bibinfo {author} {\bibfnamefont {Fangzhao~Alex}\ \bibnamefont
  {An}}, \bibinfo {author} {\bibfnamefont {Alexandre}\ \bibnamefont {Dauphin}},
  \bibinfo {author} {\bibfnamefont {Maria}\ \bibnamefont {Maffei}}, \bibinfo
  {author} {\bibfnamefont {Pietro}\ \bibnamefont {Massignan}}, \bibinfo
  {author} {\bibfnamefont {Taylor~L.}\ \bibnamefont {Hughes}}, \ and\ \bibinfo
  {author} {\bibfnamefont {Bryce}\ \bibnamefont {Gadway}},\ }\bibfield  {title}
  {\enquote {\bibinfo {title} {Observation of the topological anderson
  insulator in disordered atomic wires},}\ }\href {\doibase
  10.1126/science.aat3406} {\bibfield  {journal} {\bibinfo  {journal}
  {Science}\ }\textbf {\bibinfo {volume} {362}},\ \bibinfo {pages} {929--933}
  (\bibinfo {year} {2018})},\ \Eprint
  {http://arxiv.org/abs/https://www.science.org/doi/pdf/10.1126/science.aat3406}
  {https://www.science.org/doi/pdf/10.1126/science.aat3406} \BibitemShut
  {NoStop}%
\bibitem [{\citenamefont {NIU}(1991)}]{NIU91}%
  \BibitemOpen
  \bibfield  {author} {\bibinfo {author} {\bibfnamefont {QIAN}\ \bibnamefont
  {NIU}},\ }\bibfield  {title} {\enquote {\bibinfo {title} {Theory of the
  quantized adiabatic particle transport},}\ }\href {\doibase
  10.1142/S0217984991001155} {\bibfield  {journal} {\bibinfo  {journal} {Modern
  Physics Letters B}\ }\textbf {\bibinfo {volume} {05}},\ \bibinfo {pages}
  {923--931} (\bibinfo {year} {1991})},\ \Eprint
  {http://arxiv.org/abs/https://doi.org/10.1142/S0217984991001155}
  {https://doi.org/10.1142/S0217984991001155} \BibitemShut {NoStop}%
\bibitem [{\citenamefont {Citro}\ and\ \citenamefont
  {Aidelsburger}(2023)}]{Citro_2023}%
  \BibitemOpen
  \bibfield  {author} {\bibinfo {author} {\bibfnamefont {Roberta}\ \bibnamefont
  {Citro}}\ and\ \bibinfo {author} {\bibfnamefont {Monika}\ \bibnamefont
  {Aidelsburger}},\ }\bibfield  {title} {\enquote {\bibinfo {title} {Thouless
  pumping and topology},}\ }\href {\doibase 10.1038/s42254-022-00545-0}
  {\bibfield  {journal} {\bibinfo  {journal} {Nature Reviews Physics}\ }\textbf
  {\bibinfo {volume} {5}},\ \bibinfo {pages} {87–101} (\bibinfo {year}
  {2023})}\BibitemShut {NoStop}%
\bibitem [{\citenamefont {Kumar}\ and\ \citenamefont {Saha}(2021)}]{Kumar2021}%
  \BibitemOpen
  \bibfield  {author} {\bibinfo {author} {\bibfnamefont {Abhishek}\
  \bibnamefont {Kumar}}\ and\ \bibinfo {author} {\bibfnamefont {Kush}\
  \bibnamefont {Saha}},\ }\bibfield  {title} {\enquote {\bibinfo {title}
  {Unconventional pumping in non-hermitian rice-mele model},}\ }\href {\doibase
  10.48550/arXiv.2110.06797} {\  (\bibinfo {year} {2021}),\
  10.48550/arXiv.2110.06797}\BibitemShut {NoStop}%
\bibitem [{\citenamefont {SHEN}(2018)}]{shun2018topological}%
  \BibitemOpen
  \bibfield  {author} {\bibinfo {author} {\bibfnamefont {SHUN-QING.}\
  \bibnamefont {SHEN}},\ }\href@noop {} {\emph {\bibinfo {title} {Topological
  Insulators: Dirac Equation in Condensed Matter}}}\ (\bibinfo  {publisher}
  {Springer},\ \bibinfo {year} {2018})\BibitemShut {NoStop}%
\bibitem [{\citenamefont {Fukui}\ \emph {et~al.}(2005)\citenamefont {Fukui},
  \citenamefont {Hatsugai},\ and\ \citenamefont {Suzuki}}]{Fukui2005}%
  \BibitemOpen
  \bibfield  {author} {\bibinfo {author} {\bibfnamefont {Takahiro}\
  \bibnamefont {Fukui}}, \bibinfo {author} {\bibfnamefont {Yasuhiro}\
  \bibnamefont {Hatsugai}}, \ and\ \bibinfo {author} {\bibfnamefont {Hiroshi}\
  \bibnamefont {Suzuki}},\ }\bibfield  {title} {\enquote {\bibinfo {title}
  {Chern numbers in discretized brillouin zone: Efficient method of computing
  (spin) hall conductances},}\ }\href {\doibase 10.1143/JPSJ.74.1674}
  {\bibfield  {journal} {\bibinfo  {journal} {Journal of the Physical Society
  of Japan}\ }\textbf {\bibinfo {volume} {74}} (\bibinfo {year} {2005}),\
  10.1143/JPSJ.74.1674}\BibitemShut {NoStop}%
\bibitem [{\citenamefont {Agrawal}\ and\ \citenamefont
  {Bandyopdhyay}(2022)}]{Agrawal2022}%
  \BibitemOpen
  \bibfield  {author} {\bibinfo {author} {\bibfnamefont {Aayushi}\ \bibnamefont
  {Agrawal}}\ and\ \bibinfo {author} {\bibfnamefont {Jayendra}\ \bibnamefont
  {Bandyopdhyay}},\ }\bibfield  {title} {\enquote {\bibinfo {title} {Floquet
  topological phases with high chern numbers in a periodically driven extended
  su-schrieffer-heeger model},}\ }\href {\doibase 10.1088/1361-648X/ac6eac}
  {\bibfield  {journal} {\bibinfo  {journal} {Journal of Physics: Condensed
  Matter}\ }\textbf {\bibinfo {volume} {34}} (\bibinfo {year} {2022}),\
  10.1088/1361-648X/ac6eac}\BibitemShut {NoStop}%
\bibitem [{\citenamefont {Saha}\ \emph {et~al.}(2021)\citenamefont {Saha},
  \citenamefont {Nag},\ and\ \citenamefont {Mandal}}]{Saha21}%
  \BibitemOpen
  \bibfield  {author} {\bibinfo {author} {\bibfnamefont {Sudarshan}\
  \bibnamefont {Saha}}, \bibinfo {author} {\bibfnamefont {Tanay}\ \bibnamefont
  {Nag}}, \ and\ \bibinfo {author} {\bibfnamefont {Saptarshi}\ \bibnamefont
  {Mandal}},\ }\bibfield  {title} {\enquote {\bibinfo {title} {Eightfold
  quantum hall phases in a time reversal symmetry broken tight binding
  model},}\ }\href {\doibase 10.1103/PhysRevB.103.235154} {\bibfield  {journal}
  {\bibinfo  {journal} {Phys. Rev. B}\ }\textbf {\bibinfo {volume} {103}},\
  \bibinfo {pages} {235154} (\bibinfo {year} {2021})}\BibitemShut {NoStop}%
\bibitem [{\citenamefont {Bellissard}(1995)}]{bellissard1995noncommutative}%
  \BibitemOpen
  \bibfield  {author} {\bibinfo {author} {\bibfnamefont {Jean}\ \bibnamefont
  {Bellissard}},\ }\bibfield  {title} {\enquote {\bibinfo {title}
  {Noncommutative geometry and quantum hall effect},}\ }in\ \href@noop {}
  {\emph {\bibinfo {booktitle} {Proceedings of the International Congress of
  Mathematicians: August 3--11, 1994 Z{\"u}rich, Switzerland}}}\ (\bibinfo
  {organization} {Springer},\ \bibinfo {year} {1995})\ pp.\ \bibinfo {pages}
  {1238--1246}\BibitemShut {NoStop}%
\bibitem [{\citenamefont {Prodan}(2010)}]{Prodan_2010}%
  \BibitemOpen
  \bibfield  {author} {\bibinfo {author} {\bibfnamefont {Emil}\ \bibnamefont
  {Prodan}},\ }\bibfield  {title} {\enquote {\bibinfo {title} {Non-commutative
  tools for topological insulators},}\ }\href {\doibase
  10.1088/1367-2630/12/6/065003} {\bibfield  {journal} {\bibinfo  {journal}
  {New Journal of Physics}\ }\textbf {\bibinfo {volume} {12}},\ \bibinfo
  {pages} {065003} (\bibinfo {year} {2010})}\BibitemShut {NoStop}%
\bibitem [{\citenamefont {Prodan}(2011)}]{Prodan_2011}%
  \BibitemOpen
  \bibfield  {author} {\bibinfo {author} {\bibfnamefont {Emil}\ \bibnamefont
  {Prodan}},\ }\bibfield  {title} {\enquote {\bibinfo {title} {Disordered
  topological insulators: a non-commutative geometry perspective},}\ }\href
  {\doibase 10.1088/1751-8113/44/11/113001} {\bibfield  {journal} {\bibinfo
  {journal} {Journal of Physics A: Mathematical and Theoretical}\ }\textbf
  {\bibinfo {volume} {44}},\ \bibinfo {pages} {113001} (\bibinfo {year}
  {2011})}\BibitemShut {NoStop}%
\bibitem [{\citenamefont {Bianco}\ and\ \citenamefont
  {Resta}(2011)}]{Bianco11}%
  \BibitemOpen
  \bibfield  {author} {\bibinfo {author} {\bibfnamefont {Raffaello}\
  \bibnamefont {Bianco}}\ and\ \bibinfo {author} {\bibfnamefont {Raffaele}\
  \bibnamefont {Resta}},\ }\bibfield  {title} {\enquote {\bibinfo {title}
  {Mapping topological order in coordinate space},}\ }\href {\doibase
  10.1103/PhysRevB.84.241106} {\bibfield  {journal} {\bibinfo  {journal} {Phys.
  Rev. B}\ }\textbf {\bibinfo {volume} {84}},\ \bibinfo {pages} {241106}
  (\bibinfo {year} {2011})}\BibitemShut {NoStop}%
\bibitem [{\citenamefont {Loring}\ and\ \citenamefont
  {Hastings}(2011)}]{Loring_2010}%
  \BibitemOpen
  \bibfield  {author} {\bibinfo {author} {\bibfnamefont {T.~A.}\ \bibnamefont
  {Loring}}\ and\ \bibinfo {author} {\bibfnamefont {M.~B.}\ \bibnamefont
  {Hastings}},\ }\bibfield  {title} {\enquote {\bibinfo {title} {Disordered
  topological insulators via c*-algebras},}\ }\href {\doibase
  10.1209/0295-5075/92/67004} {\bibfield  {journal} {\bibinfo  {journal}
  {Europhysics Letters}\ }\textbf {\bibinfo {volume} {92}},\ \bibinfo {pages}
  {67004} (\bibinfo {year} {2011})}\BibitemShut {NoStop}%
\bibitem [{\citenamefont {Hastings}\ and\ \citenamefont
  {Loring}(2010)}]{hastings2010almost}%
  \BibitemOpen
  \bibfield  {author} {\bibinfo {author} {\bibfnamefont {Matthew~B}\
  \bibnamefont {Hastings}}\ and\ \bibinfo {author} {\bibfnamefont {Terry~A}\
  \bibnamefont {Loring}},\ }\bibfield  {title} {\enquote {\bibinfo {title}
  {Almost commuting matrices, localized wannier functions, and the quantum hall
  effect},}\ }\href@noop {} {\bibfield  {journal} {\bibinfo  {journal} {Journal
  of mathematical physics}\ }\textbf {\bibinfo {volume} {51}} (\bibinfo {year}
  {2010})}\BibitemShut {NoStop}%
\bibitem [{\citenamefont {Tan}\ \emph {et~al.}(2019)\citenamefont {Tan},
  \citenamefont {Zhang}, \citenamefont {Yang}, \citenamefont {Chu},
  \citenamefont {Zhu}, \citenamefont {Li}, \citenamefont {Yang}, \citenamefont
  {Song}, \citenamefont {Han}, \citenamefont {Li}, \citenamefont {Dong},
  \citenamefont {Yu}, \citenamefont {Yan}, \citenamefont {Zhu},\ and\
  \citenamefont {Yu}}]{Tan2019}%
  \BibitemOpen
  \bibfield  {author} {\bibinfo {author} {\bibfnamefont {Xinsheng}\
  \bibnamefont {Tan}}, \bibinfo {author} {\bibfnamefont {Dan-Wei}\ \bibnamefont
  {Zhang}}, \bibinfo {author} {\bibfnamefont {Zhen}\ \bibnamefont {Yang}},
  \bibinfo {author} {\bibfnamefont {Ji}~\bibnamefont {Chu}}, \bibinfo {author}
  {\bibfnamefont {Yan-Qing}\ \bibnamefont {Zhu}}, \bibinfo {author}
  {\bibfnamefont {Danyu}\ \bibnamefont {Li}}, \bibinfo {author} {\bibfnamefont
  {Xiaopei}\ \bibnamefont {Yang}}, \bibinfo {author} {\bibfnamefont {Shuqing}\
  \bibnamefont {Song}}, \bibinfo {author} {\bibfnamefont {Zhikun}\ \bibnamefont
  {Han}}, \bibinfo {author} {\bibfnamefont {Zhiyuan}\ \bibnamefont {Li}},
  \bibinfo {author} {\bibfnamefont {Yuqian}\ \bibnamefont {Dong}}, \bibinfo
  {author} {\bibfnamefont {Hai-Feng}\ \bibnamefont {Yu}}, \bibinfo {author}
  {\bibfnamefont {Hui}\ \bibnamefont {Yan}}, \bibinfo {author} {\bibfnamefont
  {Shi-Liang}\ \bibnamefont {Zhu}}, \ and\ \bibinfo {author} {\bibfnamefont
  {Yang}\ \bibnamefont {Yu}},\ }\bibfield  {title} {\enquote {\bibinfo {title}
  {Experimental measurement of the quantum metric tensor and related
  topological phase transition with a superconducting qubit},}\ }\href
  {\doibase 10.1103/PhysRevLett.122.210401} {\bibfield  {journal} {\bibinfo
  {journal} {Phys. Rev. Lett.}\ }\textbf {\bibinfo {volume} {122}},\ \bibinfo
  {pages} {210401} (\bibinfo {year} {2019})}\BibitemShut {NoStop}%
\bibitem [{\citenamefont {Panahiyan}\ \emph {et~al.}(2020)\citenamefont
  {Panahiyan}, \citenamefont {Chen},\ and\ \citenamefont
  {Fritzsche}}]{Panahiyan2020}%
  \BibitemOpen
  \bibfield  {author} {\bibinfo {author} {\bibfnamefont {S.}~\bibnamefont
  {Panahiyan}}, \bibinfo {author} {\bibfnamefont {W.}~\bibnamefont {Chen}}, \
  and\ \bibinfo {author} {\bibfnamefont {S.}~\bibnamefont {Fritzsche}},\
  }\bibfield  {title} {\enquote {\bibinfo {title} {Fidelity susceptibility near
  topological phase transitions in quantum walks},}\ }\href {\doibase
  10.1103/PhysRevB.102.134111} {\bibfield  {journal} {\bibinfo  {journal}
  {Phys. Rev. B}\ }\textbf {\bibinfo {volume} {102}},\ \bibinfo {pages}
  {134111} (\bibinfo {year} {2020})}\BibitemShut {NoStop}%
\bibitem [{\citenamefont {Ma}\ \emph {et~al.}(2013)\citenamefont {Ma},
  \citenamefont {Gu}, \citenamefont {Chen}, \citenamefont {Fan},\ and\
  \citenamefont {Liu}}]{Ma_2013}%
  \BibitemOpen
  \bibfield  {author} {\bibinfo {author} {\bibfnamefont {Yu-Quan}\ \bibnamefont
  {Ma}}, \bibinfo {author} {\bibfnamefont {Shi-Jian}\ \bibnamefont {Gu}},
  \bibinfo {author} {\bibfnamefont {Shu}\ \bibnamefont {Chen}}, \bibinfo
  {author} {\bibfnamefont {Heng}\ \bibnamefont {Fan}}, \ and\ \bibinfo {author}
  {\bibfnamefont {Wu-Ming}\ \bibnamefont {Liu}},\ }\bibfield  {title} {\enquote
  {\bibinfo {title} {The euler number of bloch states manifold and the quantum
  phases in gapped fermionic systems},}\ }\href {\doibase
  10.1209/0295-5075/103/10008} {\bibfield  {journal} {\bibinfo  {journal} {EPL
  (Europhysics Letters)}\ }\textbf {\bibinfo {volume} {103}},\ \bibinfo {pages}
  {10008} (\bibinfo {year} {2013})}\BibitemShut {NoStop}%
\bibitem [{\citenamefont {Ma}(2014)}]{Ma2014}%
  \BibitemOpen
  \bibfield  {author} {\bibinfo {author} {\bibfnamefont {Yu-Quan}\ \bibnamefont
  {Ma}},\ }\bibfield  {title} {\enquote {\bibinfo {title} {Quantum distance and
  the euler number index of the bloch band in a one-dimensional spin model},}\
  }\href {\doibase 10.1103/PhysRevE.90.042133} {\bibfield  {journal} {\bibinfo
  {journal} {Phys. Rev. E}\ }\textbf {\bibinfo {volume} {90}},\ \bibinfo
  {pages} {042133} (\bibinfo {year} {2014})}\BibitemShut {NoStop}%
\bibitem [{\citenamefont {Ozawa}\ and\ \citenamefont {Mera}(2021)}]{Ozawa2021}%
  \BibitemOpen
  \bibfield  {author} {\bibinfo {author} {\bibfnamefont {Tomoki}\ \bibnamefont
  {Ozawa}}\ and\ \bibinfo {author} {\bibfnamefont {Bruno}\ \bibnamefont
  {Mera}},\ }\bibfield  {title} {\enquote {\bibinfo {title} {Relations between
  topology and the quantum metric for chern insulators},}\ }\href {\doibase
  10.1103/PhysRevB.104.045103} {\bibfield  {journal} {\bibinfo  {journal}
  {Phys. Rev. B}\ }\textbf {\bibinfo {volume} {104}},\ \bibinfo {pages}
  {045103} (\bibinfo {year} {2021})}\BibitemShut {NoStop}%
\bibitem [{\citenamefont {Cheng}\ \emph {et~al.}(2024)\citenamefont {Cheng},
  \citenamefont {Batchelor},\ and\ \citenamefont {Cocks}}]{Cheng_2024}%
  \BibitemOpen
  \bibfield  {author} {\bibinfo {author} {\bibfnamefont {Eve}\ \bibnamefont
  {Cheng}}, \bibinfo {author} {\bibfnamefont {Murray~T}\ \bibnamefont
  {Batchelor}}, \ and\ \bibinfo {author} {\bibfnamefont {Danny}\ \bibnamefont
  {Cocks}},\ }\bibfield  {title} {\enquote {\bibinfo {title} {Topological
  analysis of the complex ssh model using the quantum geometric tensor},}\
  }\href {\doibase 10.1088/1751-8121/ad5d2e} {\bibfield  {journal} {\bibinfo
  {journal} {Journal of Physics A: Mathematical and Theoretical}\ }\textbf
  {\bibinfo {volume} {57}},\ \bibinfo {pages} {305001} (\bibinfo {year}
  {2024})}\BibitemShut {NoStop}%
\bibitem [{\citenamefont {Zeng}\ \emph {et~al.}(2024)\citenamefont {Zeng},
  \citenamefont {Lai}, \citenamefont {Wei},\ and\ \citenamefont
  {Ma}}]{Zeng_2024}%
  \BibitemOpen
  \bibfield  {author} {\bibinfo {author} {\bibfnamefont {Xiang-Long}\
  \bibnamefont {Zeng}}, \bibinfo {author} {\bibfnamefont {Wen-Xi}\ \bibnamefont
  {Lai}}, \bibinfo {author} {\bibfnamefont {Yi-Wen}\ \bibnamefont {Wei}}, \
  and\ \bibinfo {author} {\bibfnamefont {Yu-Quan}\ \bibnamefont {Ma}},\
  }\bibfield  {title} {\enquote {\bibinfo {title} {Quantum geometric tensor and
  the topological characterization of the extended su–schrieffer–heeger
  model},}\ }\href {\doibase 10.1088/1674-1056/ad1170} {\bibfield  {journal}
  {\bibinfo  {journal} {Chinese Physics B}\ }\textbf {\bibinfo {volume} {33}},\
  \bibinfo {pages} {030310} (\bibinfo {year} {2024})}\BibitemShut {NoStop}%
\bibitem [{\citenamefont {Zhang}(2022)}]{Zhang_mar}%
  \BibitemOpen
  \bibfield  {author} {\bibinfo {author} {\bibfnamefont {Anwei}\ \bibnamefont
  {Zhang}},\ }\bibfield  {title} {\enquote {\bibinfo {title} {Revealing chern
  number from quantum metric},}\ }\href {\doibase 10.1088/1674-1056/ac2f2c}
  {\bibfield  {journal} {\bibinfo  {journal} {Chinese Physics B}\ }\textbf
  {\bibinfo {volume} {31}},\ \bibinfo {pages} {040201} (\bibinfo {year}
  {2022})}\BibitemShut {NoStop}%
\bibitem [{\citenamefont {Maffei}\ \emph {et~al.}(2018)\citenamefont {Maffei},
  \citenamefont {Dauphin}, \citenamefont {Cardano}, \citenamefont
  {Lewenstein},\ and\ \citenamefont {Massignan}}]{Maffei_2018}%
  \BibitemOpen
  \bibfield  {author} {\bibinfo {author} {\bibfnamefont {Maria}\ \bibnamefont
  {Maffei}}, \bibinfo {author} {\bibfnamefont {Alexandre}\ \bibnamefont
  {Dauphin}}, \bibinfo {author} {\bibfnamefont {Filippo}\ \bibnamefont
  {Cardano}}, \bibinfo {author} {\bibfnamefont {Maciej}\ \bibnamefont
  {Lewenstein}}, \ and\ \bibinfo {author} {\bibfnamefont {Pietro}\ \bibnamefont
  {Massignan}},\ }\bibfield  {title} {\enquote {\bibinfo {title} {Topological
  characterization of chiral models through their long time dynamics},}\ }\href
  {\doibase 10.1088/1367-2630/aa9d4c} {\bibfield  {journal} {\bibinfo
  {journal} {New Journal of Physics}\ }\textbf {\bibinfo {volume} {20}},\
  \bibinfo {pages} {013023} (\bibinfo {year} {2018})}\BibitemShut {NoStop}%
\bibitem [{\citenamefont {Han}\ \emph {et~al.}(2021)\citenamefont {Han},
  \citenamefont {Liu},\ and\ \citenamefont {Liu}}]{Han_2021}%
  \BibitemOpen
  \bibfield  {author} {\bibinfo {author} {\bibfnamefont {Y~Z}\ \bibnamefont
  {Han}}, \bibinfo {author} {\bibfnamefont {J~S}\ \bibnamefont {Liu}}, \ and\
  \bibinfo {author} {\bibfnamefont {C~S}\ \bibnamefont {Liu}},\ }\bibfield
  {title} {\enquote {\bibinfo {title} {The topological counterparts of
  non-hermitian ssh models},}\ }\href {\doibase 10.1088/1367-2630/ac3e9f}
  {\bibfield  {journal} {\bibinfo  {journal} {New Journal of Physics}\ }\textbf
  {\bibinfo {volume} {23}},\ \bibinfo {pages} {123029} (\bibinfo {year}
  {2021})}\BibitemShut {NoStop}%
\bibitem [{\citenamefont {Schobert}\ \emph {et~al.}(2021)\citenamefont
  {Schobert}, \citenamefont {Berges}, \citenamefont {Wehling},\ and\
  \citenamefont {van Loon}}]{Schobert21}%
  \BibitemOpen
  \bibfield  {author} {\bibinfo {author} {\bibfnamefont {Arne}\ \bibnamefont
  {Schobert}}, \bibinfo {author} {\bibfnamefont {Jan}\ \bibnamefont {Berges}},
  \bibinfo {author} {\bibfnamefont {Tim}\ \bibnamefont {Wehling}}, \ and\
  \bibinfo {author} {\bibfnamefont {Erik}\ \bibnamefont {van Loon}},\
  }\bibfield  {title} {\enquote {\bibinfo {title} {{Downfolding the
  Su-Schrieffer-Heeger model}},}\ }\href {\doibase
  10.21468/SciPostPhys.11.4.079} {\bibfield  {journal} {\bibinfo  {journal}
  {SciPost Phys.}\ }\textbf {\bibinfo {volume} {11}},\ \bibinfo {pages} {079}
  (\bibinfo {year} {2021})}\BibitemShut {NoStop}%
\bibitem [{\citenamefont {Padhan}\ \emph {et~al.}(2024)\citenamefont {Padhan},
  \citenamefont {Mondal}, \citenamefont {Vishveshwara},\ and\ \citenamefont
  {Mishra}}]{Padhan24}%
  \BibitemOpen
  \bibfield  {author} {\bibinfo {author} {\bibfnamefont {Ashirbad}\
  \bibnamefont {Padhan}}, \bibinfo {author} {\bibfnamefont {Suman}\
  \bibnamefont {Mondal}}, \bibinfo {author} {\bibfnamefont {Smitha}\
  \bibnamefont {Vishveshwara}}, \ and\ \bibinfo {author} {\bibfnamefont
  {Tapan}\ \bibnamefont {Mishra}},\ }\bibfield  {title} {\enquote {\bibinfo
  {title} {Interacting bosons on a su-schrieffer-heeger ladder: Topological
  phases and thouless pumping},}\ }\href {\doibase 10.1103/PhysRevB.109.085120}
  {\bibfield  {journal} {\bibinfo  {journal} {Phys. Rev. B}\ }\textbf {\bibinfo
  {volume} {109}},\ \bibinfo {pages} {085120} (\bibinfo {year}
  {2024})}\BibitemShut {NoStop}%
\bibitem [{\citenamefont {Feng}\ \emph {et~al.}(2022)\citenamefont {Feng},
  \citenamefont {Xing}, \citenamefont {Poletti}, \citenamefont {Scalettar},\
  and\ \citenamefont {Batrouni}}]{Feng2022}%
  \BibitemOpen
  \bibfield  {author} {\bibinfo {author} {\bibfnamefont {Chunhan}\ \bibnamefont
  {Feng}}, \bibinfo {author} {\bibfnamefont {Bo}~\bibnamefont {Xing}}, \bibinfo
  {author} {\bibfnamefont {Dario}\ \bibnamefont {Poletti}}, \bibinfo {author}
  {\bibfnamefont {Richard}\ \bibnamefont {Scalettar}}, \ and\ \bibinfo {author}
  {\bibfnamefont {George}\ \bibnamefont {Batrouni}},\ }\bibfield  {title}
  {\enquote {\bibinfo {title} {Phase diagram of the
  su-schrieffer-heeger-hubbard model on a square lattice},}\ }\href {\doibase
  10.1103/PhysRevB.106.L081114} {\bibfield  {journal} {\bibinfo  {journal}
  {Phys. Rev. B}\ }\textbf {\bibinfo {volume} {106}},\ \bibinfo {pages}
  {L081114} (\bibinfo {year} {2022})}\BibitemShut {NoStop}%
\bibitem [{\citenamefont {Jin}\ \emph {et~al.}(2023)\citenamefont {Jin},
  \citenamefont {Ruggiero},\ and\ \citenamefont {Giamarchi}}]{Jin23}%
  \BibitemOpen
  \bibfield  {author} {\bibinfo {author} {\bibfnamefont {Tony}\ \bibnamefont
  {Jin}}, \bibinfo {author} {\bibfnamefont {Paola}\ \bibnamefont {Ruggiero}}, \
  and\ \bibinfo {author} {\bibfnamefont {Thierry}\ \bibnamefont {Giamarchi}},\
  }\bibfield  {title} {\enquote {\bibinfo {title} {Bosonization of the
  interacting su-schrieffer-heeger model},}\ }\href {\doibase
  10.1103/PhysRevB.107.L201111} {\bibfield  {journal} {\bibinfo  {journal}
  {Phys. Rev. B}\ }\textbf {\bibinfo {volume} {107}},\ \bibinfo {pages}
  {L201111} (\bibinfo {year} {2023})}\BibitemShut {NoStop}%
\bibitem [{\citenamefont {Obana}\ \emph {et~al.}(2019)\citenamefont {Obana},
  \citenamefont {Liu},\ and\ \citenamefont {Wakabayashi}}]{Obana19}%
  \BibitemOpen
  \bibfield  {author} {\bibinfo {author} {\bibfnamefont {Daichi}\ \bibnamefont
  {Obana}}, \bibinfo {author} {\bibfnamefont {Feng}\ \bibnamefont {Liu}}, \
  and\ \bibinfo {author} {\bibfnamefont {Katsunori}\ \bibnamefont
  {Wakabayashi}},\ }\bibfield  {title} {\enquote {\bibinfo {title} {Topological
  edge states in the su-schrieffer-heeger model},}\ }\href {\doibase
  10.1103/PhysRevB.100.075437} {\bibfield  {journal} {\bibinfo  {journal}
  {Phys. Rev. B}\ }\textbf {\bibinfo {volume} {100}},\ \bibinfo {pages}
  {075437} (\bibinfo {year} {2019})}\BibitemShut {NoStop}%
\bibitem [{\citenamefont {Lee}\ \emph {et~al.}(2022)\citenamefont {Lee},
  \citenamefont {Io},\ and\ \citenamefont {chung Kao}}]{LEE202296}%
  \BibitemOpen
  \bibfield  {author} {\bibinfo {author} {\bibfnamefont {Chen-Shen}\
  \bibnamefont {Lee}}, \bibinfo {author} {\bibfnamefont {Iao-Fai}\ \bibnamefont
  {Io}}, \ and\ \bibinfo {author} {\bibfnamefont {Hsien}\ \bibnamefont {chung
  Kao}},\ }\bibfield  {title} {\enquote {\bibinfo {title} {Winding number and
  zak phase in multi-band ssh models},}\ }\href {\doibase
  https://doi.org/10.1016/j.cjph.2022.05.007} {\bibfield  {journal} {\bibinfo
  {journal} {Chinese Journal of Physics}\ }\textbf {\bibinfo {volume} {78}},\
  \bibinfo {pages} {96--110} (\bibinfo {year} {2022})}\BibitemShut {NoStop}%
\bibitem [{\citenamefont {Pérez-González}\ \emph {et~al.}(2018)\citenamefont
  {Pérez-González}, \citenamefont {Bello}, \citenamefont {Álvaro
  Gómez-León},\ and\ \citenamefont {Platero}}]{Beatriz2018}%
  \BibitemOpen
  \bibfield  {author} {\bibinfo {author} {\bibfnamefont {Beatriz}\ \bibnamefont
  {Pérez-González}}, \bibinfo {author} {\bibfnamefont {Miguel}\ \bibnamefont
  {Bello}}, \bibinfo {author} {\bibnamefont {Álvaro Gómez-León}}, \ and\
  \bibinfo {author} {\bibfnamefont {Gloria}\ \bibnamefont {Platero}},\ }\href
  {https://arxiv.org/abs/1802.03973} {\enquote {\bibinfo {title} {Ssh model
  with long-range hoppings: topology, driving and disorder},}\ } (\bibinfo
  {year} {2018}),\ \Eprint {http://arxiv.org/abs/1802.03973} {arXiv:1802.03973
  [cond-mat.mes-hall]} \BibitemShut {NoStop}%
\bibitem [{\citenamefont {Huang}\ and\ \citenamefont
  {Liu}(2018{\natexlab{a}})}]{Huang18}%
  \BibitemOpen
  \bibfield  {author} {\bibinfo {author} {\bibfnamefont {Huaqing}\ \bibnamefont
  {Huang}}\ and\ \bibinfo {author} {\bibfnamefont {Feng}\ \bibnamefont {Liu}},\
  }\bibfield  {title} {\enquote {\bibinfo {title} {Theory of spin bott index
  for quantum spin hall states in nonperiodic systems},}\ }\href {\doibase
  10.1103/PhysRevB.98.125130} {\bibfield  {journal} {\bibinfo  {journal} {Phys.
  Rev. B}\ }\textbf {\bibinfo {volume} {98}},\ \bibinfo {pages} {125130}
  (\bibinfo {year} {2018}{\natexlab{a}})}\BibitemShut {NoStop}%
\bibitem [{\citenamefont {Huang}\ and\ \citenamefont
  {Liu}(2018{\natexlab{b}})}]{Huang18b}%
  \BibitemOpen
  \bibfield  {author} {\bibinfo {author} {\bibfnamefont {Huaqing}\ \bibnamefont
  {Huang}}\ and\ \bibinfo {author} {\bibfnamefont {Feng}\ \bibnamefont {Liu}},\
  }\bibfield  {title} {\enquote {\bibinfo {title} {Quantum spin hall effect and
  spin bott index in a quasicrystal lattice},}\ }\href {\doibase
  10.1103/PhysRevLett.121.126401} {\bibfield  {journal} {\bibinfo  {journal}
  {Phys. Rev. Lett.}\ }\textbf {\bibinfo {volume} {121}},\ \bibinfo {pages}
  {126401} (\bibinfo {year} {2018}{\natexlab{b}})}\BibitemShut {NoStop}%
\bibitem [{\citenamefont {Yoshii}\ \emph {et~al.}(2021)\citenamefont {Yoshii},
  \citenamefont {Kitamura},\ and\ \citenamefont {Morimoto}}]{Yoshii21}%
  \BibitemOpen
  \bibfield  {author} {\bibinfo {author} {\bibfnamefont {Mao}\ \bibnamefont
  {Yoshii}}, \bibinfo {author} {\bibfnamefont {Sota}\ \bibnamefont {Kitamura}},
  \ and\ \bibinfo {author} {\bibfnamefont {Takahiro}\ \bibnamefont
  {Morimoto}},\ }\bibfield  {title} {\enquote {\bibinfo {title} {Topological
  charge pumping in quasiperiodic systems characterized by the bott index},}\
  }\href {\doibase 10.1103/PhysRevB.104.155126} {\bibfield  {journal} {\bibinfo
   {journal} {Phys. Rev. B}\ }\textbf {\bibinfo {volume} {104}},\ \bibinfo
  {pages} {155126} (\bibinfo {year} {2021})}\BibitemShut {NoStop}%
\bibitem [{\citenamefont {Toniolo}(2022)}]{toniolo2022bott}%
  \BibitemOpen
  \bibfield  {author} {\bibinfo {author} {\bibfnamefont {Daniele}\ \bibnamefont
  {Toniolo}},\ }\bibfield  {title} {\enquote {\bibinfo {title} {On the bott
  index of unitary matrices on a finite torus},}\ }\href@noop {} {\bibfield
  {journal} {\bibinfo  {journal} {Letters in Mathematical Physics}\ }\textbf
  {\bibinfo {volume} {112}},\ \bibinfo {pages} {126} (\bibinfo {year}
  {2022})}\BibitemShut {NoStop}%
\bibitem [{\citenamefont {Zeng}\ \emph {et~al.}(2020)\citenamefont {Zeng},
  \citenamefont {Yang},\ and\ \citenamefont {Xu}}]{Zeng2020}%
  \BibitemOpen
  \bibfield  {author} {\bibinfo {author} {\bibfnamefont {Qi-Bo}\ \bibnamefont
  {Zeng}}, \bibinfo {author} {\bibfnamefont {Yan-Bin}\ \bibnamefont {Yang}}, \
  and\ \bibinfo {author} {\bibfnamefont {Yong}\ \bibnamefont {Xu}},\ }\bibfield
   {title} {\enquote {\bibinfo {title} {Topological phases in non-hermitian
  aubry-andr\'e-harper models},}\ }\href {\doibase 10.1103/PhysRevB.101.020201}
  {\bibfield  {journal} {\bibinfo  {journal} {Phys. Rev. B}\ }\textbf {\bibinfo
  {volume} {101}},\ \bibinfo {pages} {020201} (\bibinfo {year}
  {2020})}\BibitemShut {NoStop}%
\bibitem [{\citenamefont {Ghosh}\ \emph {et~al.}(2022)\citenamefont {Ghosh},
  \citenamefont {Nag},\ and\ \citenamefont {Saha}}]{Ghosh22_dyn}%
  \BibitemOpen
  \bibfield  {author} {\bibinfo {author} {\bibfnamefont {Arnob~Kumar}\
  \bibnamefont {Ghosh}}, \bibinfo {author} {\bibfnamefont {Tanay}\ \bibnamefont
  {Nag}}, \ and\ \bibinfo {author} {\bibfnamefont {Arijit}\ \bibnamefont
  {Saha}},\ }\bibfield  {title} {\enquote {\bibinfo {title} {Dynamical
  construction of quadrupolar and octupolar topological superconductors},}\
  }\href {\doibase 10.1103/PhysRevB.105.155406} {\bibfield  {journal} {\bibinfo
   {journal} {Phys. Rev. B}\ }\textbf {\bibinfo {volume} {105}},\ \bibinfo
  {pages} {155406} (\bibinfo {year} {2022})}\BibitemShut {NoStop}%
\bibitem [{\citenamefont {Mondal}\ \emph {et~al.}(2023)\citenamefont {Mondal},
  \citenamefont {Ghosh}, \citenamefont {Nag},\ and\ \citenamefont
  {Saha}}]{Mondal23_dyn}%
  \BibitemOpen
  \bibfield  {author} {\bibinfo {author} {\bibfnamefont {Debashish}\
  \bibnamefont {Mondal}}, \bibinfo {author} {\bibfnamefont {Arnob~Kumar}\
  \bibnamefont {Ghosh}}, \bibinfo {author} {\bibfnamefont {Tanay}\ \bibnamefont
  {Nag}}, \ and\ \bibinfo {author} {\bibfnamefont {Arijit}\ \bibnamefont
  {Saha}},\ }\bibfield  {title} {\enquote {\bibinfo {title} {Topological
  characterization and stability of floquet majorana modes in rashba
  nanowires},}\ }\href {\doibase 10.1103/PhysRevB.107.035427} {\bibfield
  {journal} {\bibinfo  {journal} {Phys. Rev. B}\ }\textbf {\bibinfo {volume}
  {107}},\ \bibinfo {pages} {035427} (\bibinfo {year} {2023})}\BibitemShut
  {NoStop}%
\bibitem [{\citenamefont {Rodriguez-Vega}\ \emph {et~al.}(2018)\citenamefont
  {Rodriguez-Vega}, \citenamefont {Lentz},\ and\ \citenamefont
  {Seradjeh}}]{Rodriguez-Vega_2018}%
  \BibitemOpen
  \bibfield  {author} {\bibinfo {author} {\bibfnamefont {M}~\bibnamefont
  {Rodriguez-Vega}}, \bibinfo {author} {\bibfnamefont {M}~\bibnamefont
  {Lentz}}, \ and\ \bibinfo {author} {\bibfnamefont {B}~\bibnamefont
  {Seradjeh}},\ }\bibfield  {title} {\enquote {\bibinfo {title} {Floquet
  perturbation theory: formalism and application to low-frequency limit},}\
  }\href {\doibase 10.1088/1367-2630/aade37} {\bibfield  {journal} {\bibinfo
  {journal} {New Journal of Physics}\ }\textbf {\bibinfo {volume} {20}},\
  \bibinfo {pages} {093022} (\bibinfo {year} {2018})}\BibitemShut {NoStop}%
\bibitem [{\citenamefont {Eckardt}\ and\ \citenamefont
  {Anisimovas}(2015)}]{Eckardt_2015}%
  \BibitemOpen
  \bibfield  {author} {\bibinfo {author} {\bibfnamefont {André}\ \bibnamefont
  {Eckardt}}\ and\ \bibinfo {author} {\bibfnamefont {Egidijus}\ \bibnamefont
  {Anisimovas}},\ }\bibfield  {title} {\enquote {\bibinfo {title}
  {High-frequency approximation for periodically driven quantum systems from a
  floquet-space perspective},}\ }\href {\doibase 10.1088/1367-2630/17/9/093039}
  {\bibfield  {journal} {\bibinfo  {journal} {New Journal of Physics}\ }\textbf
  {\bibinfo {volume} {17}},\ \bibinfo {pages} {093039} (\bibinfo {year}
  {2015})}\BibitemShut {NoStop}%
\bibitem [{\citenamefont {Nag}\ \emph {et~al.}(2015)\citenamefont {Nag},
  \citenamefont {Sen},\ and\ \citenamefont {Dutta}}]{Nag15}%
  \BibitemOpen
  \bibfield  {author} {\bibinfo {author} {\bibfnamefont {Tanay}\ \bibnamefont
  {Nag}}, \bibinfo {author} {\bibfnamefont {Diptiman}\ \bibnamefont {Sen}}, \
  and\ \bibinfo {author} {\bibfnamefont {Amit}\ \bibnamefont {Dutta}},\
  }\bibfield  {title} {\enquote {\bibinfo {title} {Maximum group velocity in a
  one-dimensional model with a sinusoidally varying staggered potential},}\
  }\href {\doibase 10.1103/PhysRevA.91.063607} {\bibfield  {journal} {\bibinfo
  {journal} {Phys. Rev. A}\ }\textbf {\bibinfo {volume} {91}},\ \bibinfo
  {pages} {063607} (\bibinfo {year} {2015})}\BibitemShut {NoStop}%
\bibitem [{\citenamefont {Ghosh}\ \emph {et~al.}(2020)\citenamefont {Ghosh},
  \citenamefont {Mukherjee},\ and\ \citenamefont {Sengupta}}]{Ghosh20_Flo}%
  \BibitemOpen
  \bibfield  {author} {\bibinfo {author} {\bibfnamefont {Roopayan}\
  \bibnamefont {Ghosh}}, \bibinfo {author} {\bibfnamefont {Bhaskar}\
  \bibnamefont {Mukherjee}}, \ and\ \bibinfo {author} {\bibfnamefont
  {K.}~\bibnamefont {Sengupta}},\ }\bibfield  {title} {\enquote {\bibinfo
  {title} {Floquet perturbation theory for periodically driven weakly
  interacting fermions},}\ }\href {\doibase 10.1103/PhysRevB.102.235114}
  {\bibfield  {journal} {\bibinfo  {journal} {Phys. Rev. B}\ }\textbf {\bibinfo
  {volume} {102}},\ \bibinfo {pages} {235114} (\bibinfo {year}
  {2020})}\BibitemShut {NoStop}%
\bibitem [{\citenamefont {Mikami}\ \emph {et~al.}(2016)\citenamefont {Mikami},
  \citenamefont {Kitamura}, \citenamefont {Yasuda}, \citenamefont {Tsuji},
  \citenamefont {Oka},\ and\ \citenamefont {Aoki}}]{Mikami16}%
  \BibitemOpen
  \bibfield  {author} {\bibinfo {author} {\bibfnamefont {Takahiro}\
  \bibnamefont {Mikami}}, \bibinfo {author} {\bibfnamefont {Sota}\ \bibnamefont
  {Kitamura}}, \bibinfo {author} {\bibfnamefont {Kenji}\ \bibnamefont
  {Yasuda}}, \bibinfo {author} {\bibfnamefont {Naoto}\ \bibnamefont {Tsuji}},
  \bibinfo {author} {\bibfnamefont {Takashi}\ \bibnamefont {Oka}}, \ and\
  \bibinfo {author} {\bibfnamefont {Hideo}\ \bibnamefont {Aoki}},\ }\bibfield
  {title} {\enquote {\bibinfo {title} {Brillouin-wigner theory for
  high-frequency expansion in periodically driven systems: Application to
  floquet topological insulators},}\ }\href {\doibase
  10.1103/PhysRevB.93.144307} {\bibfield  {journal} {\bibinfo  {journal} {Phys.
  Rev. B}\ }\textbf {\bibinfo {volume} {93}},\ \bibinfo {pages} {144307}
  (\bibinfo {year} {2016})}\BibitemShut {NoStop}%
\bibitem [{\citenamefont {Rudner}\ \emph {et~al.}(2013)\citenamefont {Rudner},
  \citenamefont {Lindner}, \citenamefont {Berg},\ and\ \citenamefont
  {Levin}}]{Rudner13}%
  \BibitemOpen
  \bibfield  {author} {\bibinfo {author} {\bibfnamefont {Mark~S.}\ \bibnamefont
  {Rudner}}, \bibinfo {author} {\bibfnamefont {Netanel~H.}\ \bibnamefont
  {Lindner}}, \bibinfo {author} {\bibfnamefont {Erez}\ \bibnamefont {Berg}}, \
  and\ \bibinfo {author} {\bibfnamefont {Michael}\ \bibnamefont {Levin}},\
  }\bibfield  {title} {\enquote {\bibinfo {title} {Anomalous edge states and
  the bulk-edge correspondence for periodically driven two-dimensional
  systems},}\ }\href {\doibase 10.1103/PhysRevX.3.031005} {\bibfield  {journal}
  {\bibinfo  {journal} {Phys. Rev. X}\ }\textbf {\bibinfo {volume} {3}},\
  \bibinfo {pages} {031005} (\bibinfo {year} {2013})}\BibitemShut {NoStop}%
\bibitem [{\citenamefont {Do~Carmo}(2016)}]{do2016differential}%
  \BibitemOpen
  \bibfield  {author} {\bibinfo {author} {\bibfnamefont {Manfredo~P}\
  \bibnamefont {Do~Carmo}},\ }\href@noop {} {\emph {\bibinfo {title}
  {Differential geometry of curves and surfaces: revised and updated second
  edition}}}\ (\bibinfo  {publisher} {Courier Dover Publications},\ \bibinfo
  {year} {2016})\BibitemShut {NoStop}%
\bibitem [{\citenamefont {Lee}(2006)}]{lee2006riemannian}%
  \BibitemOpen
  \bibfield  {author} {\bibinfo {author} {\bibfnamefont {John~M}\ \bibnamefont
  {Lee}},\ }\href@noop {} {\emph {\bibinfo {title} {Riemannian manifolds: an
  introduction to curvature}}},\ Vol.\ \bibinfo {volume} {176}\ (\bibinfo
  {publisher} {Springer Science \& Business Media},\ \bibinfo {year}
  {2006})\BibitemShut {NoStop}%
\bibitem [{\citenamefont {Gu}(2010)}]{gu2010fidelity}%
  \BibitemOpen
  \bibfield  {author} {\bibinfo {author} {\bibfnamefont {Shi-Jian}\
  \bibnamefont {Gu}},\ }\bibfield  {title} {\enquote {\bibinfo {title}
  {Fidelity approach to quantum phase transitions},}\ }\href@noop {} {\bibfield
   {journal} {\bibinfo  {journal} {International Journal of Modern Physics B}\
  }\textbf {\bibinfo {volume} {24}},\ \bibinfo {pages} {4371--4458} (\bibinfo
  {year} {2010})}\BibitemShut {NoStop}%
\bibitem [{\citenamefont {Thatcher}\ \emph {et~al.}(2022)\citenamefont
  {Thatcher}, \citenamefont {Fairfield}, \citenamefont {Merlo-Ramírez},\ and\
  \citenamefont {Merlo}}]{Thatcher_2022}%
  \BibitemOpen
  \bibfield  {author} {\bibinfo {author} {\bibfnamefont {Luke}\ \bibnamefont
  {Thatcher}}, \bibinfo {author} {\bibfnamefont {Parker}\ \bibnamefont
  {Fairfield}}, \bibinfo {author} {\bibfnamefont {Lázaro}\ \bibnamefont
  {Merlo-Ramírez}}, \ and\ \bibinfo {author} {\bibfnamefont {Juan~M}\
  \bibnamefont {Merlo}},\ }\bibfield  {title} {\enquote {\bibinfo {title}
  {Experimental observation of topological phase transitions in a mechanical
  1d-ssh model},}\ }\href {\doibase 10.1088/1402-4896/ac4ed2} {\bibfield
  {journal} {\bibinfo  {journal} {Physica Scripta}\ }\textbf {\bibinfo {volume}
  {97}},\ \bibinfo {pages} {035702} (\bibinfo {year} {2022})}\BibitemShut
  {NoStop}%
\bibitem [{\citenamefont {Xie}\ \emph {et~al.}(2019)\citenamefont {Xie},
  \citenamefont {Gou}, \citenamefont {Xiao}, \citenamefont {Gadway},\ and\
  \citenamefont {Yan}}]{xie2019topological}%
  \BibitemOpen
  \bibfield  {author} {\bibinfo {author} {\bibfnamefont {Dizhou}\ \bibnamefont
  {Xie}}, \bibinfo {author} {\bibfnamefont {Wei}\ \bibnamefont {Gou}}, \bibinfo
  {author} {\bibfnamefont {Teng}\ \bibnamefont {Xiao}}, \bibinfo {author}
  {\bibfnamefont {Bryce}\ \bibnamefont {Gadway}}, \ and\ \bibinfo {author}
  {\bibfnamefont {Bo}~\bibnamefont {Yan}},\ }\bibfield  {title} {\enquote
  {\bibinfo {title} {Topological characterizations of an extended
  su--schrieffer--heeger model},}\ }\href@noop {} {\bibfield  {journal}
  {\bibinfo  {journal} {npj Quantum Information}\ }\textbf {\bibinfo {volume}
  {5}},\ \bibinfo {pages} {55} (\bibinfo {year} {2019})}\BibitemShut {NoStop}%
\bibitem [{\citenamefont {Meier}\ \emph
  {et~al.}(2016{\natexlab{b}})\citenamefont {Meier}, \citenamefont {An},\ and\
  \citenamefont {Gadway}}]{meier2016observation}%
  \BibitemOpen
  \bibfield  {author} {\bibinfo {author} {\bibfnamefont {Eric~J}\ \bibnamefont
  {Meier}}, \bibinfo {author} {\bibfnamefont {Fangzhao~Alex}\ \bibnamefont
  {An}}, \ and\ \bibinfo {author} {\bibfnamefont {Bryce}\ \bibnamefont
  {Gadway}},\ }\bibfield  {title} {\enquote {\bibinfo {title} {Observation of
  the topological soliton state in the su--schrieffer--heeger model},}\
  }\href@noop {} {\bibfield  {journal} {\bibinfo  {journal} {Nature
  communications}\ }\textbf {\bibinfo {volume} {7}},\ \bibinfo {pages} {13986}
  (\bibinfo {year} {2016}{\natexlab{b}})}\BibitemShut {NoStop}%
\bibitem [{\citenamefont {C\'aceres-Aravena}\ \emph {et~al.}(2022)\citenamefont
  {C\'aceres-Aravena}, \citenamefont {Real}, \citenamefont {Guzm\'an-Silva},
  \citenamefont {Amo}, \citenamefont {Foa~Torres},\ and\ \citenamefont
  {Vicencio}}]{Gabriel2022}%
  \BibitemOpen
  \bibfield  {author} {\bibinfo {author} {\bibfnamefont {Gabriel}\ \bibnamefont
  {C\'aceres-Aravena}}, \bibinfo {author} {\bibfnamefont {Basti\'an}\
  \bibnamefont {Real}}, \bibinfo {author} {\bibfnamefont {Diego}\ \bibnamefont
  {Guzm\'an-Silva}}, \bibinfo {author} {\bibfnamefont {Alberto}\ \bibnamefont
  {Amo}}, \bibinfo {author} {\bibfnamefont {Luis E.~F.}\ \bibnamefont
  {Foa~Torres}}, \ and\ \bibinfo {author} {\bibfnamefont {Rodrigo~A.}\
  \bibnamefont {Vicencio}},\ }\bibfield  {title} {\enquote {\bibinfo {title}
  {Experimental observation of edge states in ssh-stub photonic lattices},}\
  }\href {\doibase 10.1103/PhysRevResearch.4.013185} {\bibfield  {journal}
  {\bibinfo  {journal} {Phys. Rev. Res.}\ }\textbf {\bibinfo {volume} {4}},\
  \bibinfo {pages} {013185} (\bibinfo {year} {2022})}\BibitemShut {NoStop}%
\bibitem [{\citenamefont {Peng}\ \emph {et~al.}(2016)\citenamefont {Peng},
  \citenamefont {Qin}, \citenamefont {Zhao}, \citenamefont {Shen},
  \citenamefont {Xu}, \citenamefont {Bao}, \citenamefont {Jia},\ and\
  \citenamefont {Zhu}}]{peng2016experimental}%
  \BibitemOpen
  \bibfield  {author} {\bibinfo {author} {\bibfnamefont {Yu-Gui}\ \bibnamefont
  {Peng}}, \bibinfo {author} {\bibfnamefont {Cheng-Zhi}\ \bibnamefont {Qin}},
  \bibinfo {author} {\bibfnamefont {De-Gang}\ \bibnamefont {Zhao}}, \bibinfo
  {author} {\bibfnamefont {Ya-Xi}\ \bibnamefont {Shen}}, \bibinfo {author}
  {\bibfnamefont {Xiang-Yuan}\ \bibnamefont {Xu}}, \bibinfo {author}
  {\bibfnamefont {Ming}\ \bibnamefont {Bao}}, \bibinfo {author} {\bibfnamefont
  {Han}\ \bibnamefont {Jia}}, \ and\ \bibinfo {author} {\bibfnamefont
  {Xue-Feng}\ \bibnamefont {Zhu}},\ }\bibfield  {title} {\enquote {\bibinfo
  {title} {Experimental demonstration of anomalous floquet topological
  insulator for sound},}\ }\href@noop {} {\bibfield  {journal} {\bibinfo
  {journal} {Nature communications}\ }\textbf {\bibinfo {volume} {7}},\
  \bibinfo {pages} {13368} (\bibinfo {year} {2016})}\BibitemShut {NoStop}%
\end{thebibliography}
%\bibliographystyle{apalike}

\end{document}